\documentclass[aps,pre,reprint,showpacs,groupedaddress,letterpaper,floatfix]{revtex4-1}
\usepackage{graphicx}
\usepackage{natbib}
\usepackage{amssymb}
\usepackage{amsmath}
\usepackage{xcolor}
\usepackage{epstopdf}
\newtheorem{theorem}{Theorem}

\newcommand{\eps}{\varepsilon}
\newcommand{\R}{\mathbb{R}}
\def\I{\mathbb{I}}
\def\pa{\partial}
\renewcommand{\vec}[1]{\mbox{\boldmath$#1$}}

\renewcommand{\d}{\mathrm d}

\begin{document}

% Invoke the following to get "accepted version"
\vspace*{-2cm}
\noindent
\textcolor{red}{\bf Submitted to Phys Rev Fluids (1 April 2016)}
%

%\preprint{}

\title{Transport between two fluids across their mutual flow interface: the streakline approach}


\author{Sanjeeva Balasuriya}
\email[]{sanjeevabalasuriya@yahoo.com}
%\thanks{}
\affiliation{School of Mathematical Sciences, University of Adelaide, SA 5005, Australia}


\date{\today}

%%%%%%%%%%%%%%%%%%%%%%%%%%%%%%%%%%%%%%%%%%%%%%%%%

\begin{abstract}
Mixing between two different miscible fluids with a mutual interface must be initiated by fluid transporting across this
fluid interface, caused for example by applying an unsteady velocity agitation.  In general, there is no necessity for this physical flow barrier between the fluids to  be associated
with extremal or exponential attraction as might be revealed by applying Lagrangian coherent structures, finite-time
Lyapunov exponents or other methods on the fluid velocity.   It is shown that streaklines are key to understanding the breaking of 
the interface under velocity agitations, and a theory for locating the relevant streaklines is presented.  Simulations of streaklines in a cross-channel
mixer and a perturbed Kirchhoff's elliptic vortex are quantitatively compared to the theoretical results.  A methodology for
quantifying the unsteady advective transport between the two fluids using streaklines is presented.
\end{abstract}

%\keywords{}

\maketitle
      
%%%%%%%%%%%%%%%%%%%%%%%%%%%%%%%%%%%%%%%%%%%%%%%%%

\section{Introduction}
\label{sec:intro}

If present in a steady nonchaotic flow, coherent blobs of two miscible fluids separated by a streamline
will tend to mix together via the typically inefficient mechanism of diffusion.  This is a
common situation in microfluidics, in which a sample and a reagent are to be mixed in order to achieve a biochemical reaction
in, say, a DNA synthesis experiment, and in which low Reynolds numbers are inevitable due to spatial dimensions and typical
velocity scales.  Accelerating the mixing can be achieved by introducing unsteady velocity agitations to impart advective transport
across the flow interface.  If this process results in fluid filamentation across/near the interface, it will enhance diffusive
mixing  in addition to causing advective intermingling between the two fluids.  Understanding this process, and being able
to quantify resulting fluid mixing, is important in flows ranging from geophysical to microfluidic, for example in assessing
how an introduced pollutant blob mixes with exterior fluid in the ocean, or how a sample and a reagent can be mixed together
effectively in micro- or nano-level bioreactors.

The role of  Lagrangian `unsteady flow barriers' in separating regions of fluid which move coherently is now well-established \cite{hallerreview,peacockfroylandhaller,siam_book}.  Ideas for these
arose from the concepts of stable and unstable manifolds for stagnation points in steady flows, which in many models can be
shown to explicitly demarcate regions which have different Lagrangian flow characteristics (see Fig.~1 in each of \cite{open,microfluid_review,eddy,siam_book}, for steady examples; these arguments have been shown to work in unsteady
flows as well \cite{kelley}.).  Thus, for example, the outer boundary  of an oceanic eddy in such a model
would contain one or more stagnation points, each of whose stable manifolds connects up as the unstable manifold of another
(or the same) stagnation point.  This is necessary to ensure a different flow topology inside the eddy/vortex from the outside.
When a stable manifold coincides with an unstable manifold it is called a {\em heteroclinic manifold} \cite{guckenheimerholmes}.
These entities satisfy a variety of other interesting characteristics, including (i) being associated with curves/surfaces of maximal attraction or
repulsion, (ii) blobs on them eventually expanding exponentially in some directions while contracting exponentially in 
complementary directions, and
(iii) being transport barriers in the sense that the fluid on the two sides remains `almost' coherent under the flow. 
By targetting exactly these characteristics in {\em unsteady} flows defined only for {\em finite} times, obtained either from
numerical solutions of governing equations or directly from experimental/observational velocity data, researchers from many 
fields attempt to determine {\em unsteady flow barriers}.  For the three characteristics mentioned, the relevant methods are respectively
(i) hyperbolic Lagrangian coherent structures \cite{hallerreview,kelley,halleryuan,karraschfarazmandhaller,onu}, (ii) finite-time Lyapunov exponent ridges \cite{shadden,he,huntley,nelsonjacobs,johnsonmeneveau,bozorgmaghamross,branickiwiggins}, and (iii) eigenvectors
associated with the Perron-Frobenius (transfer) operator \cite{froylandpadberg,froylanddetection,froylandchaos}.  (It must be noted that the term `Lagrangian Coherent 
Structures' is often used for {\em all} these techniques, and many more, to highlight the fact that these are based on {\em Lagrangian} evolution
of particles; the idea is to determine entities which
separate fluid blobs which move in some coherent fashion in a velocity field.)  Direct stable/unstable manifold definitions
for unsteady flows can also be used by appropriately extending to infinite times \cite{eigenvector,branickiwiggins,delcastillonegrete}.  These methods {\em all} 
identify the flow barriers purely by examining the velocity field, i.e., they determine curves/surfaces arising from analysis
of the Lagrangian flow equation $ \dot{\vec{x}} = \vec{u}(\vec{x},t) $, where $ \vec{u}(\vec{x},t) $ is the (potentially unsteady)
velocity field, known either explicitly or from discrete data.

\begin{figure}[t]
\includegraphics[width=0.42 \textwidth]{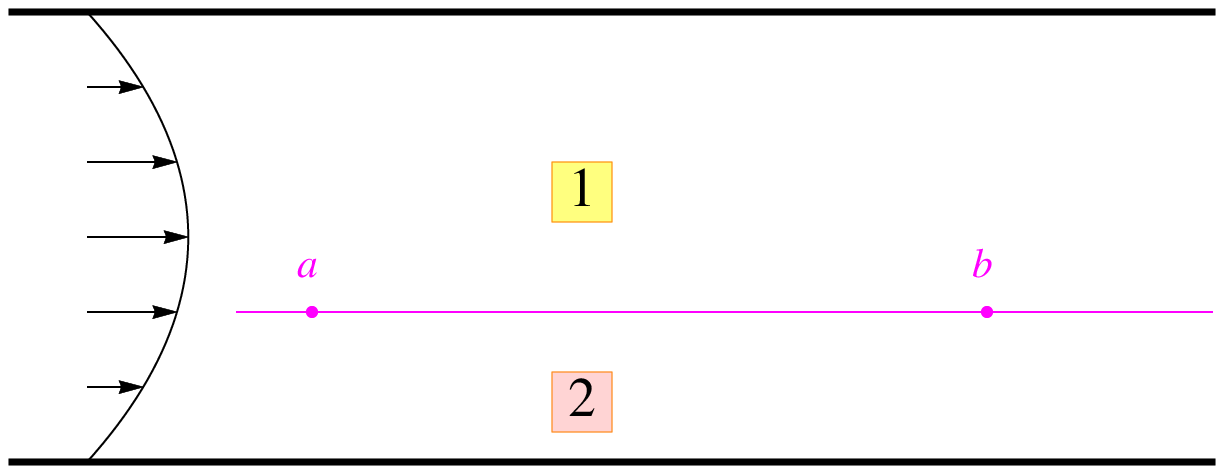} \\

\vspace*{0.4cm}
\includegraphics[width= 0.35 \textwidth]{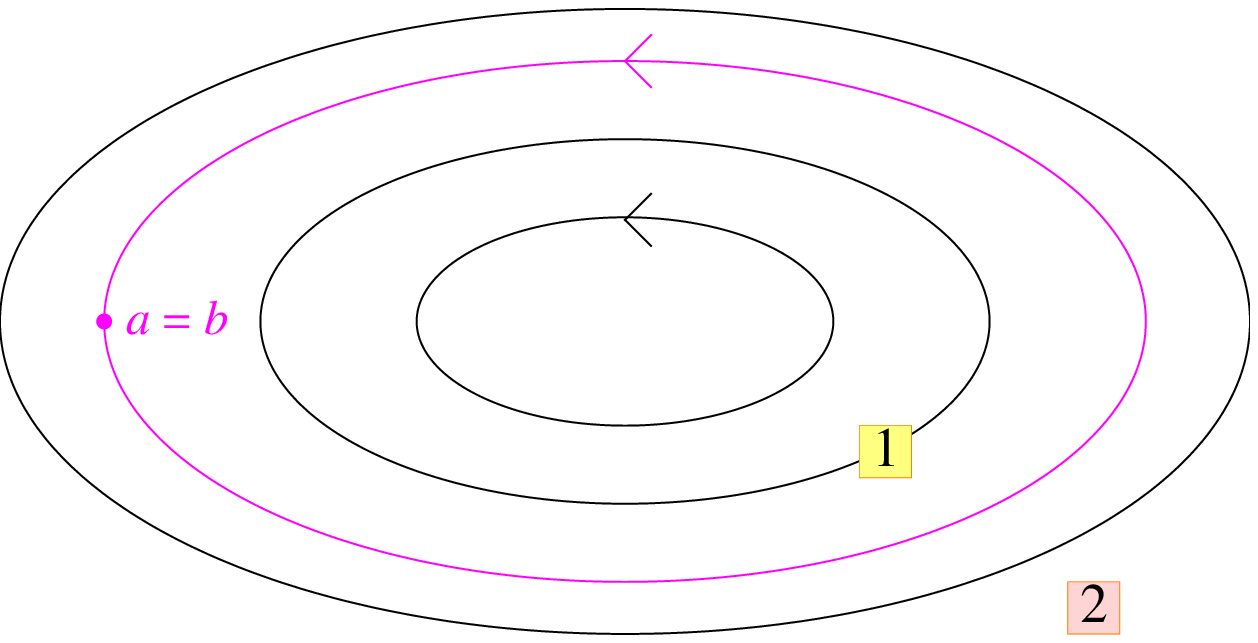}
\caption{Two situations in which a steady flow interface cannot be characterized in terms of a distinguished entity of
$ \dot{\vec{x}} = \vec{u} \left( \vec{x} \right) $, but is rather the interface (magenta) between fluids $ 1 $ and $ 2 $.}
\label{fig:2fluids}
\end{figure}

This rich suite of methods continues to grow 
\cite{allshousethiffeault,mundel,ma,mabollt,budisicthiffeault,mezicscience,levnajicmezic}, and improvements to existing methods continue to be reported \cite{peacockfroylandhaller,budisicmezic,droplet,karraschfarazmandhaller,onu,nelsonjacobs,raben}.  However,
given that these are all based directly on the velocity, {\em they ignore actual flow interfaces between two
fluids}.  Thus, they are generally inapplicable for two-phase flows of miscible fluids, as highlighted by simple examples in Fig.~\ref{fig:2fluids}.  In the upper panel, two different fluids enter a microchannel from
the left, each entering perhaps from syringes or tubes (not pictured) positioned on the upper and lower sides of the channel.
Since these fluids could for example be a sample and a reagent in a microfluidic bioreactor, there is no necessity for the two fluids to be needed in the same proportion.  This results in the fluids flowing to the right in a laminar
fashion, with their mutual fluid interface not along the centerline of the channel.   Attempting to identify the flow interface {\em purely from the
fluid velocity} is futile; there is absolutely nothing distinguished about the streamline along the flow interface (in magenta) in comparison
to other streamlines.  It is not even the streamline of maximum speed, which (if assuming the classical parabolic velocity
profile) is at the centerline \footnote{Local maxima of the speed are defined by Haller as `parabolic Lagrangian Coherent
Structures,' for which a theory has been developed \cite{hallerreview}.}.
All methods describe above will therefore fail in this simple example.  The flow interface
is something {\em physical}, and not derivable from the velocity field.  
The lower panel of Fig.~\ref{fig:2fluids} shows a situation
in which an anomalous fluid (a pollutant, nutrient, chemical, plume of higher temperature, etc) has intruded into the center of
a vortex.  The flow interface between the interior and exterior fluids here is a streamline, but once again there is nothing
distinguished about this streamline based on the velocity field.  There are closed streamlines both inside and outside this particular
interface which distinguishes between the inside and outside fluids.  It is such flow interfaces between miscible fluids, and determining transport across them under
the introduction of velocity agitations, that is the focus of this article. 

All methods based purely on the velocity field (such as finite-time Lyapunov exponents, curves of extremal attraction, transfer operators, or stable/unstable manifolds) fail in identifying this physical flow interface, even in these simple steady examples.
If attempting to determine transport between the fluids, one numerical method which {\em would} work is to think of evolving
the fluid density of each of the fluids according to an advection-diffusion (convection-diffusion) equation \cite{thiffeault,microfluid_review,guo,zheng,delcastillonegrete,meuniervillermaux} or other relevant dynamics \cite{fernandez,guo,zheng,camassa}.
Rather than limiting attention to the fluid velocity and resulting Lagrangian trajectories, the trick would be to consider the
two evolving fluid density fields. One might define the flow interface as a front of one of these density fields, and under the 
operation of advection and diffusion (assuming that there is a known way to parametrize the P\'{e}clet number
which characterizes the strength of the diffusion), one might be able to {\em computationally} make progress on this issue.
However, this will not help in obtaining a broader conceptual understanding of the process.  How does the fluid interface
evolve?  Is it possible to characterize how `complicated' it gets?   How can the interchange of fluid be quantified?  Is there an
optimal velocity agitation to maximize cross-interface transport?  Having a theory would give the ability to analyze questions
such as these, with the longer term goal of being able to maximize or limit transport \cite{mathewoptimal,linthiffeaultdoering,cortelezzi,vikhanskyoptimal,lin,mixer,l2mixer,frequency}
according to our wishes.

Studying interfaces between two miscible fluids is not new, and includes much recent work \cite{guo,zheng,sussman,fernandez,lyndenbell,young,camassa,sahu}. 
The approach followed here, in which the Lagrangian particle evolution is directly used in conjunction with techniques
inspired by dynamical systems theory, is however a novel approach, which moreover provides tools for answering the questions
posed above. 
The key to proceeding is in determining how one might identify the flow interface under unsteady velocity agitations.  It will 
be argued that the concept of a {\em streakline} is the most appropriate to use, under the condition
that the velocity agitation is confined to a certain region.  (The streaklines used here are {\em not} associated with stable/unstable
manifolds, as might be the case if considering the interface between a fluid and a bluff body \cite{roca,shariff,ziemniak,alam}.)
Briefly, this is because if the agitation is confined to being downstream of
$ \vec{a} $ in the top panel of Fig.~\ref{fig:2fluids}, then the streakline passing through $ \vec{a} $ will demarcate the boundary
between the two fluids, as fluids arriving from upstream on the two sides of $ \vec{a} $ are different. 

This paper is organized as follows.  Section~\ref{sec:streaklines}
will develop the theory for the streaklines---the `nominal' flow interfaces when the weak velocity agitations are considered---and
their evolution with time.  Explicit analytical expressions are obtained by utilizing dynamical systems methods, and are valid
for general time-dependence in the velocity agitation, and also allow for compressibility in the fluid.
Section~\ref{sec:streakexamples} provides a validation of
these expressions in comparison to numerical simulations of streaklines, in two examples which are loosely based on 
Fig.~\ref{fig:2fluids}.  Specifically, the first example considers the impact on the flow interface as a result of introducing
flow in cross-channels (a so-called {\em cross-channel micromixer} \cite{bottausciRSoc,bottausci,niuactive,tabelingchaotic,lee,microfluid_review}), while the second concerns the impact on 
the `boundary' of Kirchhoff's elliptic vortex  \cite{kirchhoff,wang,mitchellrossi,friedland,meleshkovanheijst} due to weak external 
strain.
In Section~\ref{sec:transport}, a theory for quantifying the transport between the two fluids as a result
of the agitation is developed.  Of particular interest here is the question of `transport across what?' since when the flow is
unsteady, there is ambiguity in defining the fluid interface.  If the interface is thought of in terms of a timeline (i.e., particles 
seeded on the interface, and evolved with time), since the timeline is a material surface, there can be no transport across it.  
Therefore, an appropriate way of defining a `nominal' flow barrier, which respects the streakline ideas, needs to be 
formulated.  An explicit approximation for the transport between the two fluids is obtained; this shall be useful in future work
in, for example, determining forms of velocity agitations which maximize transport (as in the similar developments for
heteroclinic situations \cite{optimal,mixer,frequency,l2mixer}).  The theory is once again validated by numerical simulations
of the same two examples in Section~\ref{sec:transportexamples}.  The second of these offers a novel way of examining the oft-studied problem of vortices in an external straining field \cite{romkedarjfm,kida,delcastillonegrete,meuniervillermaux,leweke,mula,turner,bassomgilbert,zhmur}; here, the Lagrangian fluid interchange between the interior
and exterior fluids caused by weak external strain is quantified.   Finally, directions of future work are outlined in Section~\ref{sec:conclusions}.

%%%%%%%%%%%%%%%%%%%%%%%%%%%%%%%%%%%%%%%%
\section{Upstream and downstream streaklines}
\label{sec:streaklines}

Consider a steady two-dimensional flow in which there is a persistent (one-dimensional) flow interface between two different miscible fluids.  The {\em persistence} of this entity is annoying from the perspective of mixing the two fluids together, and hence
the goal
is to understand how an {\em unsteady} velocity agitation affects this flow interface, and how the resulting advective transport between the fluids
can be quantified.  Firstly, to introduce notation, consider the steady flow
\begin{equation}
\dot{\vec{x}} = \vec{u} \left( \vec{x} \right) \quad , \quad \vec{x} \in \R^2 \, .
\label{eq:steady}
\end{equation}
Incompressibility is {\em not} assumed for the fluids, but $ \vec{u} $ is assumed to be smooth.
Now, a  flow interface $ \Gamma $ that persists in the steady flow (\ref{eq:steady}) must have no
fluid velocity perpendicular to $ \Gamma $.  Thus, the velocity $ \vec{u} $ is tangential to $ \Gamma $, which can be characterized as part of a streamline of the initial steady flow.  Since the velocity is steady, this can be thought of
as a streamline, streakline, or pathline, but as shall be seen shortly, when an unsteady velocity agitation is applied, thinking of
the flow barrier as a {\em streakline} is the correct approach.  Let $ \Gamma $ be such a streakline.  A velocity agitation will
be applied to the part of $ \Gamma $ which lies between the points $ \vec{a} $ and $ \vec{b} $ only, and this restricted
part of $ \Gamma $ shall be denoted $ \tilde{\Gamma} $. Thus, $ \tilde{\Gamma} $ is 
a curve which starts at the upstream anchor point $ \vec{a} $, and connects along the streamline emanating from $ \vec{a} $ and progressing to the downstream anchor point
$ \vec{b} $.  Two generic situations are possible: (i) $ \vec{a} \ne \vec{b} $, in which case
$ \tilde{\Gamma} $ is an open curve, and (ii) $ \vec{a} = \vec{b} $, in which case $ \tilde{\Gamma} $ is a closed curve.  These two situations are exactly analogous to those shown in Fig.~\ref{fig:2fluids}; however, there is no necessity for the streaklines to
be as uniform as those pictured here. 
 The streakline $ \Gamma $ is the extension of $ \tilde{\Gamma} $ along the 
streaklines passing through $ \vec{a} $ and $ \vec{b} $; it is clear that in the open case $ \Gamma $ extends beyond $
\tilde{\Gamma} $.  In the closed case, the streakline $ \Gamma $ can be thought of as retracing the closed loop
$ \tilde{\Gamma} $ repeatedly,
as will happen if dyed particles are continually ejected at $ \vec{a} = \vec{b} $.

There are two assumptions on the flow interface.  The first is that $ \tilde{\Gamma} $ be a {\em simple} (non self-intersecting) curve.  The second---which is crucial---is that $ \vec{u} \ne \vec{0} $ on $ \Gamma $.  
If $ \vec{u} =
{\bf 0} $ at some points on $ \Gamma $, then $ \Gamma $ will consist of parts of heteroclinic manifolds, and established theory
for locating these \cite{tangential,open,siam_book}, and the resulting transport time-periodic
\cite{romkedarjfm,wiggins,periodic}, aperiodic \cite{aperiodic} or impulsive \cite{impulsive} situations, applies.  Moreover, 
standard diagnostic tools such as finite-time Lyapunov exponents or curves of maximal attraction are viable
candidates for numerically determining the flow barriers.  Therefore, stagnation points will be explicitly precluded on $ \Gamma $;
it shall be {\em non-heteroclinic}.

\begin{figure}[t]
\includegraphics[width=0.5 \textwidth, height = 0.2 \textheight]{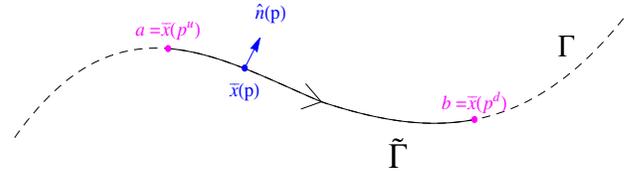} \\
\caption{The generic steady streakline $ \Gamma $ (dashed), and its restriction $ \tilde{\Gamma} $ (solid) to between
$ \vec{a} $ and $ \vec{b} $, which is 
parametrized in the form
$ \bar{\vec{x}}(p) $, with $ p \in [p^u,p^d] $.}
\label{fig:gamma}
\end{figure}

The streakline $ \Gamma $ is easily defined as a curve in $ \R^2 $, via a parametrization $ \bar{\vec{x}}(p) $  as shown in 
Fig.~\ref{fig:gamma}.  Here $ \bar{\vec{x}}(p) $ is a solution to (\ref{eq:steady})---where the parameter $ p $ can be thought
of as time---which obeys $ \bar{\vec{x}}(p^u) = \vec{a} $ and $ \bar{\vec{x}}(p^d) = \vec{b} $.  The superscript $ u $ is to
be identified with `{\em u}pstream,' and $ d $ with `{\em d}ownstream' throughout this article.  The restriction $ p \in [p^u,p^d] $
identifies  $ \tilde{\Gamma} $, the part lying between $ \vec{a} $ and $ \vec{b} $, to which the velocity agitation will be
confined.  If $ \tilde{\Gamma} $ is closed, then the velocity agitation will occur throughout $ \Gamma $ (the periodic repetition of
$ \tilde{\Gamma} $, except at the anchor point $ \vec{a} = \vec{b} $.  Indeed, $ \bar{\vec{x}}(p) $ is a periodic function of $ p $
in this instance, but the restriction to $ \tilde{\Gamma} $ acheived by setting $ p \in [p^u,p^d] $ implies that $ \bar{\vec{x}}(p^u) =
\vec{a} $ but that $ p^d $ is the next instance in which $ \bar{\vec{x}}(p) $ reaches $ \vec{a} $, which is of course $ \vec{b} $.  
 In either the open or closed situation, of interest is the fact that
the {\em upstream streakline} emanating from $ \vec{a}  $ is identical to the {\em downstream streakline} emanating 
from $ \vec{b} $ in this steady situation, and moreover these are each identical to $ \Gamma $.

In preparation for introducing time-dependence in the velocity, it pays to understand how the upstream streakline {\em evolves with
time}.  The time-variation of this upstream streakline can be quantified by the definition
\begin{eqnarray}
\Gamma_0^u(t) & := & \bigcup_{p \in \R} \Big\{ \vec{x}_0^u(p,t)  ~ {\mathrm{which~solves}} \, (\ref{eq:steady}) \, {\mathrm{with}} 
\nonumber \\
& & {\mathrm{condition}} ~ \vec{x}_0^u(p,t-p+p^u) = 
\vec{a} \Big\} \, .
\label{eq:upstreamsteady}
\end{eqnarray}
The $ p $ above provides a parametrization of the upstream streakline at each fixed time instance $ t $, where the $ p $ can
be thought of as {\em identifying a particle}.  The particle which is at the location $ \bar{\vec{x}}(p) $ at time $ t $ is the one
which passed through $ \vec{a} $ at time $ t - p + p^u $ (i.e., a time $ p^u - p $ prior to $ t $).   Note that the upstream streakline
here is not restricted to $ \tilde{\Gamma} $, since the $ p $-values go beyond $ [p^u, p^d] $.  This therefore incorporates
parts of $ \Gamma $ before $ \vec{a} $ (these are particles which will go through $ \vec{a} $ in the future), 
and beyond $ \vec{b} $ (these have gone through $ \vec{a} $ in the past) shown by the dashed curve  in 
Fig.~\ref{fig:gamma}.  The upstream streakline encapsulates all trajectories that will go through $ \vec{a} $, in their entirety.
Analogously, the downstream
streakline
\begin{eqnarray}
\Gamma_0^d(t) & := & \bigcup_{p \in \R} \Big\{ \vec{x}_0^d(p,t)  ~ {\mathrm{which~solves}} \, (\ref{eq:steady}) \, {\mathrm{with}} 
\nonumber \\
& & {\mathrm{condition}} ~ \vec{x}_0^d(p,t-p+p^d) = 
\vec{b} \Big\} \, 
\label{eq:downstreamsteady}
\end{eqnarray}
identifies all particles which go through the downstream location $ \vec{b} $ at some time.  For closed $ \tilde{\Gamma} $, 
since $ \vec{a} = \vec{b} $, the upstream and downstream streaklines coincide.  Thus, in this case, it suffices to simply
use one of the definitions.

The parametrization above---with $ p $ being a fixed {\em p}article along the streakline and $ t $ the {\em t}ime at which the streakline is
being observed---shall be retained when the flow is subject to an unsteady velocity agitation in the form
\begin{equation}
\dot{\vec{x}} = \vec{u} \left( \vec{x} \right) + \vec{v} \left( \vec{x}, t \right) \, . 
\label{eq:unsteady}
\end{equation} 
The agitation velocity on $ \Gamma $ shall be confined to be between $ \vec{a} $ 
and $ \vec{b} $ {\em only}.  This shall specifically be stated as
\begin{equation}
\renewcommand{\arraystretch}{1.6}
\vec{v}(\bar{\vec{x}}(p),t) = \left\{ \begin{array}{ll} 
0 ~~{\mathrm{for}} ~p \le p^u ~{\mathrm{or}}~p \ge p^d & ({\mathrm{open}}~\tilde{\Gamma}) \\
0~~{\mathrm{when}}~ \bar{\vec{x}}(p) = \vec{a}  & ({\mathrm{closed}}~\tilde{\Gamma})
\end{array} \right.
\label{eq:agitation}
\end{equation}
For open $ \tilde{\Gamma} $ this means that
 the unsteady agitation $ \vec{v} $ is
 zero up to (and including) the point $ \vec{a} $, which enables the understanding that the $ \vec{a} $ continues to be
 at the interface of the two fluids, as they come in towards $ \vec{a} $.  Thus, $ \vec{a} $ continues to be an anchor point
 on the flow interface in forward time.  Similarly, $ \vec{b} $ would be an anchor point on the interface in backward time.
 For closed $ \tilde{\Gamma} $, $ \vec{a} = \vec{b} $, and 
 $ \bar{\vec{x}}(p) $ periodically traverses $ \tilde{\Gamma} $.  Thus, once going `beyond' $ \vec{b} $ on $ \tilde{\Gamma} $,
 one returns to points in which a velocity agitation continues to exist, and it cannot be `turned off' as in the open situation.
 To enable an anchor point on the flow interface to continue to be defined under the agitation, it is necessary to fix the agitation
 to be zero at $ \vec{a} $, which shall be the release point of the streakline.   Moreover---
in representing this as an agitation on a dominant steady flow---it shall be assumed that 
\[
\left| \vec{v} \left( \vec{x},t \right) \right| \le \eps \left| \vec{u} \left( \vec{x} \right) \right| \quad {\mathrm{for}} \, 
\vec{x} \in \Gamma \, {\mathrm{and}} \, t \in \R \, ,
\]
where $ \eps \ll 1 $, and $ \vec{v} $ is smooth in $ \vec{x} $.  Note however that $ \vec{v} $ is otherwise arbitrary for the
theory to follow: it may 
satisfy $ \vec{\nabla} \cdot  \vec{v} \ne 0 $, possess aperiodic time-dependence, etc.
Now, exactly analogous to the definitions of the steady upstream and downstream streaklines, the
{\em unsteady} ones associated with (\ref{eq:unsteady}) can be defined by
\begin{eqnarray}
\Gamma_\eps^u(t) & := & \bigcup_{p \in [-P,P]} \Big\{ \vec{x}_\eps^u(p,t)  ~ {\mathrm{which~solves}} \, (\ref{eq:unsteady}) \, {\mathrm{with}} 
\nonumber \\
& & {\mathrm{condition}} ~ \vec{x}_\eps^u(p,t-p+p^u) = 
\vec{a} \Big\} \,
\label{eq:upstreamunsteady}
\end{eqnarray}
and
\begin{eqnarray}
\Gamma_\eps^d(t) & := & \bigcup_{p \in [-P,P]} \Big\{ \vec{x}_\eps^d(p,t)  ~ {\mathrm{which~solves}} \, (\ref{eq:unsteady}) \, {\mathrm{with}} 
\nonumber \\
& & {\mathrm{condition}} ~ \vec{x}_\eps^d(p,t-p+p^d) = 
\vec{b} \Big\} \, .
\label{eq:downstreamunsteady}
\end{eqnarray}
A mild technicality in comparison with the steady streakline definitions is the replacement of the $ \infty $ by $ P $, which
is any finite positive number.  This because it is not possible to maintain control for all time for particles which have gone
through the velocity agitation region, but this can only be accomplished for finite times, as large as required.

\begin{figure}[t]
\includegraphics[width=0.5 \textwidth, height = 0.2 \textheight]{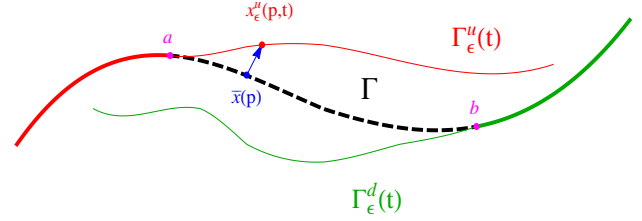} \\
\caption{The unsteady upstream ($ \Gamma_\eps^u(t) $, in red) and downstream ($ \Gamma_\eps^d(t) $, in green) 
streaklines at an instance in time $ t $, defined
according to (\ref{eq:upstreamunsteady}) and (\ref{eq:downstreamunsteady}).}
\label{fig:unsteady}
\end{figure}

These streaklines at a fixed time $ t $ are shown in Fig.~\ref{fig:unsteady}, to be viewed in conjunction with
the steady (non-agitated) streakline picture of Fig.~\ref{fig:gamma}.  The steady $ \Gamma $ of Fig.~\ref{fig:unsteady}
is shown by the thick curves: it consists of the thick red curve upstream of $ \vec{a} $, the thick dashed black curve between
$ \vec{a} $ and $ \vec{b} $ (i.e., $ \tilde{\Gamma} $), and the thick green curve downstream of $ \vec{b} $.  The unsteady
upstream steakline $ \Gamma_\eps^u(t) $ is shown in red, and consists of a thick part which coincides with $ \Gamma $
(upstream $ \vec{a} $), and then the extension which need not.  Similarly, $ \Gamma_\eps^d(t) $, shown in green, coincides
with $ \Gamma $ downstream of $ \vec{b} $ (shown by the thick green curve), while not necessarily so upstream of $ \vec{b} $.
As time progresses, (the thin portions of) $ \Gamma_\eps^{u,d}(t) $ will wiggle around due to the velocity agitation.  They may
even intersect in various ways.  Since the upstream and downstream streaklines are not necessarily coincident, the previously 
clear flow interface $ \Gamma $ between the two fluids has broken.
Fluid arriving towards $ \vec{a} $ on the two sides of $ \Gamma $, 
and then subsequently advecting in forward time, are separated by the upstream streakline $ \Gamma_\eps^u $.  On the other
hand, fluid which is `separated' by the part of $ \Gamma $ downstream of $ \vec{b} $ is separated by $ \Gamma_\eps^d $ in 
backward time.  The intermingling of $ \Gamma_\eps^u $
and $ \Gamma_\eps^d $ results in fluid which was separated in backward time not being identical to the fluid which is separated
in forward time.  Transport is achieved between the two fluids as a result of this, and shall be explored in more detail in 
Section~\ref{sec:transport}.  At this point, the focus shall be on determining the time-varying locations of the upstream and
downstream streaklines.

In preparation for stating the characterization of the unsteady streaklines, the notation
\[
J := \left( \begin{array}{cc}
0 & -1 \\ 1 & 0
\end{array} \right) \quad , \quad \I_{T}(t) := \left\{ \begin{array}{ll} 1 & {\mathrm{if}} ~ t \in T \\
0 & {\mathrm{if}} ~ t \ne T \end{array} \right.\, ,
\]
will be useful. Notice that $ J $ 
rotates vectors by $ + \pi/2 $, and from Fig.~\ref{fig:gamma}, 
\begin{equation}
\hat{\vec{n}}(p) := \frac{J \vec{u} \left( \bar{\vec{x}}(p) \right)}{\left|  \vec{u} \left( \bar{\vec{x}}(p) \right) \right|}
\label{eq:normal}
\end{equation}
is a
unit normal vector to $ \Gamma $ at the parametric location $ p $.  The unsteady modifications to the upstream and downstream streaklines in this normal
direction can now be quantified:

\begin{table}[t]
\setlength\tabcolsep{20pt}
\renewcommand{\arraystretch}{1.6}
\begin{tabular}{| c | c | c |} \hline
\mbox{} & $ \tilde{\Gamma} $ open & $ \tilde{\Gamma} $ closed \\ \hline
$ p_+^u $ & $ \min \left\{ p, p^d \right\} $ & $ p $ \\
$ p_-^u $ & $  p^u $ & $ -P $ \\
$ p_-^d $ & $ \max \left\{ p, p^u \right\} $ & $ p $ \\ 
$ p_+^d $ & $ p^d $ & $ P $ \\ \hline
\end{tabular}
\caption{Definitions of $ p_{\pm}^{u,d} $ for use in Theorems~\ref{theorem:upstream} and \ref{theorem:downstream}.}
\label{table:p}
\end{table}

\begin{theorem}[Upstream streakline]
\label{theorem:upstream}
Under the definitions of Table~\ref{table:p}, 
the parametric representation  $ \vec{x}_\eps^u(p,t) $ of $ \Gamma_\eps^u(t) $ satisfies
\begin{equation}
\left[ \vec{x}_\eps^u(p,t) - \bar{\vec{x}}(p) \right] \cdot 
\hat{\vec{n}}(p)
=  \frac{M^u(p,t)}{\left| \vec{u} \left( \bar{\vec{x}}(p) \right) \right|} + {\mathcal O}(\eps^2)
\label{eq:upstreamnormal}
\end{equation}
where
\begin{widetext}
\begin{equation}
M^u(p,t):= \I_{[p_-^u,P]}(p) \int_{p^u}^{p_+^u}  \exp \left[ \int_\tau^p  \left[  
\vec{\nabla}  \cdot \vec{u} \right] \left( \bar{\vec{x}}(\xi) \right) \d \xi \right] \left[ J \vec{u} \left( \bar{\vec{x}}(\tau) \right) \right] \cdot 
\vec{v} \left( \bar{\vec{x}}(\tau), \tau +  t -  p \right) \d \tau \, . 
\label{eq:mupstream}
\end{equation}
\end{widetext}
\end{theorem}

For the proof, the reader is referred to Appendix~\ref{app:proof}.  The crux of this theorem is in characterizing the
normal displacement from $ \bar{\vec{x}}(p) $ to a point $ \vec{x}_\eps^u(p,t) $ on $ \Gamma_\eps^u(p,t) $, as
indicated by the arrow in Fig.~\ref{fig:unsteady}.  This is therefore given for $ p \in [-P,P] $ by
\begin{equation}
\vec{x}_\eps^u(p,t) = \bar{\vec{x}}(p) + \frac{M^u(p,t)}{\left| \vec{u} \left( \bar{\vec{x}}(p) \right) \right|} \hat{\vec{n}}(p) + 
{\mathcal O}(\eps^2)
\label{eq:upstream}
\end{equation}
if the tangential displacement is ignored \footnote{Ignoring the tangential displacement is reasonable in the sense that
the streakline has $ p $ as its parameter varying in precisely the tangential direction, and when plotting $ \vec{x}_\eps^u(p,t) $
for a range of $ p $ to obtain the streakline curve, any tangential displacement will be barely visible \cite{tangential}. This is
indeed reflected in the examples presented here.  If needed,
tangential displacements can be quantified \cite{tangential}, but this is unwieldy.}.  This allows for the streakline
to be located and tracked theoretically to $ {\mathcal O}(\eps) $, since $ M_\eps^u = {\mathcal O}(\eps) $ due to the presence
of $ \vec{v} $ in the integral (\ref{eq:mupstream}).

The prefactor in (\ref{eq:mupstream}) simply ensures that the displacement is zero upstream of $ \vec{a} $ if $ \tilde{\Gamma} $ is open, but if
$ \tilde{\Gamma} $ is closed, `upstream of $ \vec{a} $' is once again in the velocity agitation region and thus the displacement
incurred is not zero.
The subtlety in the upper limit $ p_+^u $ is because in 
the open situation, once a particle has gone beyond $ \vec{b} $ (i.e., beyond  $ \bar{\vec{x}}(p^d) $), there is no longer
any velocity agitation applying.  Hence there will be no additional (leading-order) change in each particle's position, and so
$ p_+^u $ will be set to $ p^d $ as shown in Table~\ref{table:p}.  Before arriving at $ \vec{b} $, it will be $ p $; hence the
expression $ \min \left\{ p, p^d \right\} $.
If $ \tilde{\Gamma} $ is closed, when a particle approaches $ \bar{\vec{x}}(p^d) $ it has once again arrived at $ \vec{a} $,
and it will experience the velocity agitation once again as it repeatedly traverses a path which is $ {\mathcal O}(\eps) $ close to
$ \tilde{\Gamma} $.  Hence the setting of $ p_+^u = p $.  There is {\em no} necessity for $ \Gamma_\eps^u(t) $ to consist of a closed loop, however, since
when a streakline arrives back to near $ \vec{a} $, it will generically {\em not} be at $ \vec{a} $.  This streakline will
then wrap around repeatedly.  In computing the displacement
of the streakline in the closed $ \tilde{\Gamma} $ situation, the terms in (\ref{eq:mupstream}) involving $ \vec{u} $ will 
therefore periodically repeat, but the presence of the general time-dependence in $ \vec{v} $ in (\ref{eq:mupstream}) ensures
that the normal displacement is {\em not} generally periodic in $ p $ or $ t $.

If only interested in forward time, i.e., the time-varying curve which separates the two different fluids arriving at $ \vec{a} $ 
and progressing beyond, then the upstream streakline is what is needed.  
However, the downstream streakline is relevant to
understanding the origins of the fluid which are separated {\em beyond} $ \vec{b} $.  The separating curve is the
streakline passing through $ \vec{b} $, i.e., the downstream streakline.  As will be seen in Section~\ref{sec:transport}, the
downstream streakline also becomes important when attempting to quantify the transport across the (now broken) fluid
interface.
A similar quantification of the modification, in the direction normal to $ \Gamma $, of the downstream streakline is possible:

\begin{theorem}[Downstream streakline]
\label{theorem:downstream}
Under the definitions of Table~\ref{table:p}, 
the parametric representation  $ \vec{x}_\eps^d(p,t) $ of $ \Gamma_\eps^d(t) $ satisfies
\begin{equation}
\left[ \vec{x}_\eps^d(p,t) - \bar{\vec{x}}(p) \right] \cdot 
\hat{\vec{n}}(p) =  \frac{M^d(p,t)}{\left| \vec{u} \left( \bar{\vec{x}}(p) \right) \right|} + {\mathcal O}(\eps^2)
\label{eq:downstreamnormal}
\end{equation}
where
\begin{widetext}
\begin{equation}
M^d(p,t):= - \I_{[-P,p_+^d]}(p) \int_{p_-^d}^{p^d}  \exp \left[ \int_\tau^p  \left[  
\vec{\nabla}  \cdot \vec{u} \right] \left( \bar{\vec{x}}(\xi) \right)  \d \xi \right] \left[ J \vec{u} \left( \bar{\vec{x}}(\tau) \right) \right] \cdot 
\vec{v} \left( \bar{\vec{x}}(\tau), \tau +  t -  p \right) \d \tau \, . 
\label{eq:mdownstream}
\end{equation}
\end{widetext}
\end{theorem}

The proof for Theorem~\ref{theorem:downstream} is similar to that of Theorem~\ref{theorem:upstream}, and shall be skipped.
The use of this theorem is the possibility of representing $ \Gamma_\eps^d(t) $ parametrically by
\begin{equation}
\vec{x}_\eps^d(p,t) = \bar{\vec{x}}(p) + \frac{M^d(p,t)}{\left| \vec{u} \left( \bar{\vec{x}}(p) \right) \right|} \hat{\vec{n}}(p) + 
{\mathcal O}(\eps^2)
\label{eq:downstream}
\end{equation}
for $ p \in [-P,P] $.  A point which is perhaps not obvious is that when drawn at a time $ t $, an upstream streakline {\em may}
intersect a downstream streakline, in either a transverse or tangential fashion.  This is because such intersections correspond 
to fluid particles which went through $ \vec{a} $ in the past, and will go through $ \vec{b} $ in the future.  This is in constrast
to `standard' streakline approaches which might consider releasing particles from both points continuously in time, and viewing
the resulting evolving curves forward in time; in this case, intersections are prohibited unless the streakline through $ \vec{a} $
has at some intermediate time gone through $ \vec{b} $.  

%%%%%%%%%%%%%%%%%%%%%%%%%
%%%%%%%%%%%%%%%%%%%%%%%%
\section{Streakline validation}
\label{sec:streakexamples}

In this section, streaklines will be obtained by numerical simulation, and compared with the theoretical expressions of
derived previously, in two examples: two fluids in a channel, and an anomalous fluid inside an elliptic vortex.  These same
examples will be examined subsequently, in Section~\ref{sec:transportexamples}, in computing the associated fluid transport.

%%%%%%%%%%%%%%%%%%%%
\subsection{Two fluids in a microchannel}

\begin{figure}
\includegraphics[width=0.45 \textwidth]{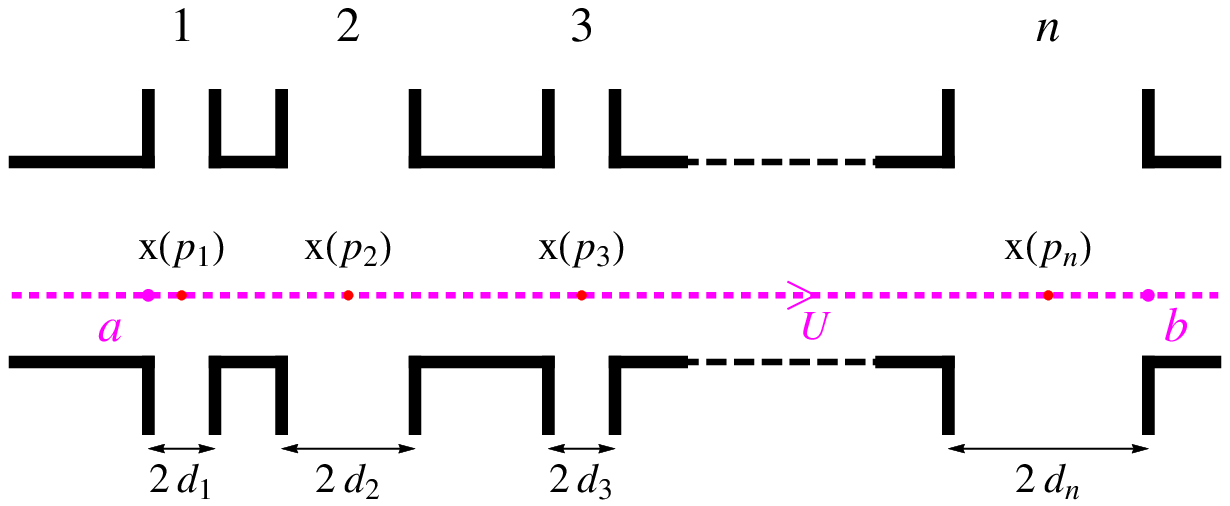}
\caption{Channel flow with cross-channels.}
\label{fig:crosschannel}
\end{figure}

As the first example, consider two incompressible
fluids travelling along a straight channel.  At the microfluidic level, it is well-known that these will
tend not to mix across their flow interface, and sloshing fluid in the direction normal to this interface via cross-channels is a
standard strategy which is used \cite{bottausciRSoc,bottausci,niuactive,tabelingchaotic,lee,microfluid_review}.  This interface is shown by the 
dashed curve in Fig.~\ref{fig:crosschannel}, which need not be centered since the volume flow rates of the upper and lower
fluids need not be the same.  For well-developed steady flow (with no flow in the cross-channels), fluid along
the interface will at a constant speed, $ U $.   The steady streakline is therefore
$ \bar{\vec{x}}(\tau) = \left( x(\tau), 0 \right)  = \left( U \tau, 0 \right)  $.  To account
for the many possibilities which are available in the literature, a {\em general} geometry consisting of $ n $ cross-channels
shall be assumed.  The $ j $th cross-channel is centered at the $ x $ location $ x(p_j) $, and is assumed to have width
$ 2 d_j $, where for consistency it is necessary that $ x(p_j) + d_j < x(p_{j+1}) - d_{j+1} $.
In this case,  $ x(p_j) = U p_j $, and the upstream point can be taken to be any point upstream of $  (U p_1-d_1, 0) $ and the downstream
point any point downstream of $  (U p_n+d_n, 0) $.  Thus, $ p^u \le p_1 - d_1/U $ and $ p^d \ge p_n + d_n/U $.  Experimental evidence  \cite{tabelingchaotic,niuactive,lee,frequency}  suggests that 
the flow in the cross-channels takes
on a parabolic profile, which can be modeled by
\begin{equation}
\vec{v}_j(x,y,t) = \frac{v_j}{d_j^2} \left[ \left( x \! - \! U p_j \right)^2 \! - \! d_j^2 \right]  \! \cos \left( \omega  t + \phi_j \right) 
\hat{\vec{j}}
\, , \,  \label{eq:crosschannelvelocity}
\end{equation}
for $ U p_j - d_j \le x \le U p_j + d_j $,
where $ v_j > 0 $ is a velocity scale representing the speed at the
center of the cross-channel, $ \omega > 0 $ is the frequency
of fluid sloshing, and $ \phi_j $ enables the specification of how the cross-channels are
operating in relation to one another.  For example, if all cross-channels are in phase, then $ \phi_j \equiv 0 $,
and if adjacent ones are exactly out of phase, then $ \phi_j = j \pi $.    Therefore,
the geometry and velocity specification can account for very general cross-channel configurations.  It is assumed that
$ \eps = \max_j \left| v_j \right| / U \ll 1 $.

\begin{figure}
\vspace*{-0.3cm}
\includegraphics[width=0.5 \textwidth, height=0.18 \textheight]{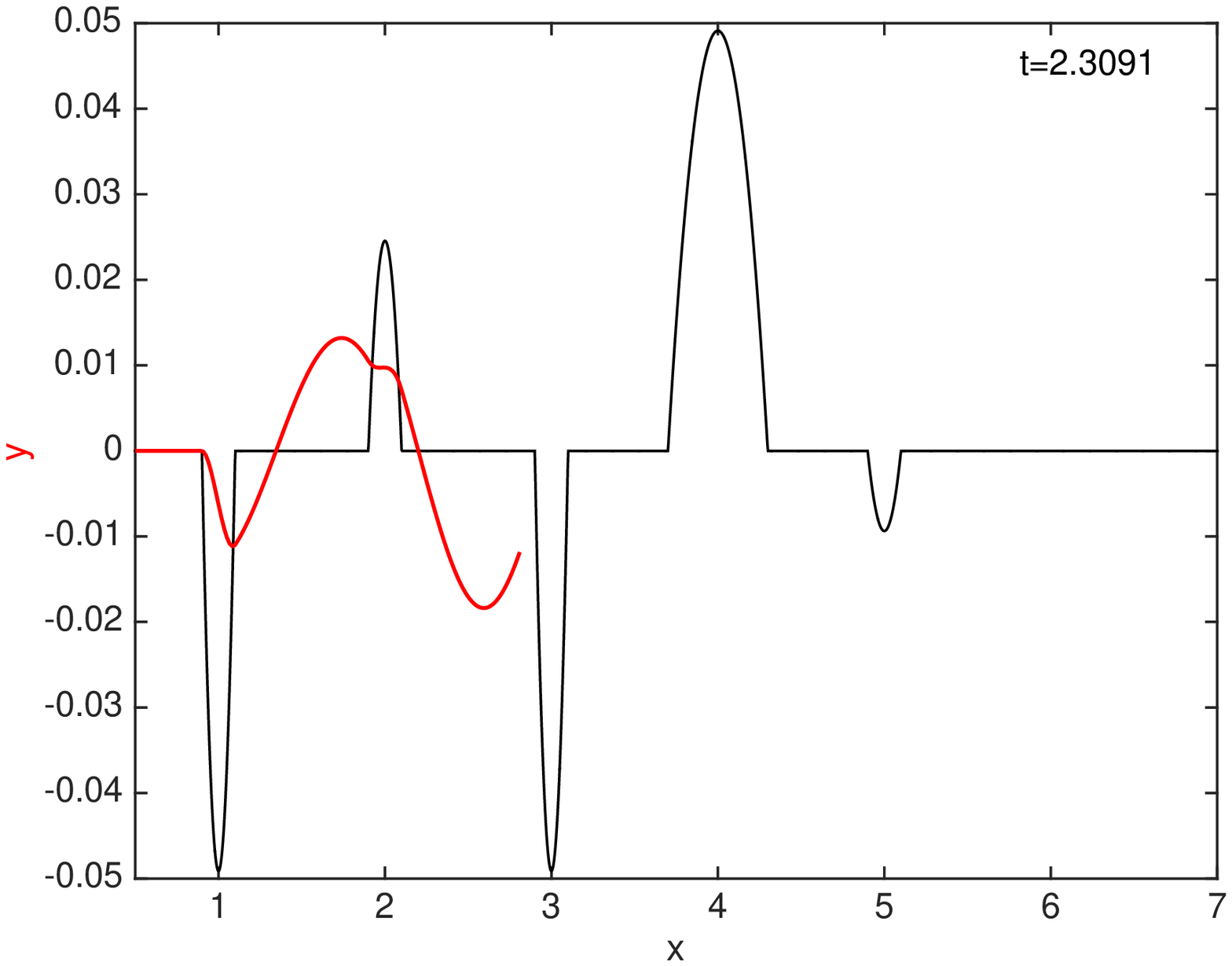} 
\includegraphics[width=0.5 \textwidth,  height=0.18 \textheight]{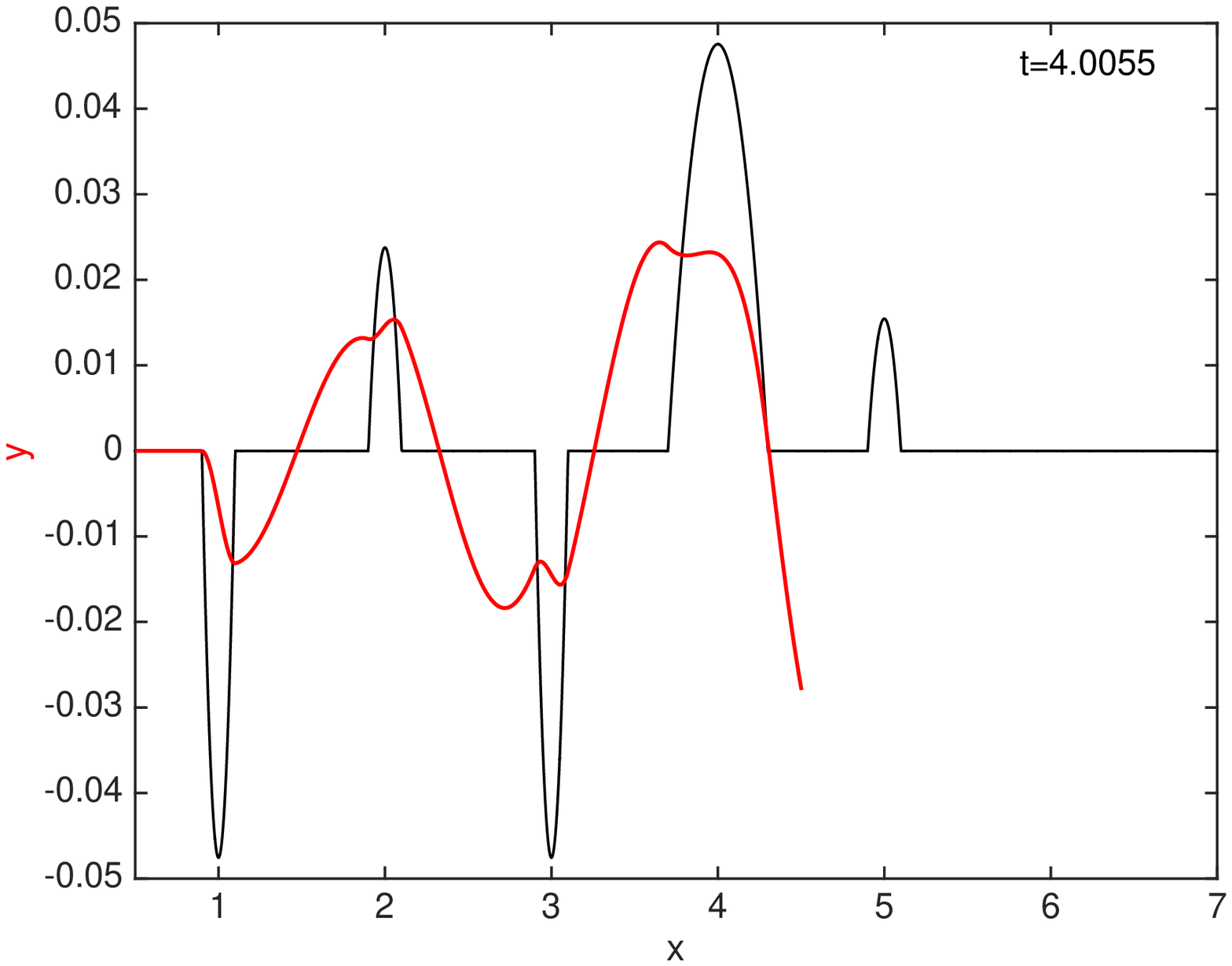} 
\includegraphics[width=0.5 \textwidth,  height=0.18 \textheight]{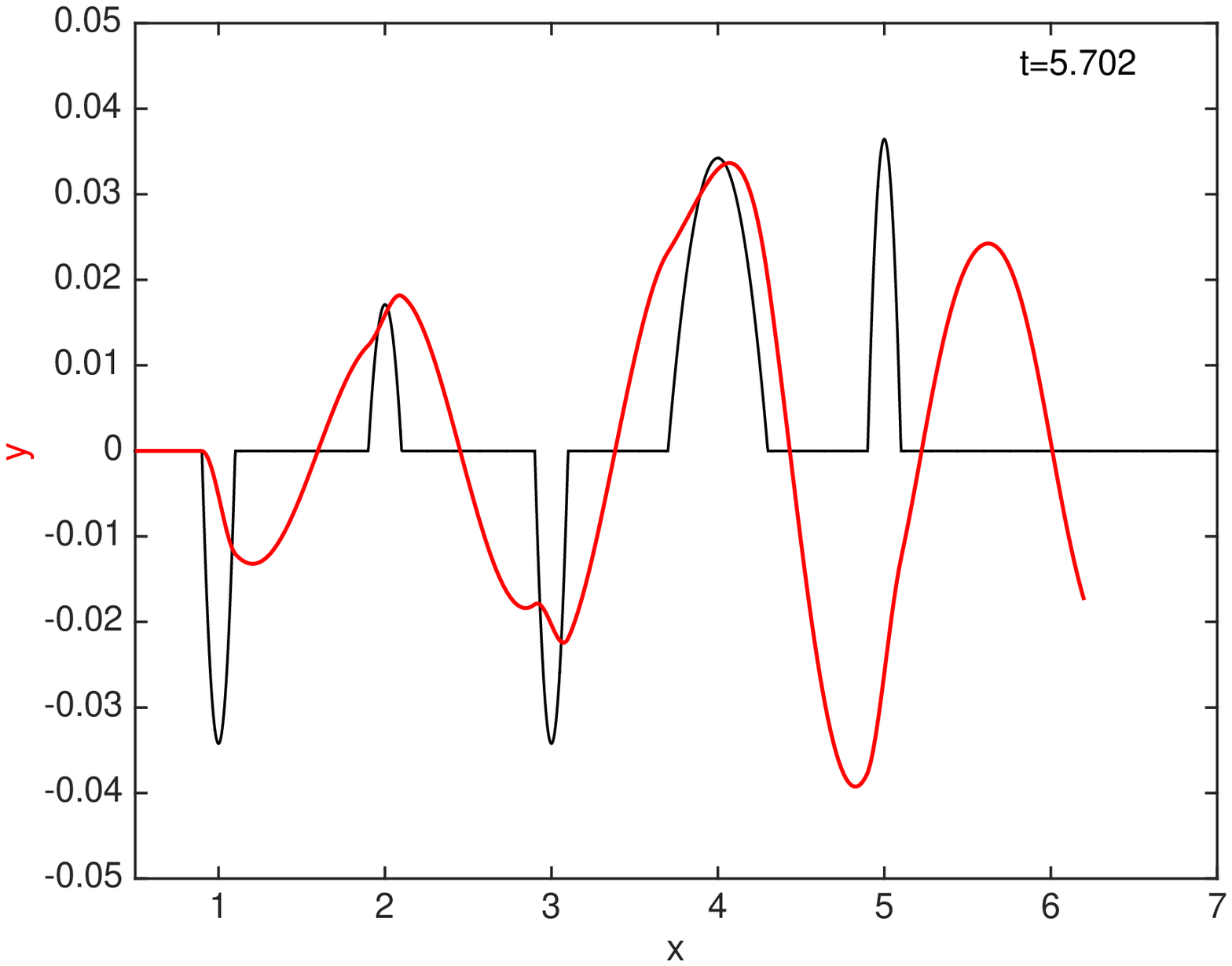} 
\includegraphics[width=0.5 \textwidth,  height=0.18 \textheight]{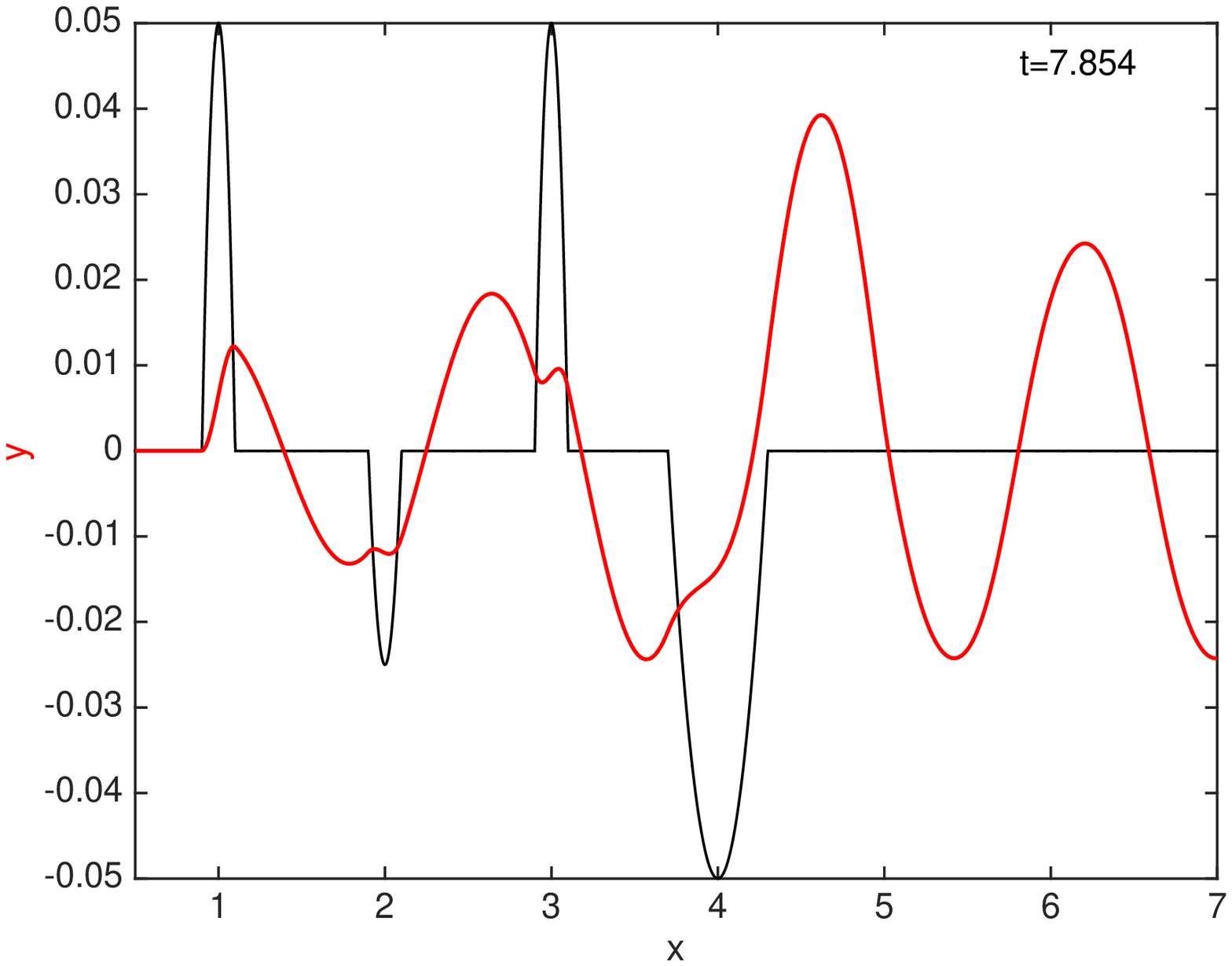}
\caption{Evolution of upstream streakline for channel flow, with dye released on the fluid interface at $ (0.5,0) $ from time $ 0 $ onwards.
The evolving streakline, representing the perturbed fluid interface, is shown in red, with the instantaneous cross-velocity
(with channel configuration as described in the text) shown by the black curves.}
\label{fig:ccstreak}
\end{figure}

The observations $ \left| \vec{u} \right| = U $, $ \vec{\nabla} \cdot \vec{u}  = 0 $ and $ J \vec{u} =
U \hat{\vec{j}} $ are useful in computing the upstream streakline as given in (\ref{eq:upstream}).  Thus, to leading-order
\begin{equation}
\vec{x}_\eps^u(p,t) = U p \, \hat{\vec{i}} + \frac{M_c^u(p,t)}{U} \hat{\vec{j}} \, , 
\label{eq:crosschannelupstream}
\end{equation}
where, from (\ref{eq:mupstream}), 
\begin{widetext}
\begin{equation}
M_c^u(p,t)  \! = \! \I_{[p_1 \! - \! d_1/U,P]}(p)  \! \int_{p_1 \! -\! d_1/U}^{\min \left\{ p, p_n \! + \! d_n/U \right\} }  \sum_{j=1}^n \I_{[ p_j-d_j/U,p_j \! + \! d_j/U]}(\tau) 
U \frac{v_j}{d_j^2} \left[ U^2 \left( \tau \! - \! p_j \right)^2 \! - \! d_j^2 \right]  \! \cos \left[ 
\omega  \left( \tau \! + \! t \! - \! p \right) \! + \! \phi_j \right] 
 \d \tau \, . 
\label{eq:mcrosschannelupstream}
\end{equation}
\end{widetext}
(The subscript $ c $ is used for `{\em c}hannel,' to contrast with the vortex example to be presented subsequently.)
For given parameter values, the integral above can be explicitly computed, thereby providing the time-variation of the
upstream streakline via an explicit expression for this general geometry.  To compare with numerics, choose a
situation where $ U = 1 $, $ \omega = 4 $, $ \eps = 0.1 $ and $ n = 5 $, with channels specified by $ 
\left\{ p_j \right\}  = \left\{ 1, 2, 3, 4, 5 \right\} $, 
$ \left\{ v_j \right\}  = \eps \left\{ 1, 0.5, 1, 1, 1 \right\} $, $ \left\{ d_j \right\} = \left\{ 0.1, 0.1, 0.1, 0.3, 0.1 \right\} $ and
$ \left\{ \phi_j \right\}  = \left\{ \pi, 2 \pi, 3 \pi, 4 \pi, 7 \pi/2 \right\} $.  Thus, the second cross-channel has a smaller maximum
speed than the others, the fourth is triple the width of the others, and the fifth has a phase which is at odds with the exactly-out-of-phase nature of the other channels.  Initially, $ y = 0 $ is taken to be the fluid interface.  Numerical simulations with red dye released
at $ \vec{a} = (0.5,0) $ (upstream of velocity agitations) on the interface at time $ 0 $ are shown in Fig.~\ref{fig:ccstreak} at several
instances in time.  The black curves illustrate the instantaneous velocities in the cross directions, scaled so that they fit into this
picture.  It should be noted that beyond $ (5.1,0) $ (the final point at which the velocity agitation applies with these
parameter values, the streakline is {\em not} simply along $ y = 0 $.   The curves in the streakline caused by the velocity
agitations will be swept along, with no additional agitation.  A video of the upstream streakline (i.e., unsteady fluid interface) evolution is provided with the Supplementary Materials.

\begin{figure}
\includegraphics[width=0.44 \textwidth,  height=0.175 \textheight]{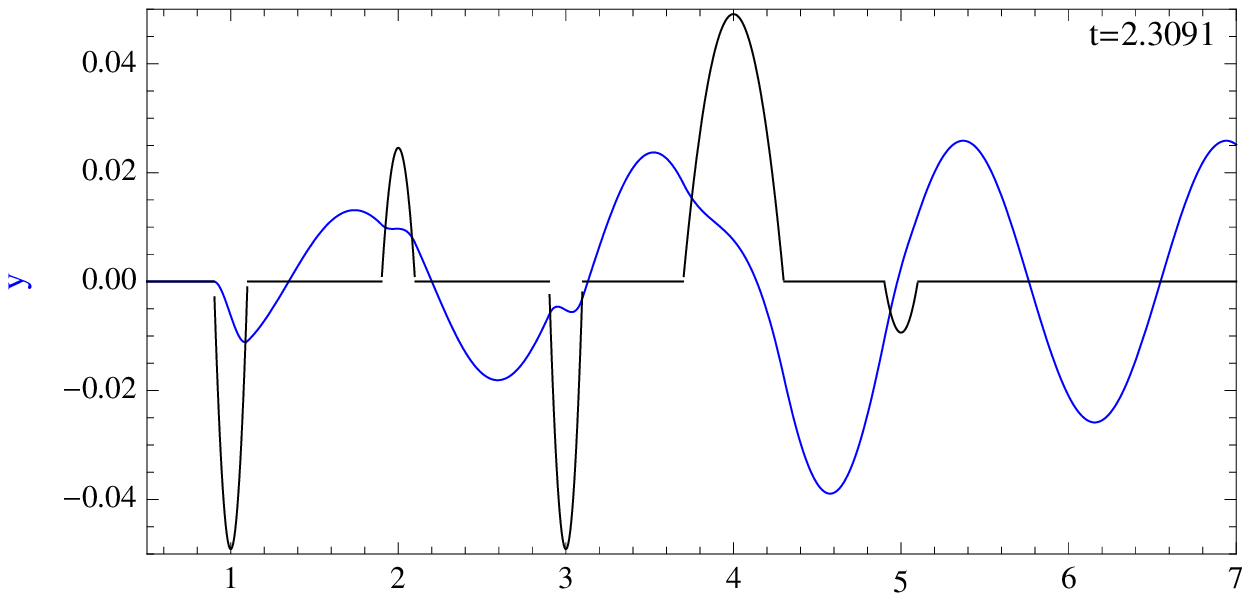}
\includegraphics[width=0.44 \textwidth, height=0.175 \textheight]{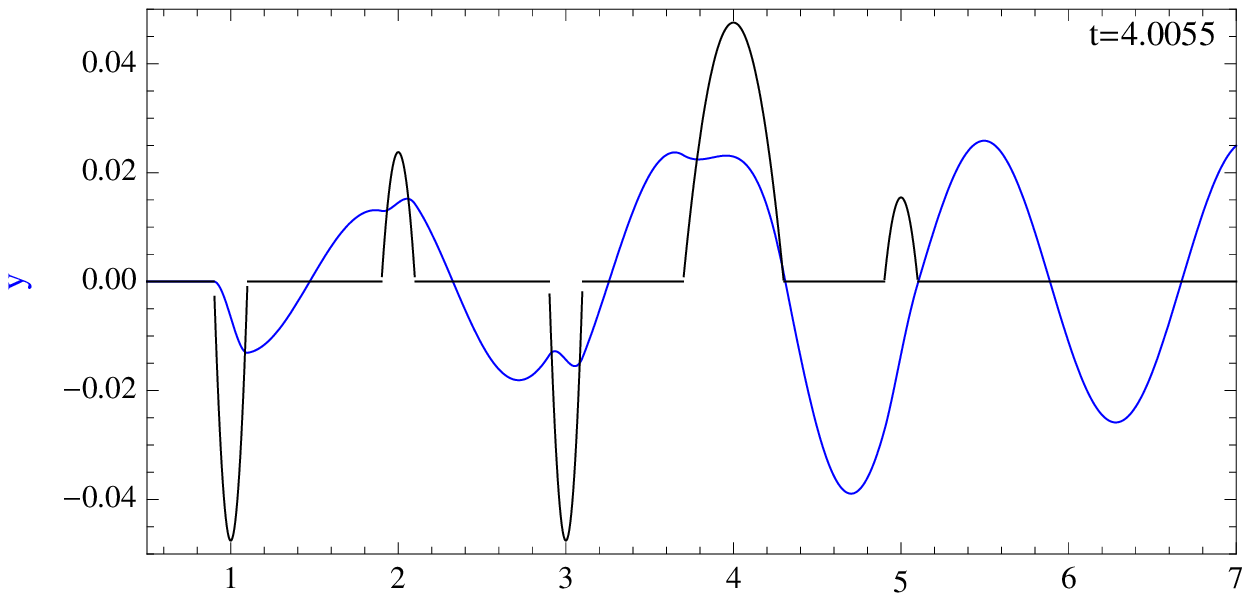}
\includegraphics[width=0.44 \textwidth, height=0.175 \textheight]{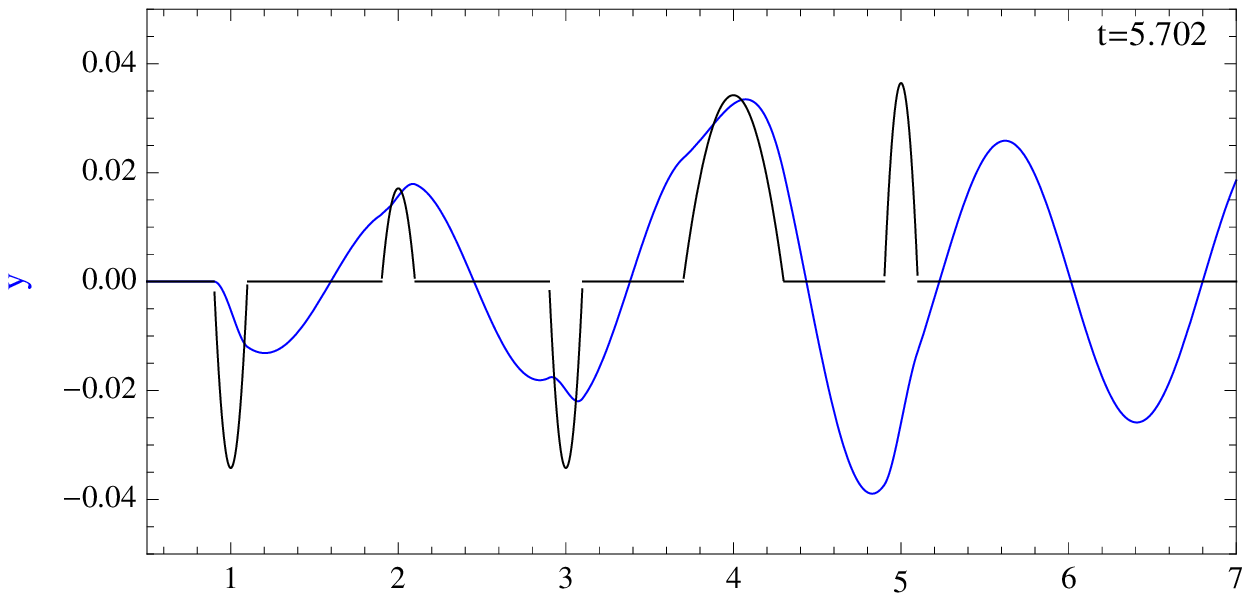} 
\includegraphics[width=0.44 \textwidth, height=0.175 \textheight]{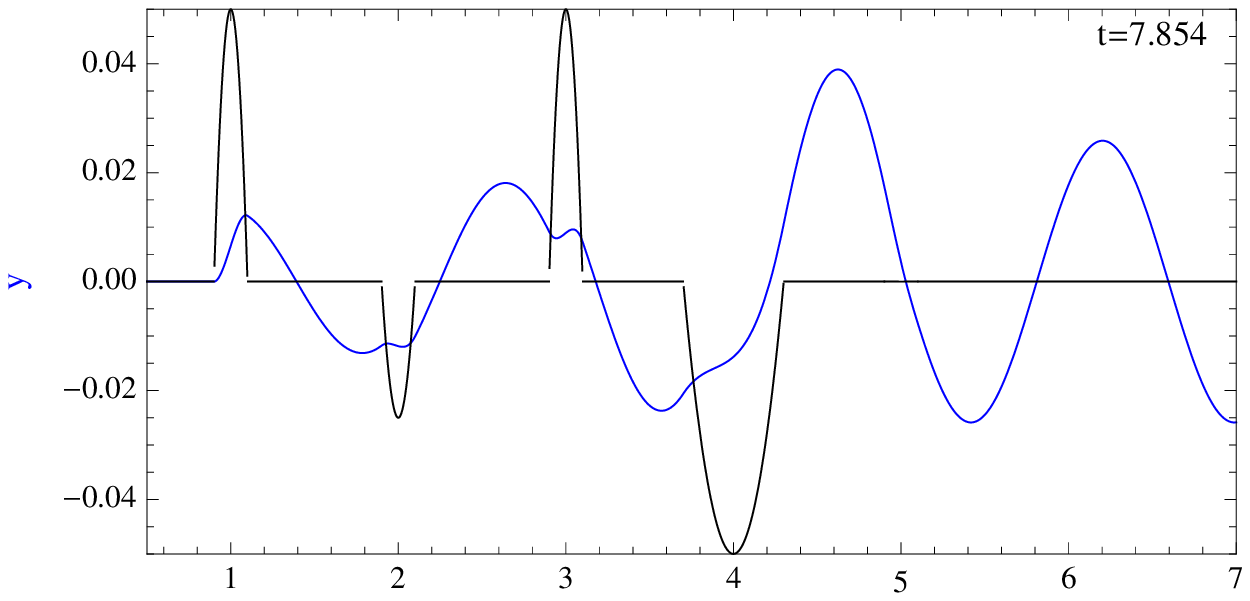}
\caption{The upstream streakline computed using (\ref{eq:crosschannelupstream}) and (\ref{eq:mcrosschannelupstream}), for
exactly the same parameters and times associated with the numerically obtained Fig.~\ref{fig:ccstreak}.}
\label{fig:ccmstreak}
\end{figure}

In Fig.~\ref{fig:ccmstreak}, the identical parameter and time values associated with the numerical streakline calculations
were used, but now the {\em theoretical} leading-order upstream streakline (expressions (\ref{eq:crosschannelupstream}) and
(\ref{eq:mcrosschannelupstream})) is plotted.  The agreement between Figs.~\ref{fig:ccstreak} and \ref{fig:ccmstreak} is
excellent.  One slight difference is that in the theoretical streakline as shown in Fig.~\ref{fig:ccmstreak}, the streakline extends
all the way across.  This is because the theoretical streakline has been computed 
by considering fluid particles going through $ \vec{a} = (0.5,0) $ at all times in the past.  In contrast, the numerical streaklines
shown in Fig.~\ref{fig:ccstreak} were obtained by synthetically releasing red dye at $ \vec{a} $ from time $ 0 $ onwards.  
While making a decision of this sort is inevitable in a numerical simulation, the theoretical expressions enable the {\em full}
streakline, associated with particles released at $ \vec{a} $ in the distant past, to be obtained.

\begin{figure}
\includegraphics[width=0.45 \textwidth]{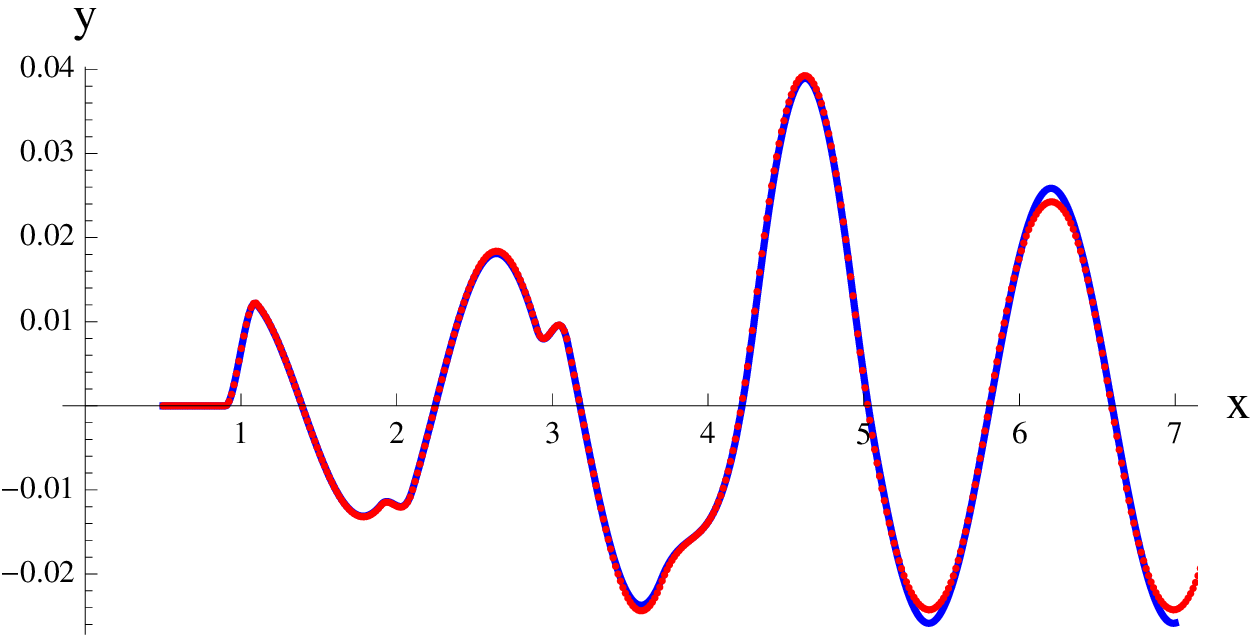}
\caption{Upstream streakline computed using (\ref{eq:mcrosschannelupstream}) at $ t = 10 \pi / \omega $ (blue curve), 
compared with the streakline computed using direct numerical simulation (red dots), for the cross-channel flow with channel
configuration as specified in the text.}
\label{fig:crosschannelcompare}
\end{figure}

\begin{figure}
\includegraphics[width=0.42 \textwidth]{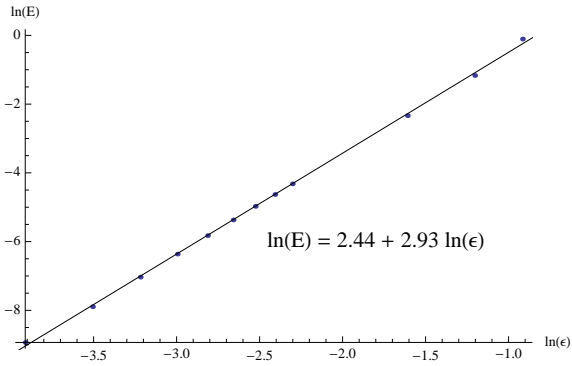}
\caption{The variation of the $ {\mathrm{L}}^2 $-error between the numerically simulated and the explicit approximation
with $ \eps $ (dots) for the cross-channel micromixer, in a log-log plot.}
\label{fig:ccerror}
\end{figure}

To compare the differences in more detail, Fig.~\ref{fig:crosschannelcompare} shows the
numerical simulations (red dots) and the explicit
approximation (blue curve) in one plot, at the time corresponding to the last panel in Figs.~\ref{fig:ccstreak} and \ref{fig:ccmstreak}.
This is five times the period $ 2 \pi / \omega $ of the flow (allowing for the numerically simulated streakline to approach the right-end of the figure),
and indeed it should be noted that the streakline $ \vec{x}_\eps^u(p,t) $ must also be periodic in $ t $ with exactly this period.
In general, however, the theory of the previous section does {\em not} require such periodicity.  The red dots are virtually
on top of the blue curve, while it should be noted that $ \eps = 0.1 $ is of moderate size.  The
error is further investigated in Fig.~\ref{fig:ccerror}, in which the square-sum ($ {\mathrm{L}}^2 $) error along the streakline, $ E $,  between the numerical
and explicit streaklines is computed at $ t = 10 \pi / \omega $ for different values of $ \eps $, and shown by the dots.  The linear fit in the log-log
plot indicates that the error goes as $ \eps^{2.9} $, which is consistent with the $ {\mathcal O}(\eps^2) $ prediction of the theory.

%%%%%%%%%%%%%%%%%%%
\subsection{Anomalous fluid in a vortex}

\begin{figure}
\includegraphics[width=0.45 \textwidth]{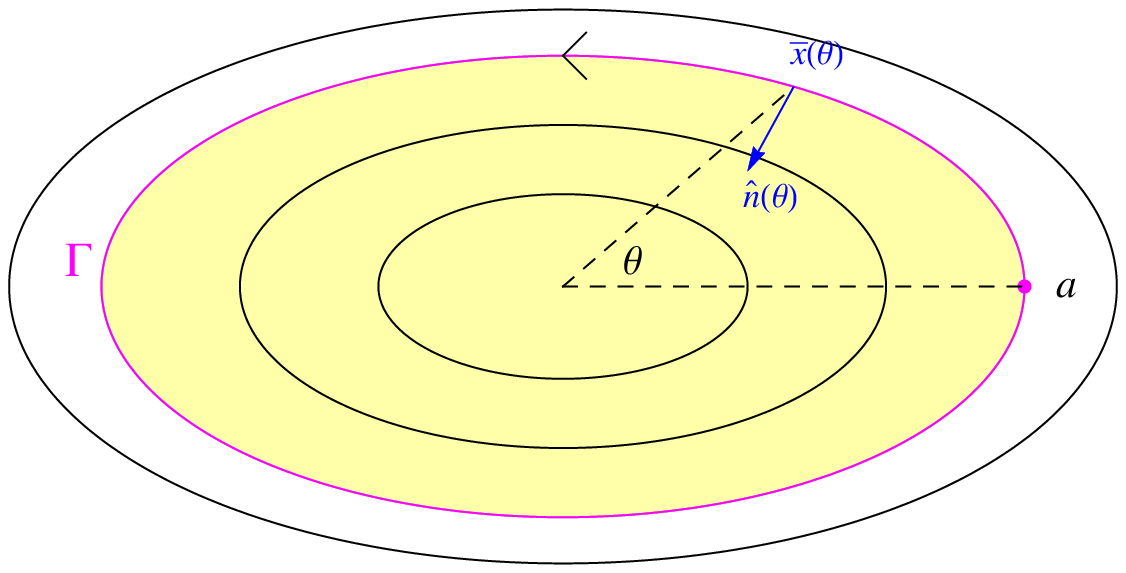} \\
\caption{Kirchhoff's elliptic vortex with a different fluid inside the streakline $ \Gamma $ (magenta).}
\label{fig:elliptic}
\end{figure}

For this example, the attitude adopted by Turner \cite{turner} of modeling the interaction of a coherent vortex with its
surroundings (consisting possibly of many other vortices  distant to it, and also the effect of boundaries) by using a weak
external strain field is adopted.  While it would be convenient to use a line or Gaussian vortex with circular streamlines
(on which particles flow at a constant speed) as the base flow, the utility of the method will be illustrated by using the
more complicated 
Kirchhoff's classical elliptic vortex \cite{kirchhoff,wang,mitchellrossi,friedland,meleshkovanheijst} as the prototype.
In nondimensional coordinates in 2D, this has the flow given by
\begin{equation}
\renewcommand{\arraystretch}{1.6}
\left. \begin{array}{l}
\dot{x} = - 2 y / m^2 \\
\dot{y} = 2 x / l^2 
\end{array} \right\} \, , 
\label{eq:elliptic}
\end{equation}
with $ m, l > 0 $, which consists of nested elliptical streamlines centered at the origin.  Suppose there are two different fluids inside and
outside the elliptic streamline $ \Gamma $ defined by
\[
\frac{x^2}{l^2} + \frac{y^2}{m^2} = 1 \, ,
\]
as shown in Fig.~\ref{fig:elliptic}, and take $ \vec{a} = (l,0) $.  Another rationalization for the choice of this particular streamline
could be that it is associated with a critical angular momentum value
as dictated by an outer flow \cite{turner}, thereby defining the `boundary' of the vortex; however, the `two-fluid' paradigm as
illustrated in Fig.~\ref{fig:elliptic} will be the main motivation which drives the analysis to follow.
Before introducing an external strain field as an unsteady velocity agitation, a useful parametrization shall be presented.
If $ \theta $ is the standard polar angle, then the ellipse has a parametrization $ \bar{\vec{x}}(\theta) = (\bar{x},\bar{y}) = (l \cos \theta, m \sin \theta) $ with
$ \theta = 0 $ being $ \vec{a} $.  So $ \theta $ will be used instead of $ p $ to identify location/particle along the streakline.
Therefore
\[
\left| \vec{u}(\vec{\bar{x}}(\theta) \right| = \sqrt{\frac{4 \bar{y}^2}{m^4} \! + \! \frac{4 \bar{x}^2}{l^4}} = \frac{2}{ml} \sqrt{l^2 \sin^2 \theta \! + \! 
m^2 \cos^2 \theta} \, .
\]
The rotation is anticlockwise around $ \Gamma $, and thus the relevant normal unit vector is
\[
\hat{\vec{n}}(\theta) \! := \! \frac{J \vec{u}\left(\bar{\vec{x}}(\theta)\right)}{\left| \vec{u}(\vec{\bar{x}}(\theta) \right|} 
\! = \! \frac{-1}{\sqrt{l^2 \! \sin^2 \theta \! + \! 
m^2 \! \cos^2 \theta}} \! \left( \begin{array}{c} \! \! m \cos \theta \\ l \sin \theta \end{array} \! \! \right) \, 
\]
which points `inwards' as shown in Fig.~\ref{fig:elliptic}.
If $ \tau $ is the time variation as a particle traverses $ \Gamma $, then
\[
\left| \vec{u}(\vec{\bar{x}}(\theta) \right| \d \tau = \sqrt{\bar{x}^2(\theta) + \bar{y}^2(\theta)} \, \d \theta \, , 
\]
and so the relationship between the time $ \tau $ and the polar location is 
\begin{equation}
\tau(\theta) = \frac{m l}{2} \int_0^\theta \sqrt{ \frac{m^2 \tan^2 \alpha + l^2}{l^2 \tan^2 \alpha + m^2}} \, \d \alpha
\label{eq:tautheta}
\end{equation}
where the choice $ \tau = 0 $ when $ \theta = 0 $ (i.e., at $ \vec{a} $) has been made. 

Now, following a commonly modeled idea \cite{romkedarjfm,kida,delcastillonegrete,meuniervillermaux,leweke,mula,turner,bassomgilbert,zhmur}, suppose the vortex is
placed in an unsteady strain field.  Here, this is modeled by the inclusion of a weak  unsteady velocity
agitation $ \vec{v} $ added to (\ref{eq:elliptic}), subject to the constraint that $ \vec{v}(\vec{a},t) = 0 $ for all $ t $.
Then, Theorem~\ref{theorem:upstream} gives the fact that the unsteady streakline at a location $ \theta $
and time $ t $ perturbs in the direction $ \hat{\vec{n}}(\theta) $   by an amount $ M_v^u(\theta,t) / \left|
\vec{u} \left( \bar{\vec{x}}(\theta) \right) \right| $ (see also
Fig.~\ref{fig:elliptic}).  The $ v $ subscript used here is for `{\em v}ortex,' to distinguish $ M^u $ from that of the previous
example.  Thus, the $ x $- and $ y $-coordinates of the unsteady streakline to leading-order obey
\begin{equation}
x_\eps^u(\theta,t) = l \cos \theta \left[ 1 - \frac{M_v^u(\theta,t) m^2}{2 \left( l^2 \sin^2 \theta + m^2 \cos^2 \theta \right)}
\right]
\label{eq:ellipticx}
\end{equation}
and
\begin{equation}
y_\eps^u(\theta,t) = m \sin \theta \left[ 1 - \frac{M_v^u(\theta,t) l^2}{2 \left( l^2 \sin^2 \theta + m^2 \cos^2 \theta \right)}
\right] \, .
\label{eq:ellipticy}
\end{equation}
The value of $ M_v^u(\theta,t) $ can be obtained from (\ref{eq:mupstream}), with $ p $ now identified with $ \theta $, 
and by recasting the integral with respect to the polar angle as opposed to $ \tau $ using (\ref{eq:tautheta}):
\begin{widetext}
\begin{equation}
M_v^u(\theta,t) = - \int_0^\theta \left( m \cos \alpha, l \sin \alpha \right) \cdot \vec{v} \! \left( l \cos \alpha, m \sin \alpha, 
t + \frac{m l}{2} \int_\theta^\alpha \! \sqrt{ \frac{m^2 \tan^2 \beta \! + \! l^2}{l^2 \tan^2 \beta \! + \! m^2}} \, d \beta \right) \, \left( \sqrt{ \frac{m^2 \tan^2 \alpha \! + \! l^2}{l^2 \tan^2 \alpha \! +\!  m^2}} \, \d \alpha
\right) \, .
\label{eq:meluelliptic}
\end{equation}
\end{widetext}
In obtaining (\ref{eq:meluelliptic}), several factors such as the incompressibility of the flow in (\ref{eq:elliptic}), and the
rewriting using (\ref{eq:tautheta}) of  $ \tau(\alpha) - \tau(\theta) $ in the temporal argument of $ \vec{v} $, have been used.
There should be a term $  \I_{[-P,P]}(p) $ in (\ref{eq:mupstream}) which multiplies the above expression (but has not been
explicitly stated), where $ P $ is any large
number; this simply means that (\ref{eq:melelliptic}) is valid for any finite $ \theta $, but not for $ \theta = \pm \infty $.

\begin{figure}
\vspace*{-0.3cm}
\includegraphics[width=0.5 \textwidth, height=0.18 \textheight]{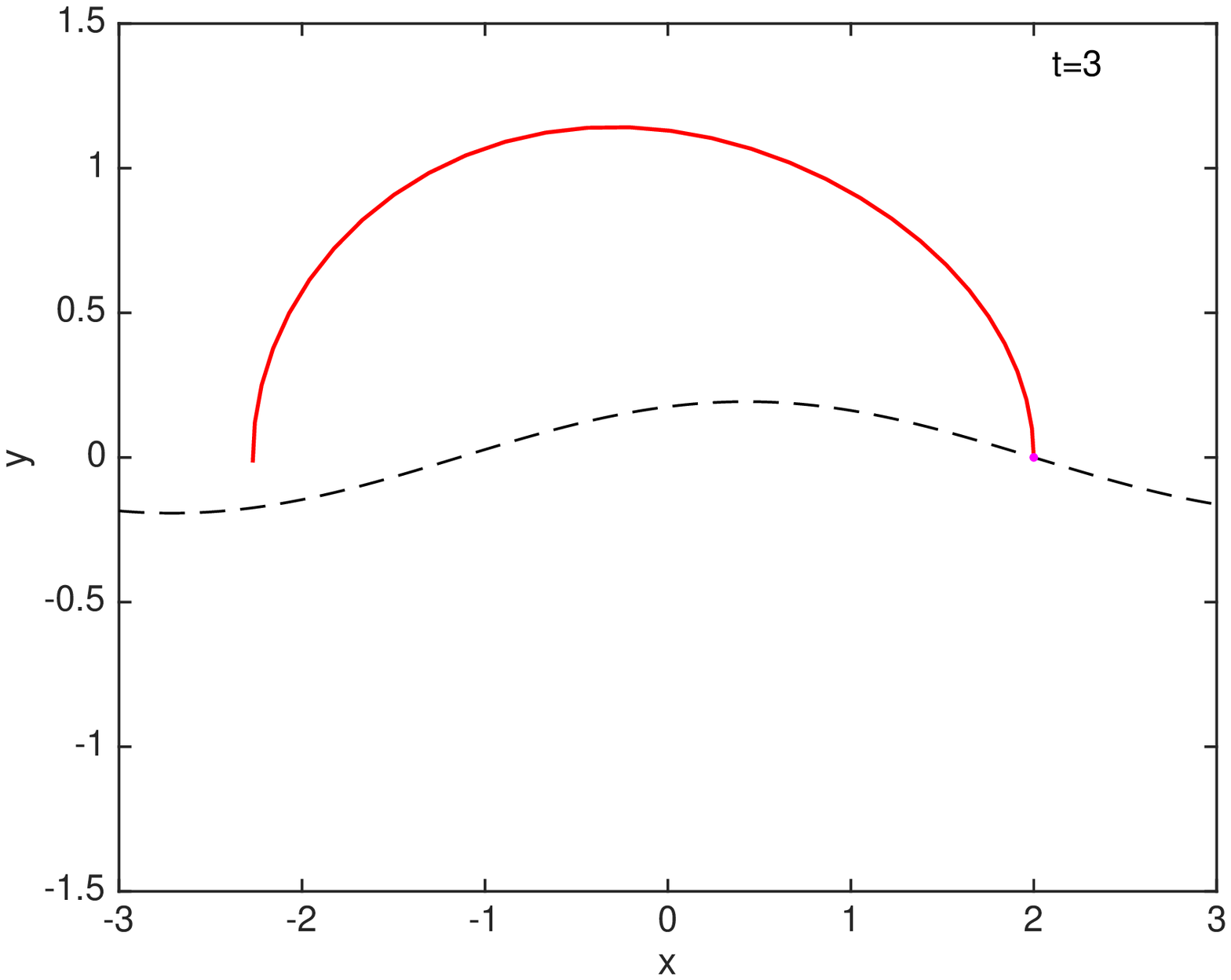} 
\includegraphics[width=0.5 \textwidth,  height=0.18 \textheight]{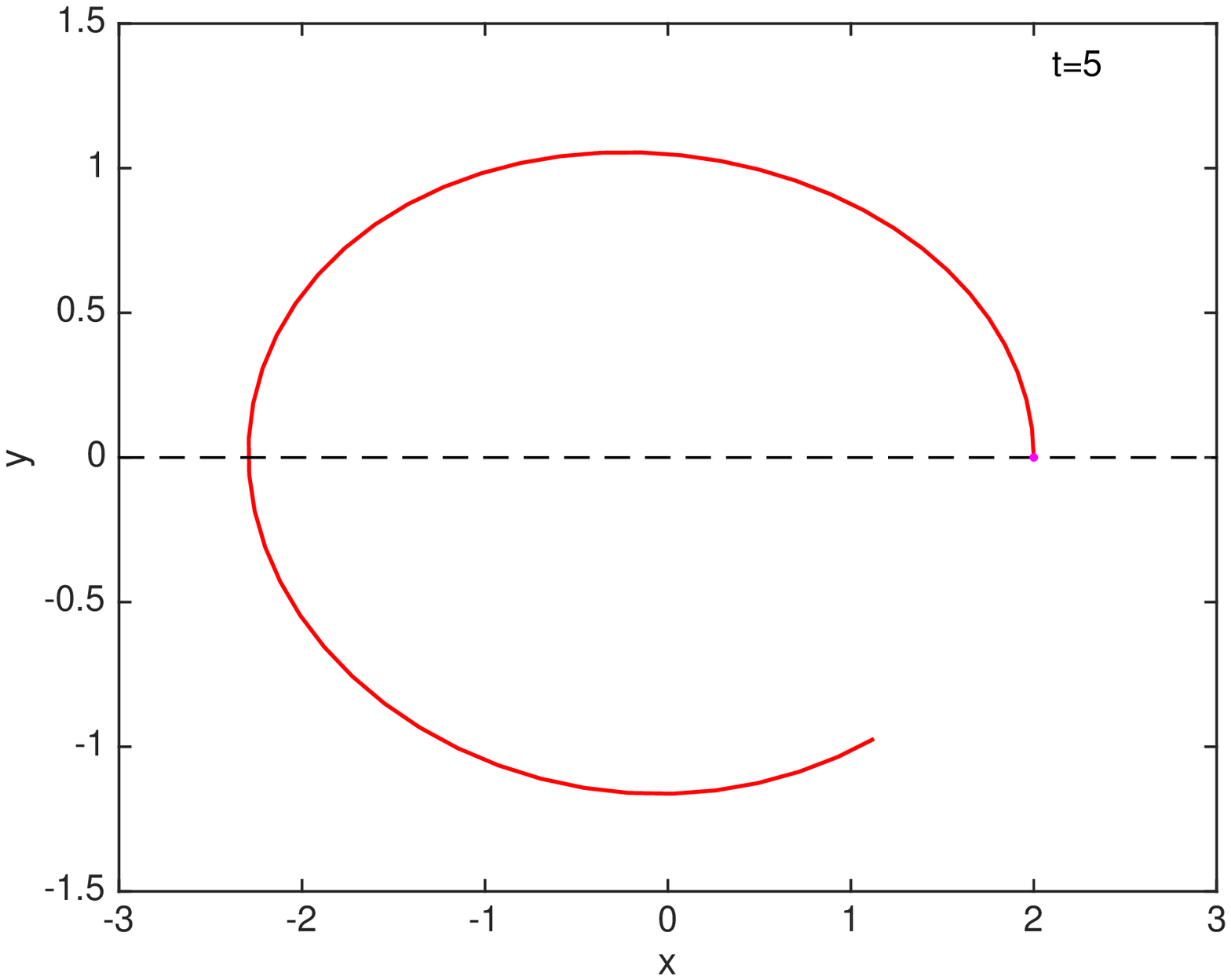} 
\includegraphics[width=0.5 \textwidth,  height=0.18 \textheight]{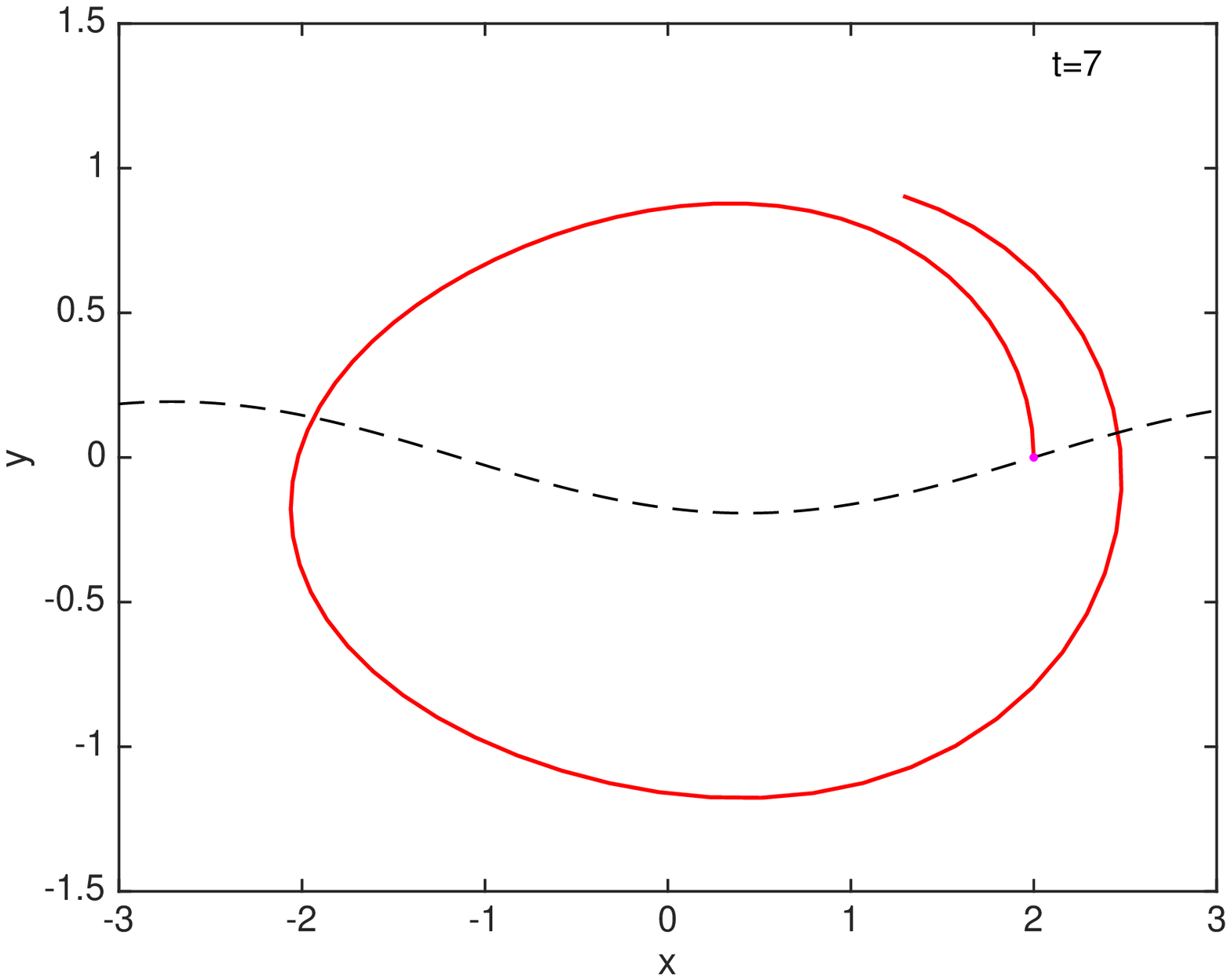} 
\includegraphics[width=0.5 \textwidth,  height=0.18 \textheight]{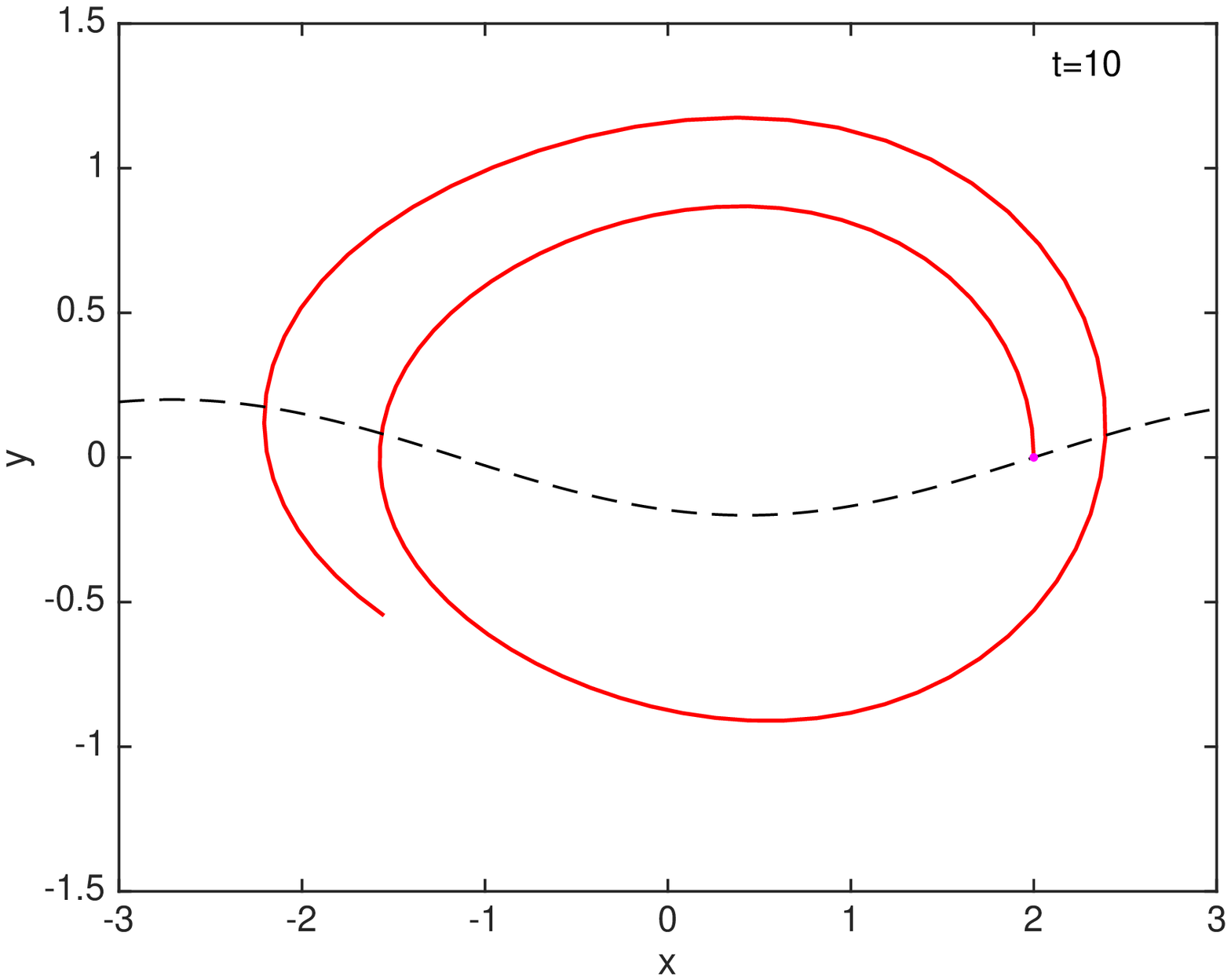}
\caption{Evolution of streakline, with dye released on the fluid interface at $ (2,0) $ from time $ 0 $ onwards.
The evolving streakline, representing the perturbed fluid interface, is shown in red, with the instantaneous velocity
agitation (\ref{eq:agitationelliptic})'s $ y $-component variation  shown by the black dashed curves.}
\label{fig:evstreak}
\end{figure}

\begin{figure}
\includegraphics[width=0.41 \textwidth,  height=0.162 \textheight]{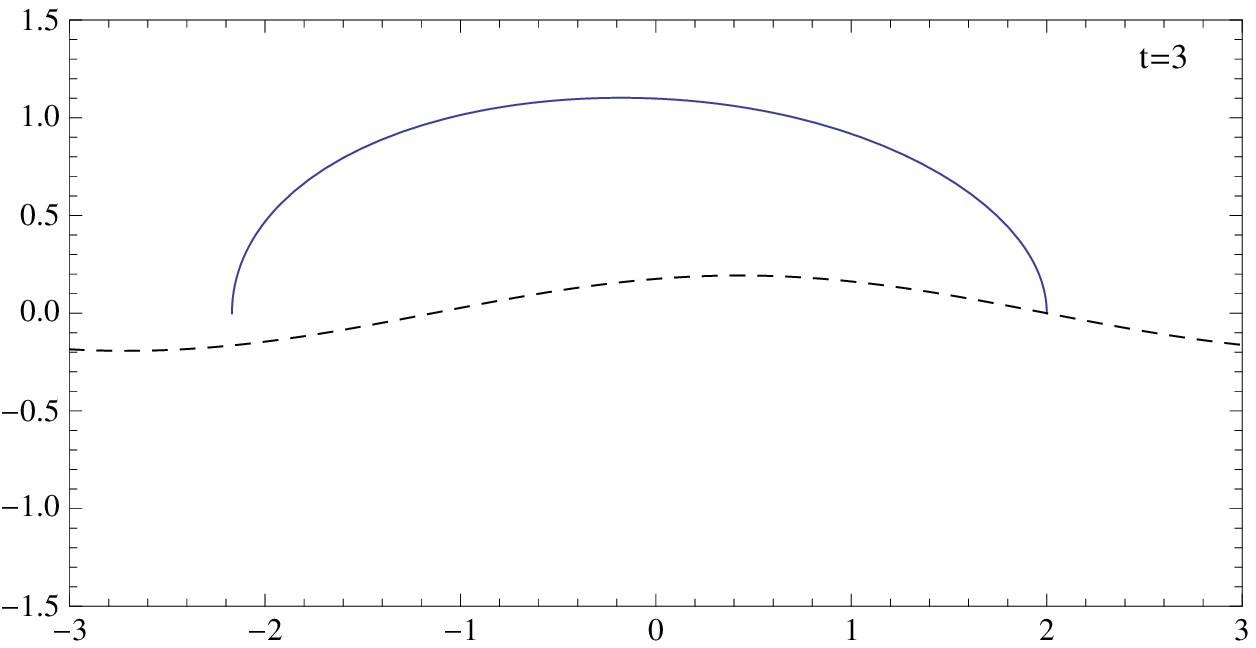} \\ \vspace*{0.4cm}
\includegraphics[width=0.41 \textwidth, height=0.162 \textheight]{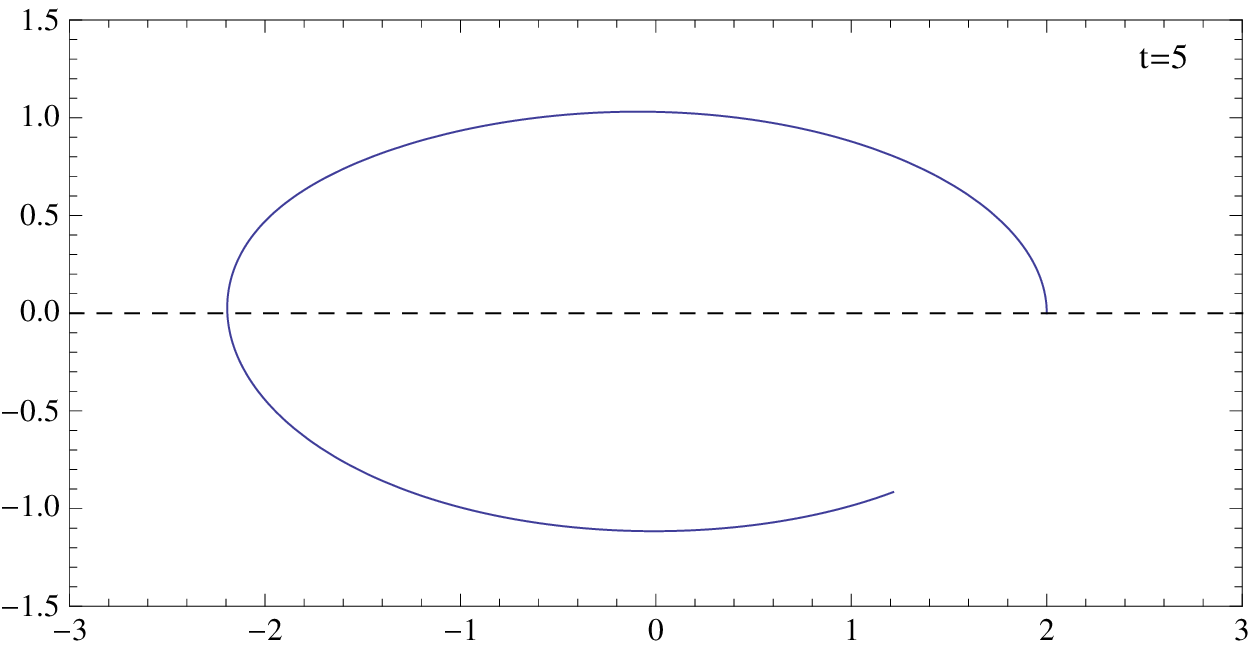} \\ \vspace*{0.4cm}
\includegraphics[width=0.41 \textwidth, height=0.162 \textheight]{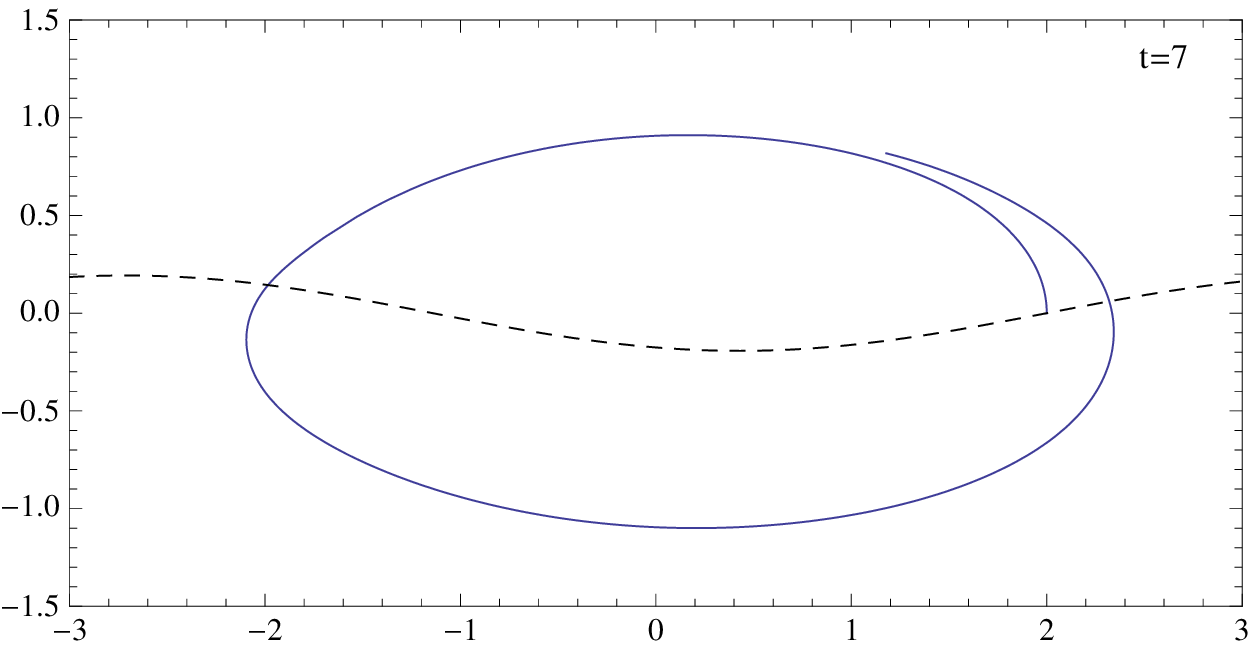} \\ \vspace*{0.4cm}
\includegraphics[width=0.41 \textwidth, height=0.162 \textheight]{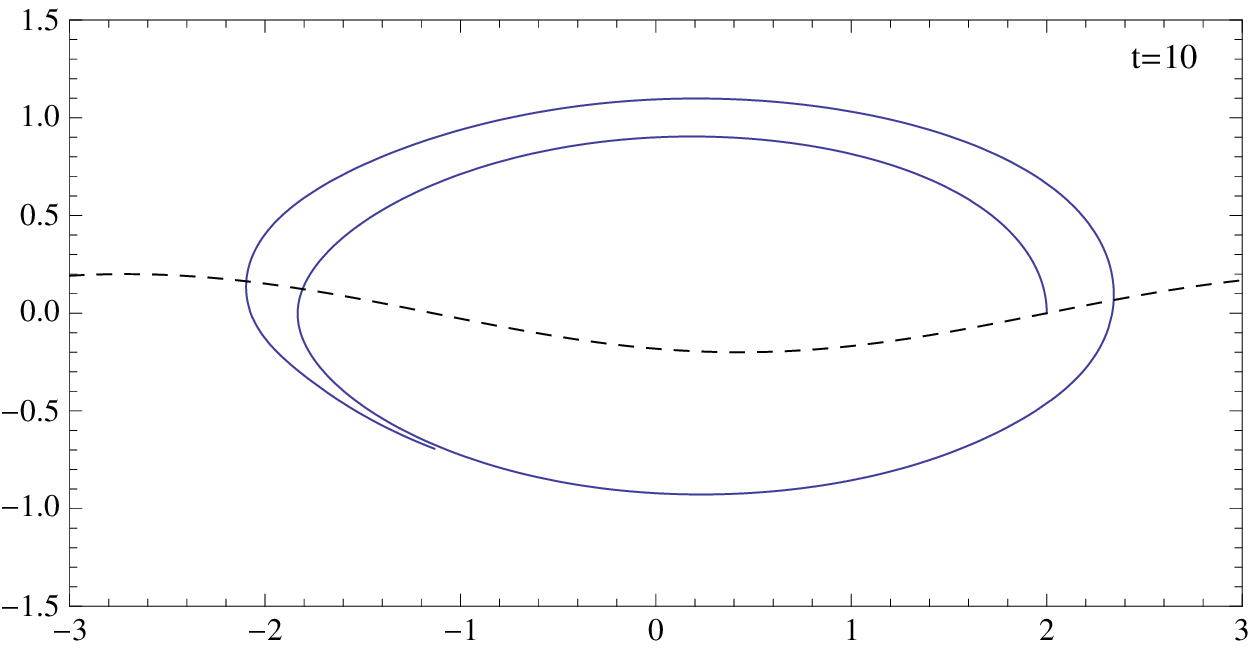}
\caption{The streakline computed using (\ref{eq:ellipticx}), (\ref{eq:ellipticy}) and (\ref{eq:meluelliptic}), for
exactly the same parameters and times associated with the numerically obtained Fig.~\ref{fig:evstreak}.}
\label{fig:evmstreak}
\end{figure}

Next, the theoretical flow interface, as characterized by the unsteady streakline expression above, shall be verified for
a particular choice of unsteady velocity agitation $ \vec{v} $. Suppose that, conforming with $ \vec{v} (\vec{a},t) = \vec{0} $
for all $ t $,
\begin{equation}
\vec{v}(x,y,t) = \eps \sin \left( x - l \right) \left( \begin{array}{c} 0 \\ 1 \end{array} \right) \tanh \left( t - 5 \right)
\label{eq:agitationelliptic}
\end{equation}
is chosen, where $ \left| \eps \right| $ is small.  This represents an agitation in the $ y $-direction which is modulated 
periodically in $ x $, but {\em aperiodically} in time.  Thus, 
the expressions in (\ref{eq:ellipticx}) and (\ref{eq:ellipticy}) will be the expressions for the streakline, with error 
$ {\mathcal O}(\eps^2) $.  The general expression (\ref{eq:meluelliptic}) becomes in this situation
\begin{widetext}
\begin{equation}
M_v^u(\theta,t) = - \eps l \int_0^\theta \sin \alpha \sin \left[ l (\cos \alpha - 1) \right] \tanh \left[ t - 5 + 
\frac{m l}{2} \int_\theta^\alpha \! \sqrt{ \frac{m^2 \tan^2 \beta \! + \! l^2}{l^2 \tan^2 \beta \! + \! m^2}} \, d \beta \right]
\, \sqrt{ \frac{m^2 \tan^2 \alpha \! + \! l^2}{l^2 \tan^2 \alpha \! +\!  m^2}} \, \d \alpha
 \, .
\label{eq:meluellipticex}
\end{equation}
\end{widetext}
Numerical simulations of the streakline passing through $ (l,0) $ are shown in Fig.~\ref{fig:evstreak}, where $ l = 2 $, 
$ m = 1 $, with red dye is released from time $ 0 $ onwards.  To accentuate the variation displayed, the relatively large
value of $ \eps = 0.2 $ is used.  It should be noted that as $ t $ 
increases, the streaklines shown are not simple retracings and extensions of previous curves; the previous curves are
themselves moving.  The velocity agitation (\ref{eq:agitationelliptic})
considered here is purely in the $ y $-direction, and displays a transition at $ t = 5 $ between two (almost) stationary states;
this is displayed by the black dashed curve (scaled in the $ y $-direction to be visible in this plot). 
A movie of the streakline evolution is provided with the Supplementary Materials.  As the streakline wraps around, in this
case the inner parts of the streakline accummulate towards an almost elliptic trajectory.  The analytical expressions given
by (\ref{eq:ellipticx}), (\ref{eq:ellipticy}) and (\ref{eq:meluelliptic}) are used to generate Fig.~\ref{fig:evmstreak}.  By choosing
the range of $ \theta $, the analytical streakline can be computed beyond the lead point of the figures in Fig.~\ref{fig:evstreak}
(representing dye released from $ (l,0) $ before $ t = 0 $), and also backwards from the point $ (l,0) $ (representing points
which will go through $ (l,0) $ in the future).  However, in producing Fig.~\ref{fig:evmstreak}, a $ \theta $ range which is 
approximately that displayed in Fig.~\ref{fig:evstreak} has been used to enable comparison.  The agreement between the
theoretical and numerical streaklines is good even at this value of $ \eps $.  However, as the streakline wraps around, the
analytical expression loses its ability to match the simulation.  The reason for this is that in the analytical expression, the
leading-order velocity {\em on $ \Gamma $} is what is being used, even as it wraps around.  In reality, however, after wrapping
around once, the streakline would have ventured into a different location.  This is still $ {\mathcal O}(\eps) $-close which 
means the analytical expression is rigorous, but the $ {\mathcal O}(\eps^2) $ error term, which is valid for $ p \in [-P,P] $
in terms of the definition of the upstream streakline (\ref{eq:upstreamunsteady}), now kicks in.  In terms of $ \theta $, this
means that the error can increase outside a domain $ \theta \in [-\Theta,\Theta] $.  This fact is being displayed in Fig.~\ref{fig:evmstreak}, with the error clearly increasing at large $ \theta $.  The matching between the curves when restricted to wrapping around just once is very good, and
smaller $ \eps $ values (not pictured), have greater accuracy (even for several wraps around).  Furtunately, as shall be 
seen in Section~\ref{sec:transport}, restricting to wrapping around less than once is sufficient {\em in evaluating transport of fluid}. 

%%%%%%%%%%%%%%%%%%%%%%%
%%%%%%%%%%%%%%%%%%%%%%%%
\section{Transport quantification}
\label{sec:transport}

Previous sections outlined theory and examples in determining the upstream and downstream streaklines under a velocity agitation.  Before the agitation, these coincided, and ran along the flow interface between the two fluids.  Thus, in that situation,
there was no transport across the flow interface.  The issue now is to attempt to quantify the transport `across the flow interface'
after the agitation.  But what exactly {\em is} the flow interface after applying the agitation?  Does it make sense to compute
a flux across a stationary (Eulerian) curve?  How can a flux in a {\em Lagrangian} sense be defined?

The difficulties here are familiar in a different situation: when the interface consisted of a coincident stable and unstable manifold
(a so-called heteroclinic manifold) before perturbation.  After the agitation, it would split into stable and unstable manifolds which are not coincident, and which
moreover move with time.  In this case, the concept of lobe dynamics \cite{romkedarjfm,wiggins} can be applied when the agitation is 
time-periodic in a specific way, or for more general perturbations it is possible to define an instantaneous transport
\cite{aperiodic,droplet}.  However, the current situation is different: the original flow interface is {\em not} a heteroclinic manifold.
Nevertheless, the ideas from \cite{aperiodic,droplet,siam_book} can be adapted to this situation.

\begin{figure}[t]
\includegraphics[width=0.5 \textwidth]{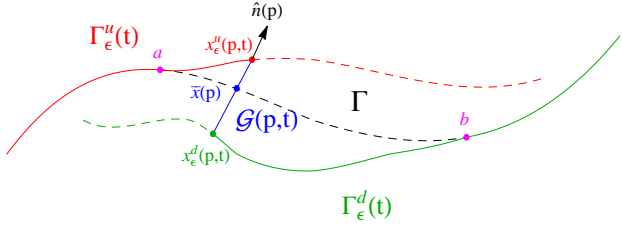} \\
\caption{Construction of the pseudo-streakline for transport assessment, from the configuration in Fig.~\ref{fig:unsteady}.}
\label{fig:pseudostreakline}
\end{figure}

Suppose that, before perturbation, the flow interface was unequivocally defined by $ \Gamma $, as shown in Fig.~\ref{fig:gamma}.
Now, when the velocity agitation is included, at each time $ t $ there will be an upstream streakline going through $ \vec{a} $,
and a downstream streakline going through $ \vec{b} $, as shown in Fig.~\ref{fig:unsteady}.  Consider fluid arriving near to 
$ \vec{a} $ from upstream.  The fluid on either side of the upstream streakline is different, and therefore will continue to be different
as time progresses forward.  In other words, the upstream streakline can be considered a flow interface between
the two fluids in {\em forward} time.  On the other hand, consider fluid on the two sides of the downstream streakline to
downstream of $ \vec{b} $.    Had there been no velocity agitation, the downstream streakline would separate the two fluids as one
progresses downstream from $ \vec{b} $.    But the presence of the velocity agitation means that this separation has been disturbed, and the relative positioning of the upstream and downstream streaklines, and their movement with time, affects 
the transfer of fluid.  How can this transfer be assessed, bearing in mind that the
streaklines shown in Fig.~\ref{fig:unsteady} are wiggling around with time, and may (or may not) intersect in various ways
in the agitation region?

The trick to computing a transfer, specifically as an instantaneous flux of fluid, is to use the idea of a gate which was
originally suggested by Poje and Haller \cite{pojehaller} (and used elsewhere \cite{miller}) 
as a strategy for flux determination across numerically determined stable and unstable manfolds, but later
adapted by Balasuriya to obtain a flux for a broken heteroclinic situation in time-aperiodic flows \cite{aperiodic,siam_book}.
Consider a time $ t $.  Take a parameter value $ p \in [p^u,p^d] $, and consider the point $ \bar{\vec{x}}(p) $ on $ \tilde{\Gamma} $, i.e., a point
on the original steady streakline but within the agitation region.  Draw a normal line to $ \tilde{\Gamma} $ at this point,
extending far enough out to intersect both the upstream $ \Gamma_\eps^u(t) $ and the downstream streakline 
$ \Gamma_\eps^d(t) $.  The intersections occur at the points $ \vec{x}_\eps^u(p,t) $ and $ \vec{x}_\eps^d(p,t) $
respectively, and these must be $ {\mathcal O}(\eps) $ near to $ \bar{\vec{x}}(p) $.  This construction is shown in Fig.~\ref{fig:pseudostreakline}, based on the streakline configuration in Fig.~\ref{fig:unsteady}.
This line between $ \vec{x}_\eps^u(p,t) $ and $ \vec{x}_\eps^d(p,t) $ is the gate $ {\mathcal G}(p,t) $.  Now 
define the {\em pseudo-streakline} to be the three connected curves (i) the upstream streakline until it hits $ {\mathcal G}(p,t) $, (ii)
the gate $ {\mathcal G}(p,t) $, and (iii) the downstream streakline continuing on from that point.  This is the collection of
solid curves in Fig.~\ref{fig:pseudostreakline}.

\begin{figure}[t]
\includegraphics[width=0.5 \textwidth]{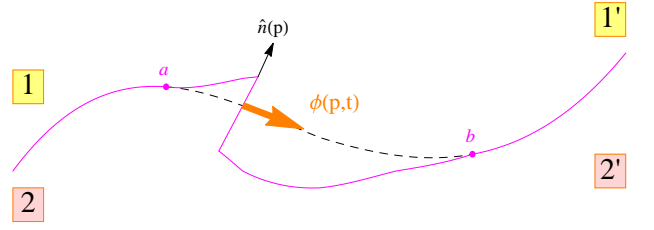} \\
\caption{Transport across the pseudo-streakline (magenta), reflecting fluid exhange across the flow interface.}
\label{fig:transport}
\end{figure}

The pseudo-streakline is a method for trying to identify a `nominal' flow interface, in this unsteady instance in which there
actually is no impermeable barrier (since transport occurs between the two fluids).  Moreover, since it is defined using segments of 
$ \Gamma_\eps^{u,d}(t) $, it specifically incorporates the Lagrangian nature of the flow, which is essential in determining
{\em transport}.  To see how transport can be quantified, refer to Fig.~\ref{fig:transport}, which retains the pseudo-streakline
as constructed in Fig.~\ref{fig:pseudostreakline}.  Suppose the fluid upstream of $ \vec{a} $ is labeled $ 1 $ and $ 2 $;
the upstream streakline $ \Gamma_\eps^u(t) $ separates these two fluids in forwards time.  On the other hand, suppose the
fluids downstream of $ \vec{b} $ were labeled $ 1' $ and $ 2' $.  In the absence of a velocity agitation, the upstream and
downstream streaklines would coincide, and lie exactly along $ \Gamma $.  Thus, fluid from region $ 1 $ will go to $ 1' $, and $ 2 $ to 
$ (2') $, with no intermingling; $ \Gamma $ is in this steady instance an unequivocal flow interface.  The agitation has broken
this interface into {\em two} entities (the upstreak and downstream streaklines), which are both moving with time, leading to
difficulty in defining an interface.  The pseudo-streakline incorporates information from both these entities.  The transport of
fluid across the pseudo-streakline will help define a fluid transport, as follows.

The wish is to quantify how fluid $ 1 $ transfers to $ 2' $, and how $ 2 $ transfers to $ 1' $; these were both zero when there
was no unsteady agitation. Now, examining Fig.~\ref{fig:transport},
the transfer of fluid $ 2 $ to $ 1' $, across the pseudo-streakline, 
at this instance in time, only occurs by fluid crossing $ {\mathcal G}(p,t) $.  This is because
portions of $ \Gamma_\eps^{u,d}(t) $ which are also part of the pseudo-streakline {\em cannot} be crossed since these are flow separators.  Therefore, the transport of fluid across the gate explicitly characterizes the transfer of fluid from $ 2 $ to $ 1' $
in the situation pictured.  In this case, note that the transfer, in relation to the unperturbed $ \Gamma $, is in the direction
of the normal vector $ \hat{\vec{n}}(p) $ across $ \Gamma $.  
If the upstream and downstream streaklines were positioned in the opposite orientation (i.e., 
$ \vec{x}_\eps^d(p,t) $ was above $ \vec{x}_\eps^u(p,t) $ along $ {\mathcal G}(p,t) $ in Fig.~\ref{fig:transport}), then the transfer of fluid
across the pseudo-streakline would be associated with fluid $ 1 $ going to $ 2' $.  This is in the direction $ - \hat{\vec{n}}(p) $
across $ \Gamma $.  In either case, this is clearly an {\em exchange} of fluid.  

Suppose the gate is parametrized in terms of the
arclength $ \ell $ along it.  Let $ U^\perp (\ell,t) $ be the normal component of the full velocity field across the gate at each
location, i.e., the component of the velocity in the direction of $ - J \hat{\vec{n}}(p) $.  Then, the instantaneous flux across the 
pseudo-streakline is by
definition
\begin{equation}
\phi(p,t) = \int_{{\mathcal G}(p,t)} U^\perp(\ell,t) \d \ell \, .
\label{eq:fluxdef}
\end{equation}
Note that the velocity normal to $ {\mathcal G} $
in general includes contributions from both the steady ($ \vec{u} $) and unsteady ($ \vec{v} $) terms, and moreover is not the same
value at all points on $ {\mathcal G} $.  The $ \phi $ above is a explicitly a flux in the sense that
it is an area of fluid per unit time which crosses the pseudo-streakline instantaneously.  If in the positive $ \hat{\vec{n}}(p) $ 
direction (as in Fig.~\ref{fig:transport}), $ \phi $ will be positive.  In the upstream/downstream streaklines have the opposite
orientation along the normal vector, then it will be negative.  When thinking of a time-varying flux, it is best to think of $ p $
as fixed, corresponding to fixing the location of the gate.  As time $ t $ varies, however, $ \phi(p,t) $ will change,
since the locations $ \vec{x}_\eps^{u,d}(p,t) $ will vary along the normal vector.  At some instances in time, $ \vec{x}_\eps^u(p,t) $
and $ \vec{x}_\eps^d(p,t) $ can interchange their relative positions.  If so, these points will `go through each other' on the gate,
at that
instance in time.  The instantaneous flux at this time will be zero.  However, once they have gone through one another, there
will be flux in the {\em opposite} direction.  The continuing time-variation of $ \phi(p,t) $ will indicate how fluid continued to
go back and forth.

If $ \tilde{\Gamma} $ were closed, the pseudo-streakline would be a closed curve consisting of parts of the upstream and downstream
streaklines going through $ \vec{a} $, capped by a gate.  (Think of glueing $ \vec{a} $ to $ \vec{b} $ in Figs.~\ref{fig:transport}
and \ref{fig:pseudostreakline}, and throwing away the parts upstream of $ \vec{a} $ and downstream of $ \vec{b} $.)
This is a nominal flow barrier between the interior and exterior fluids.
Moreover, the upstream and downstream streaklines must connect smoothly at $ \vec{a} $, since adjacent points are associated
with dye particles which went though $ \vec{a} $ a moment ago (on the upstream streakline), and which will go through $ \vec{a} $
in a moment (on the downstream streakline).  Since $ \vec{a} $ remains indubitably on the flow interface, one might consider that
the streakline (upstream and downstream) passing through $ \vec{a} $ is the flow interface.  However, in general these
do not coincide as one wraps around $ \tilde{\Gamma} $, and there will have to be a gate of finite size which connects them at
the location $ \bar{\vec{x}}(p) $.  Transport between the interior and exterior fluids---crossing the nominal flow interface---will
in general occur by fluid crossing the gate.

Thus in either the open or closed situation, (\ref{eq:fluxdef}) gives the instananeous flux across the nominal flow interface.  
A simple leading-order expression for $ \phi $ is possible:

\begin{theorem}[Instantaneous transport]
\label{theorem:transport}
Consider a pseudo-streakline at time $ t $ as defined in Fig.~\ref{fig:transport}, with the gate drawn at the location 
$ \bar{\vec{x}}(p) $.  The instantaneous flux $ \phi(p,t) $ across the pseudo-streakline is 
\begin{equation}
\phi(p,t) = M(p,t) + {\mathcal O}(\eps^2)
\label{eq:transport}
\end{equation}
where
\begin{widetext}
\begin{equation}
M(p,t):=  \int_{p^u}^{p^d}  \exp \left[ \int_\tau^p  \left[  
\vec{\nabla}  \cdot \vec{u} \right] \left( \bar{\vec{x}}(\xi) \right) \d \xi \right] \left[ J \vec{u} \left( \bar{\vec{x}}(\tau) \right) \right] \cdot 
\vec{v} \left( \bar{\vec{x}}(\tau), \tau +  t -  p \right) \d \tau \, . 
\label{eq:melnikov}
\end{equation}
\end{widetext}
\end{theorem}

For the proof, the reader is referred to Appendix~\ref{app:transport}.  The power of Theorem~\ref{theorem:transport} is that,
unlike in the definition (\ref{eq:fluxdef}), the instantaneous flux can be represented in terms of known quantities from the steady
flow, and the unsteady velocity.  Unsteady {\em trajectories} (and in particular the unsteady streaklines) are not needed.

A question that might arise is the effect of the location of the gate on the transport quantification.  This dependence occurs
precisely because a genuine flow interface does not exist after velocity agitation, and therefore a choice needs to be made
in demarcating such a interface.  This is almost exactly the same issue pondered by Rom-Kedar and collaborators \cite{romkedarjfm,wiggins} in the freedom of deciding on a `pseudo-separatrix' associated with transport across a heteroclinic
manifold in an specifically time-periodic vortical flow.  While the current problem is neither heteroclinic nor time-periodic, having
to make such a choice is inevitable in the absence of a genuine flow interface.  However, if the flow is incompressible, it turns
out that this freedom for choosing the location of the gate is actually a spurious freedom.  This is because if one takes 
(\ref{eq:melnikov}) under the hypothesis that $ \vec{\nabla} \cdot \vec{u} = 0 $, it is clear that $ p $ and $ t $ do not appear
independently in the instantaneous flux $ M(p,t) $, but together in the combination $ (t-p) $.  Thus, a shifting of $ p $ (corresponding
to choosing a different location for the gate) merely shifts the time-variation of the flux.  
 
%%%%%%%%%%%%%%%%%%%%%%
%%%%%%%%%%%%%%%%%%%%%%%
\section{Transport validation}
\label{sec:transportexamples}

The two examples examined in Section~\ref{sec:streakexamples} are now re-examined.  Here, the focus is on
determining the advective transport resulting from the interface streakline separating into upstream and downstream streaklines,
as outlined in the previous section.

%%%%%%%%%%%%%%%%%%%%
\subsection{Two fluids in a microchannel}

\begin{figure*}
\vspace*{-0.3cm}
\includegraphics[width=0.45 \textwidth, height=0.18 \textheight]{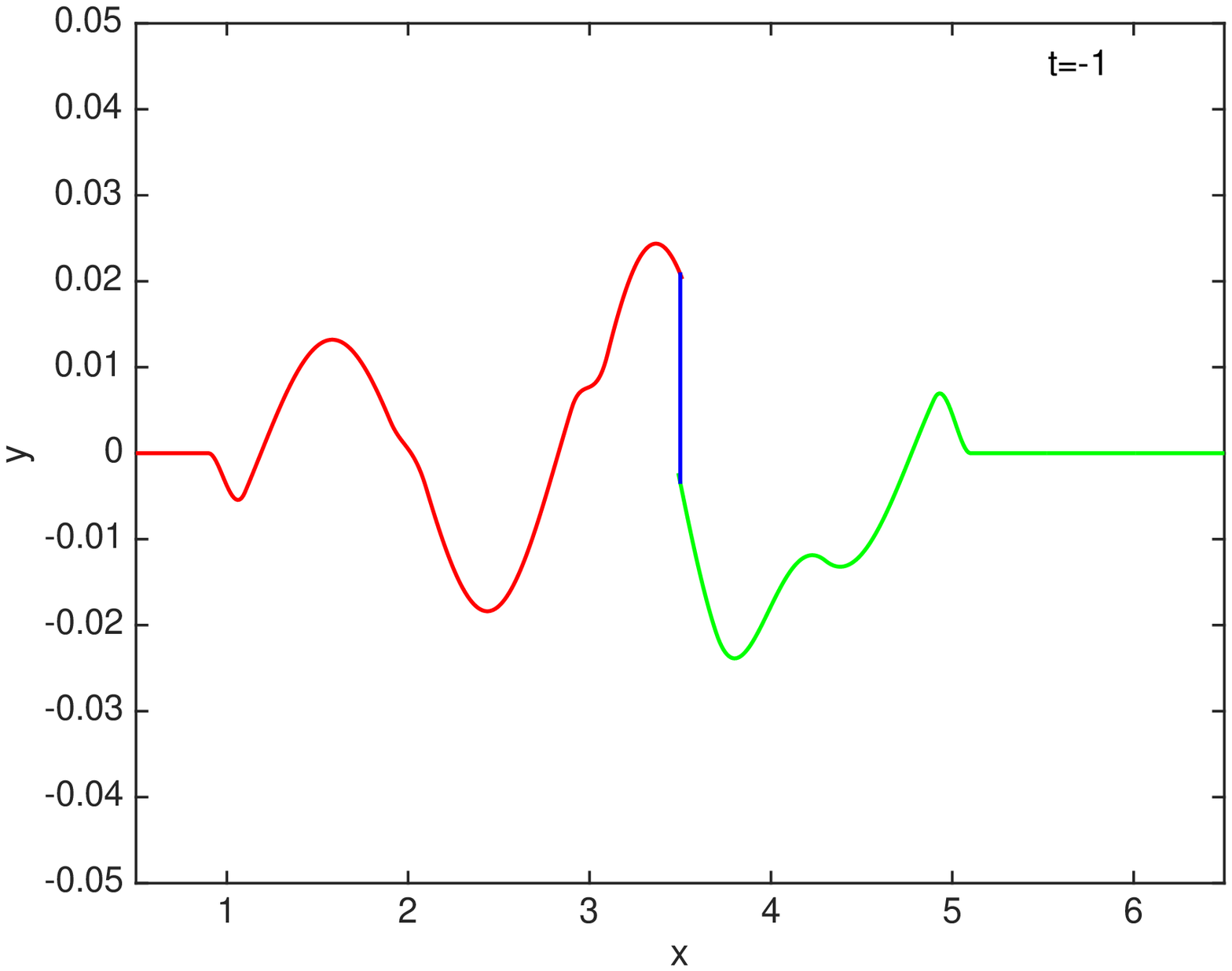} 
\includegraphics[width=0.45 \textwidth,  height=0.18 \textheight]{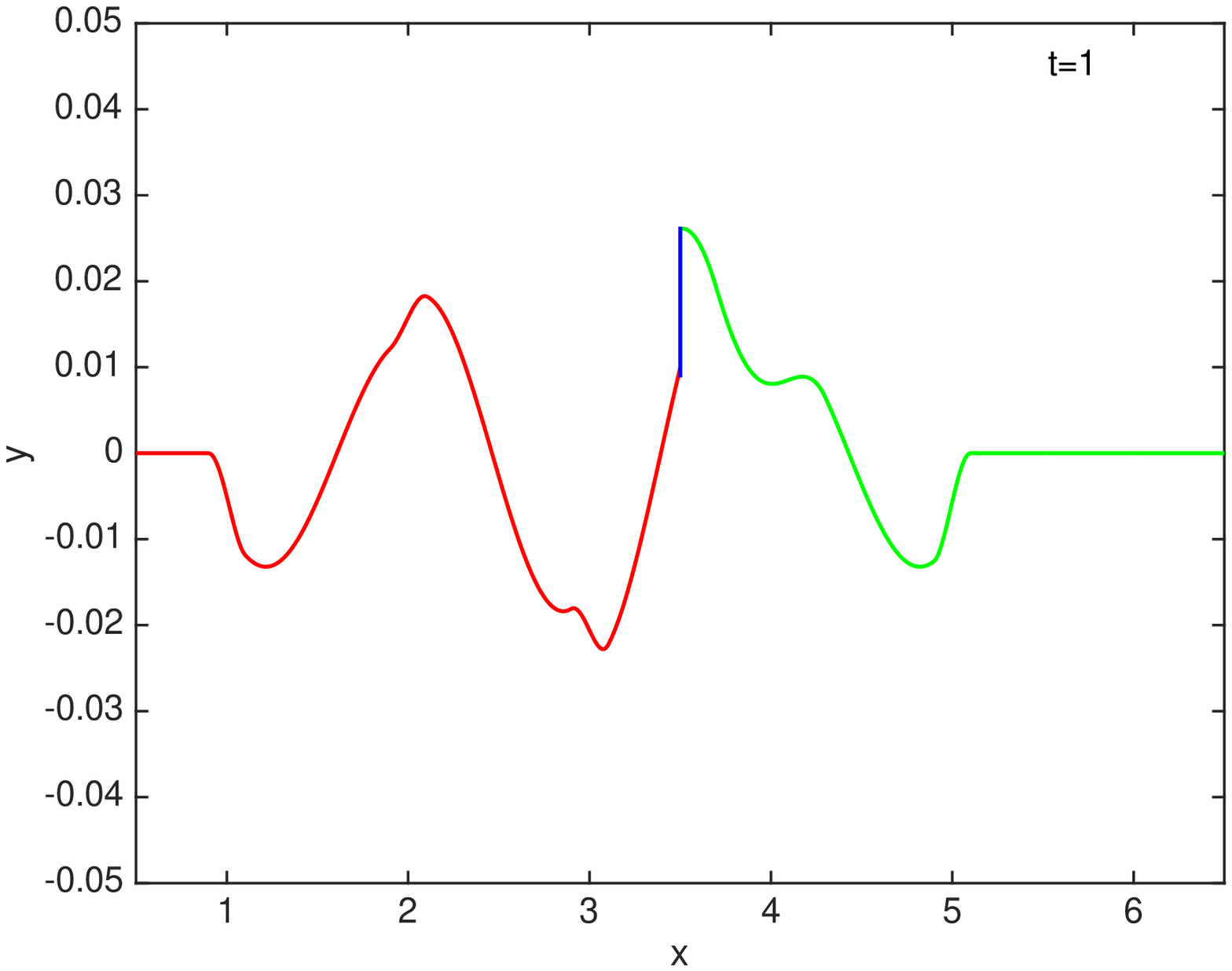} \\
\includegraphics[width=0.45 \textwidth,  height=0.18 \textheight]{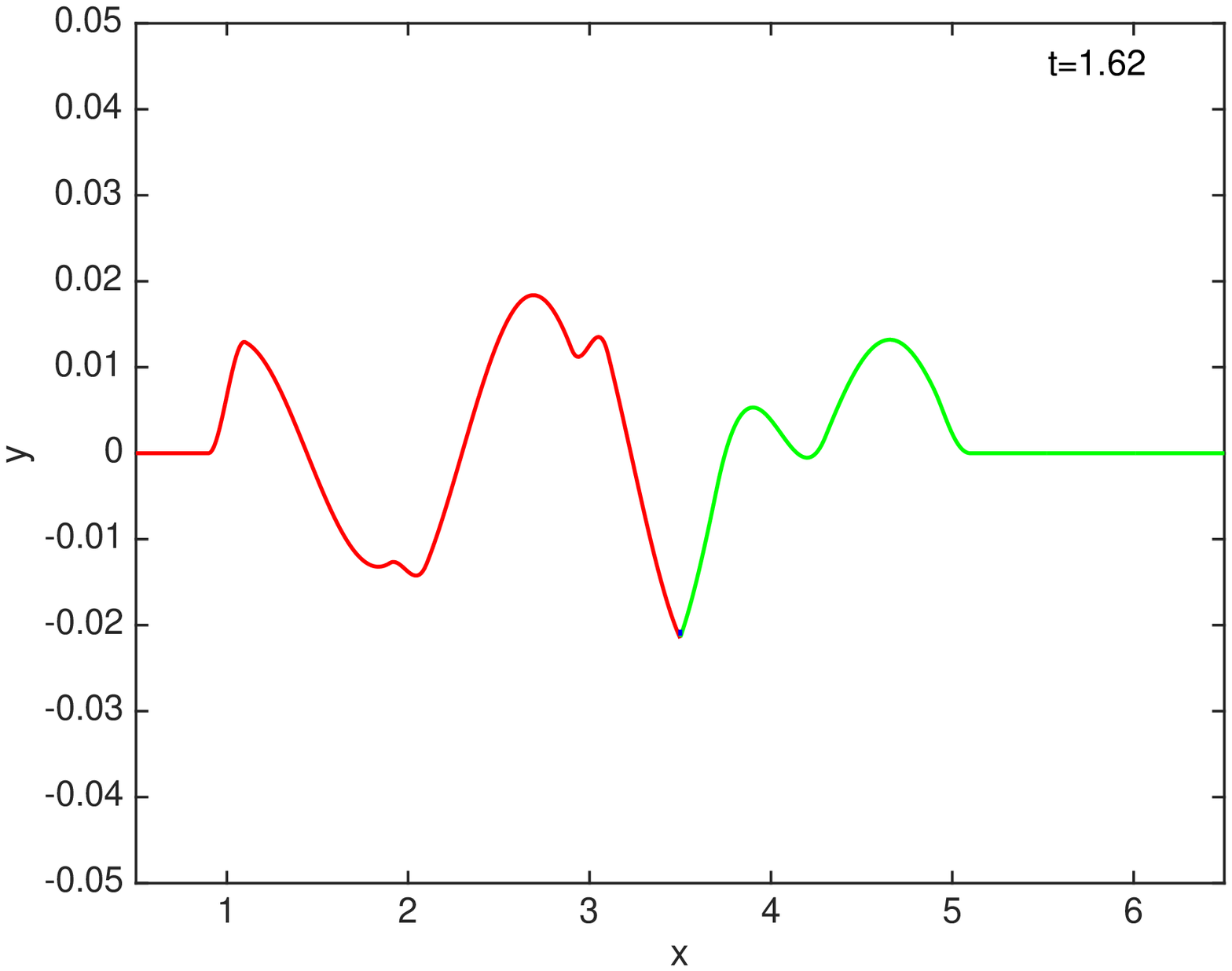} 
\includegraphics[width=0.45 \textwidth,  height=0.18 \textheight]{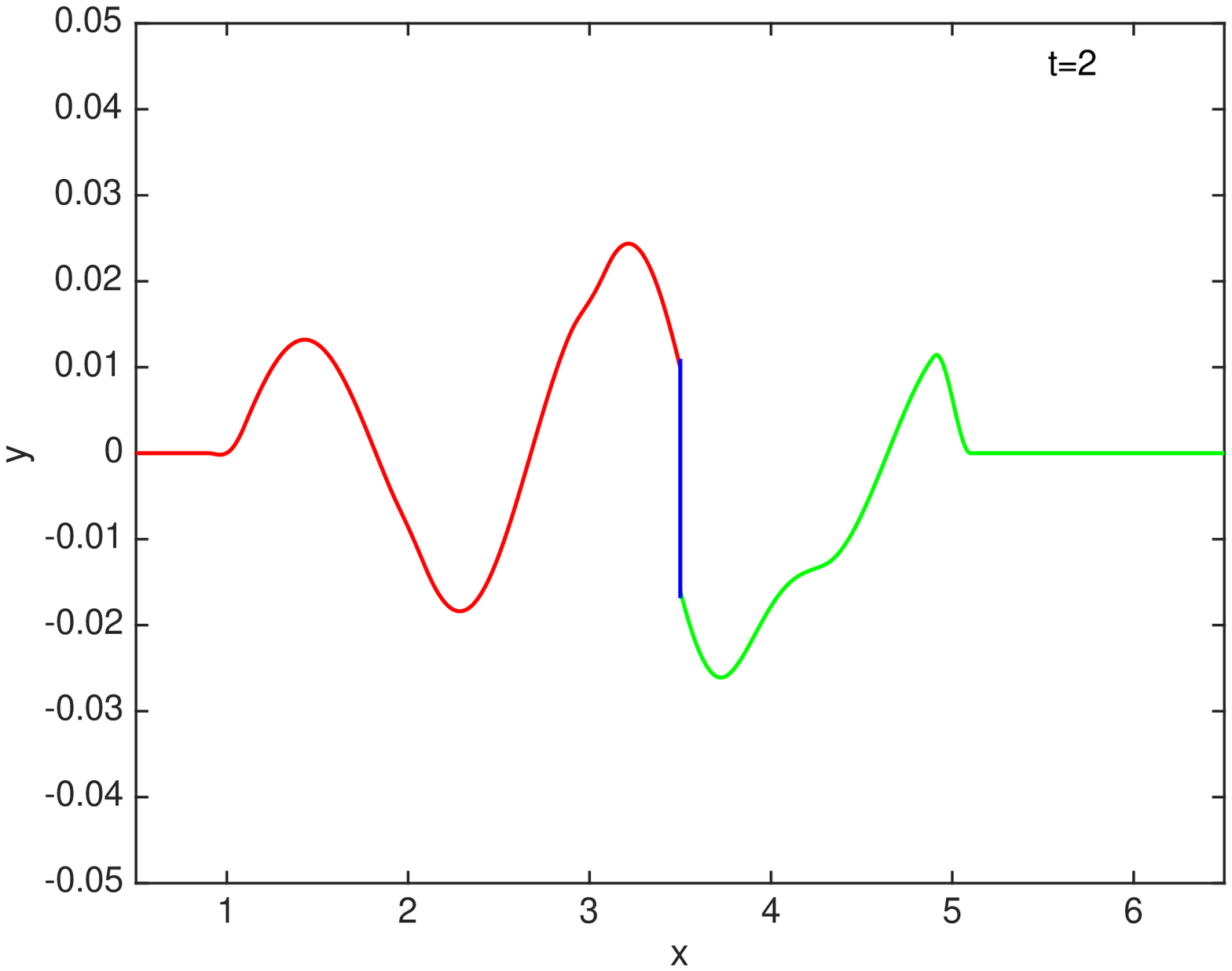}
\caption{Pseudo-streaklines for the channel flow obtained by numerical simulation, at several instances in time. The
color-coding of the pseudo-streakline is consistent with Fig.~\ref{fig:pseudostreakline}, showing the upstream streakline
(red), gate (blue) and downstream streakline (green).}
\label{fig:ccpseudostreakline}
\end{figure*}

Consider two fluids in a microchannel, with exactly the velocity agitation and parameter values as used in Section~\ref{sec:streakexamples}.  Suppose the gate is to be drawn at the value $ \bar{\vec{x}}(p) = (3.5,0) $, i.e., at
the choice $ p = 3.5 $.  To numerically simulate the upstream pseudo-streakline at some instance in time $ t $, it is therefore necessary 
to release particles from $ \vec{a} = (0.5,0) $ at times prior to $ t $, and allow the streakline to evolve until it intersects a vertical
line drawn at $ x = 3.5 $.  To plot the downstream streakline at the same instance in time, particles must be released synthetically
from $ \vec{b} $, which was unspecified in Section~\ref{sec:streakexamples} beyond the fact that it must be downstream of
$ (5.1,0) $.  Here, let us take $ \vec{b} = (6.5,0) $ to have the gate be symmetrically between $ \vec{a} $ and $ \vec{b} $.
However, particles need to be released from $ \vec{b} $ {\em after} the time $ t $, and evolved {\em backwards} in time until
the gate is intersected.  This means that, in general, using numerical simulations to determine the pseudo-streakline---which
contains simultaneous snapshots of the upstream and the downstream streakline, ending exactly on the gate---may
not be easy.  It may not be clear at what instance in time to release particles at $ \vec{a} $ and at $ \vec{b} $ such that the
streakline evolving from these points {\em precisely} intersects the gate at the {\em same} specified instance in time $ t $.
Fortunately, for this particular example, since the velocity agitation is in the $ y $-direction, and the steady velocity is a constant
$ U $ in the $ x $-direction, this can be determined.  

Fig.~\ref{fig:ccpseudostreakline} shows the pseudo-streaklines obtained by this process at four different times.  At $ t = - 1 $,
the flow through the gate indicates that fluid will get transported from the lower to the upper fluid.  Since this is in the
direction of $ \hat{\vec{n}} $, this represents a positive instantaneous flux.  At $ t = 1 $, the flux is negative (fluid transports from
the upper to the lower fluid through the gate), while it is positive once again at $ t = 2 $.  There is a value between these,
near $ t = 1.62 $, at which the upstream and downstream streaklines interchange their relative positions on the gate.
This is shown in the third panel of Fig.~\ref{fig:ccpseudostreakline}; the red and green curves meet {\em on} the gate.
This is a situation at which the instantaneous flux is zero.  As time progresses, repeated interchanges of relative positioning
along the gate implies that fluid sloshes back and forth across the
gate (and hence across the pseudo-streakline), causing advective transport between the two fluids.  The flux (\ref{eq:fluxdef})
in this case is easily obtained by multiplying the length of the gate
by the horizontal speed $ U $, since the normal velocity to the gate at {\em all} points on the gate is the same value.  
In doing this calculation,
the length of the gate should be considered a signed quantity, to reflect the correct direction of transport.

\begin{figure}
\includegraphics[width=0.5 \textwidth, height=0.18 \textheight]{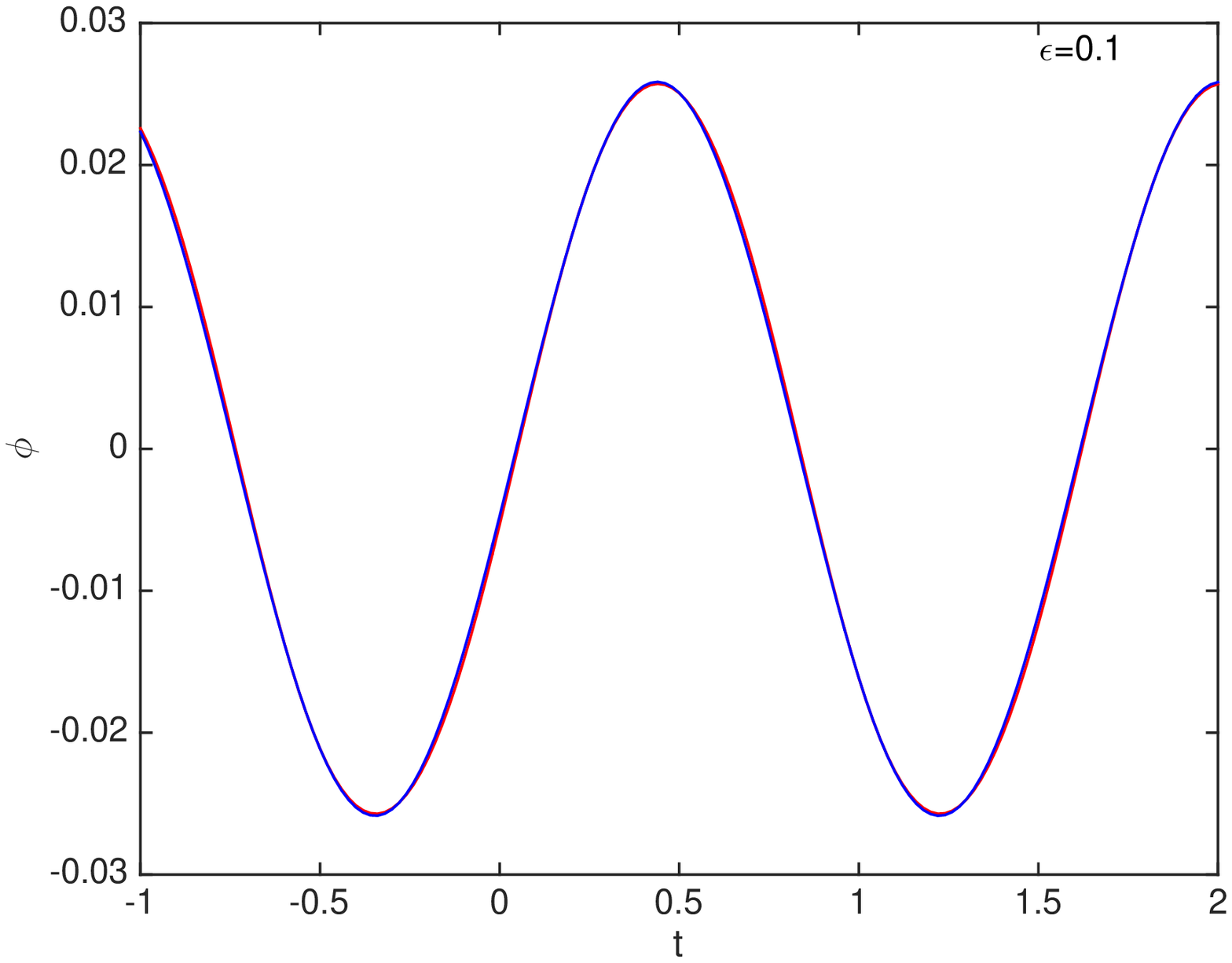} 
\includegraphics[width=0.5 \textwidth,  height=0.18 \textheight]{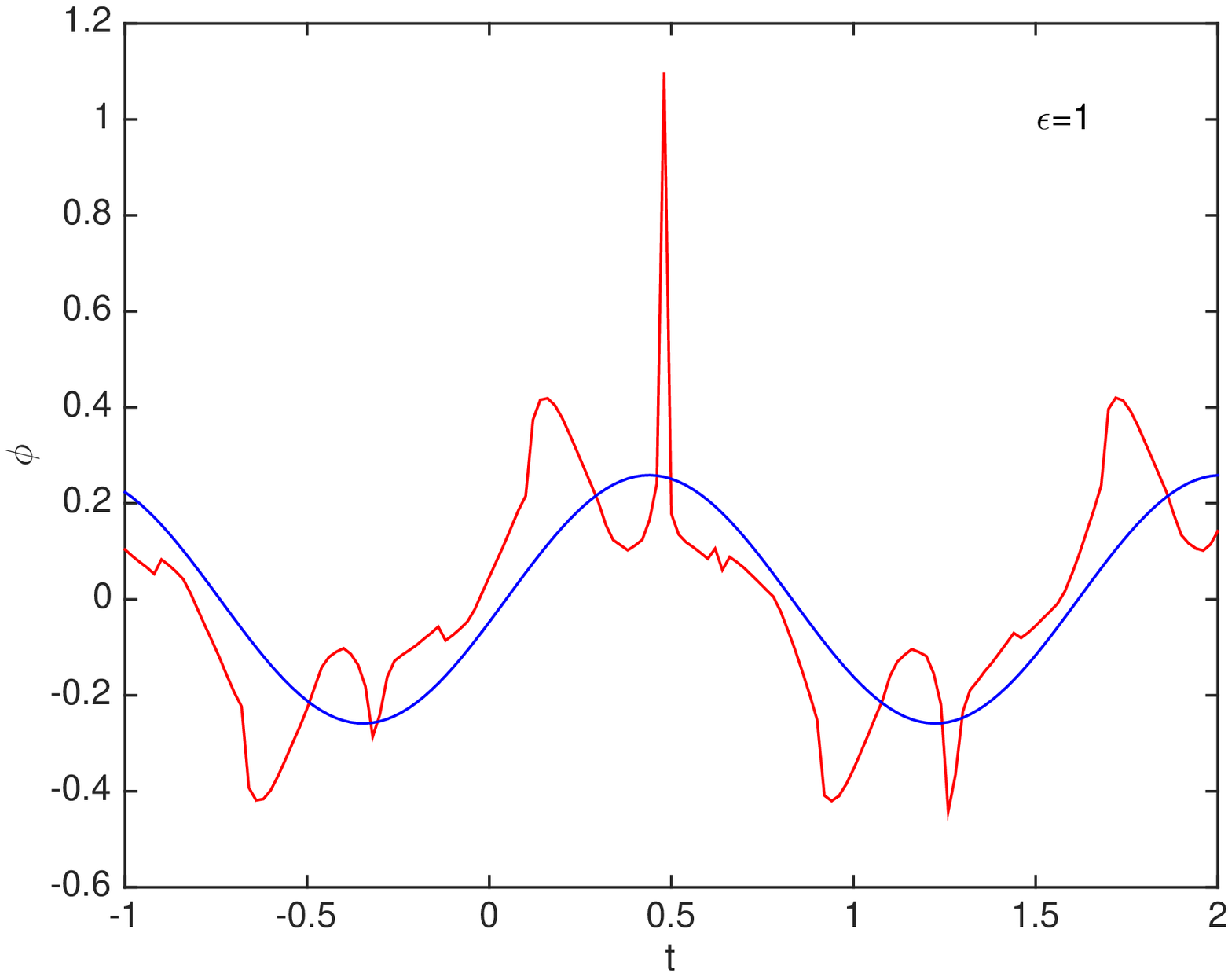} 
\caption{Transport from the lower to upper fluid in the channel using numerical simulation (red) and the analytical expression
(\ref{eq:mcrosschannel}), for the agitation and parameter values as specified in Section~\ref{sec:streakexamples} (top), and
with $ \eps $ changed to $ 1 $ (bottom).}
\label{fig:cctransport}
\end{figure}

Next, this shall be compared to the analytical expression in Theorem~\ref{theorem:transport}. Before writing the instantaneous
flux expression for this specific configuration, it can be written for a {\em general} channel configuration as described in
Section~\ref{sec:streakexamples} by
\begin{widetext}
\begin{equation}
M_c(p,t)  \! =  \int_{p_1 \! -\! d_1/U}^{p_n \! + \! d_n/U}  \sum_{j=1}^n \I_{[ p_j-d_j/U,p_j \! + \! d_j/U]}(\tau) 
U \frac{v_j}{d_j^2} \left[ U^2 \left( \tau \! - \! p_j \right)^2 \! - \! d_j^2 \right]  \! \cos \left[ 
\omega  \left( \tau \! + \! t \! - \! p \right) \! + \! \phi_j \right] 
 \d \tau \, , 
\label{eq:mcrosschannel}
\end{equation}
\end{widetext}
where $ p_1 - d_1/U < p < p_n + d_n / U $, and $ \bar{\vec{x}}(p) $ denotes the location of the gate.  For a fixed
$ p $, it is clear that the flux is periodic in $ t $ with period $ 2 \pi / \omega $.  This is easily computed for the given parameter
values.  Now, for the specific channel configuration and parameter values as examined in Section~\ref{sec:streakexamples},
the analytical expression is compared with that obtained by the direct numerical simulations inserted into the flux
definition (\ref{eq:fluxdef}) in Fig.~\ref{fig:cctransport}.  The
top panel shows $ \eps = 0.1 $, the particular value used in Section~\ref{sec:streakexamples}, with the red being the numerics
and the blue the analytics. The curves are indistinguishable.  The lower panel shows the comparison performed now with $ \eps = 1 $
(but everything else kept identical), showing that indeed the theory loses predictive ability when the agitation is comparable in
size to the main flow.  Having said that, the theorerical blue curve, obtained using perturbative methods in $ \eps $, still does an excellent job of following the trend of the flux, even for non-small velocity agitations.

%%%%%%%%%%%%%%%%%%%%
\subsection{Anomalous fluid in a vortex}

Unlike in the previous example, the flow on $ \tilde{\Gamma} $ is not of a constant speed, $ \tilde{\Gamma} $ is curved,
and the velocity agitation is not  always perpendicular to $ \tilde{\Gamma} $.  Determining the pseudo-streakline at a general
time $ t $ will therefore be more complicated.  Take $ \vec{a} = (l,0) $ as before, and let $ \vec{b} $ be the identical point
but after the vortical flow has taken the particle released from $ \vec{a} $ one round around the vortex.  Thus, when using
$ \theta $ instead of $ p $ as a parameter, $ \vec{a} $ corresponds to $ \theta = 0 $ and $ \vec{b} $ to $ 2 \pi $.  Choose the
gate to be drawn at $ \theta = \pi $.  Since the flow at $ \bar{\vec{x}}(\pi) $ is in the $ -y $-direction, the gate occurs along
the $ x $-axis at this point.  Note that $ \hat{\vec{n}} $ points {\em into} the vortex; positive instantaneous flux will therefore
entrain exterior fluid into the fluid in the interior, whereas negative flux indicates that at that instance in time the interior fluid
is escaping.

The fact that the gate is chosen such that the pseudo-streakline consists of parts of streaklines which wrap around the vortex
{\em less than once} is immediately comforting.  This avoids having to deal with accummulated errors when streaklines
wrap around more than once, as seen in Section~\ref{sec:streakexamples}. 

\begin{figure}
\vspace*{-0.3cm}
\includegraphics[width=0.5 \textwidth, height=0.18 \textheight]{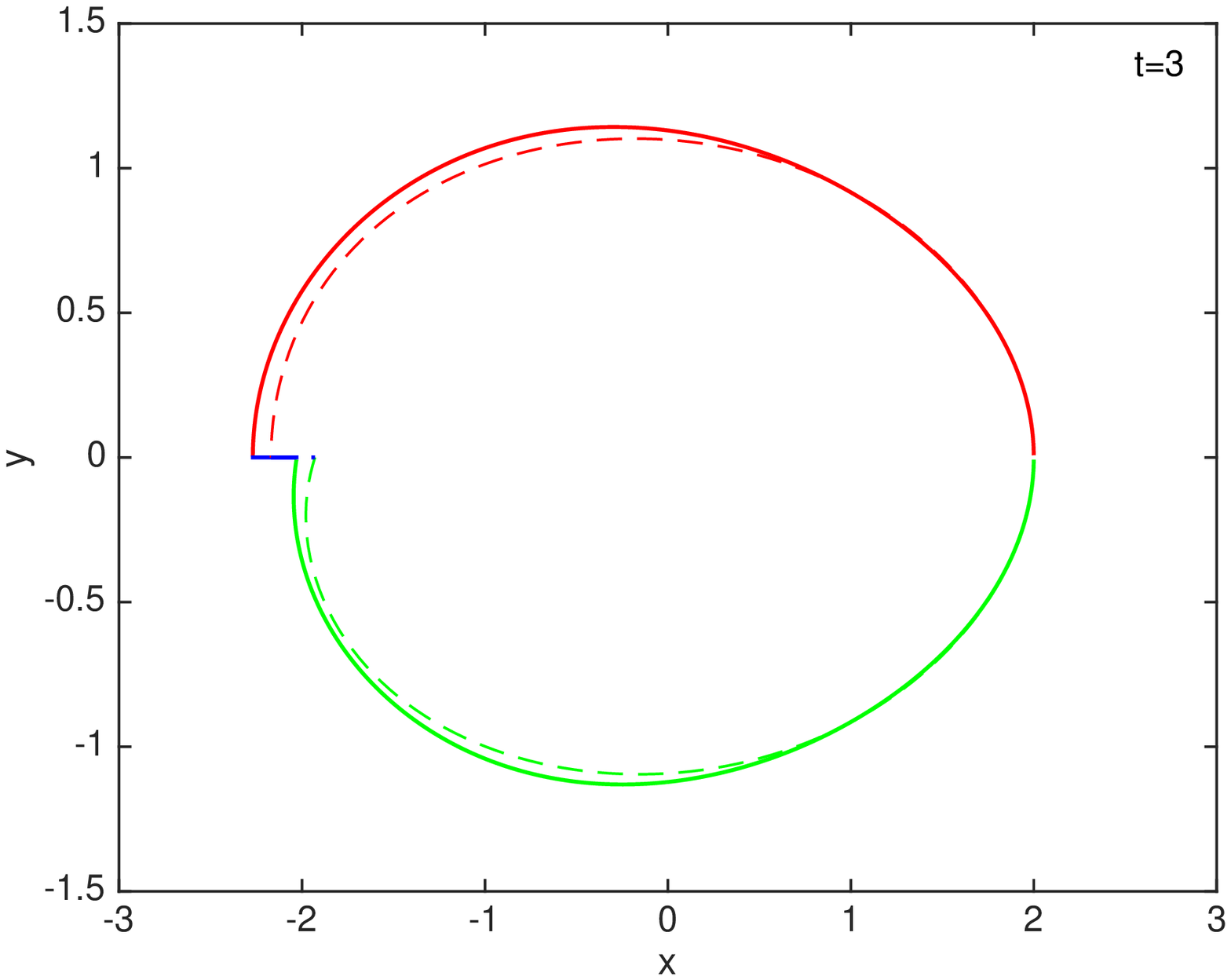} 
\includegraphics[width=0.5 \textwidth,  height=0.18 \textheight]{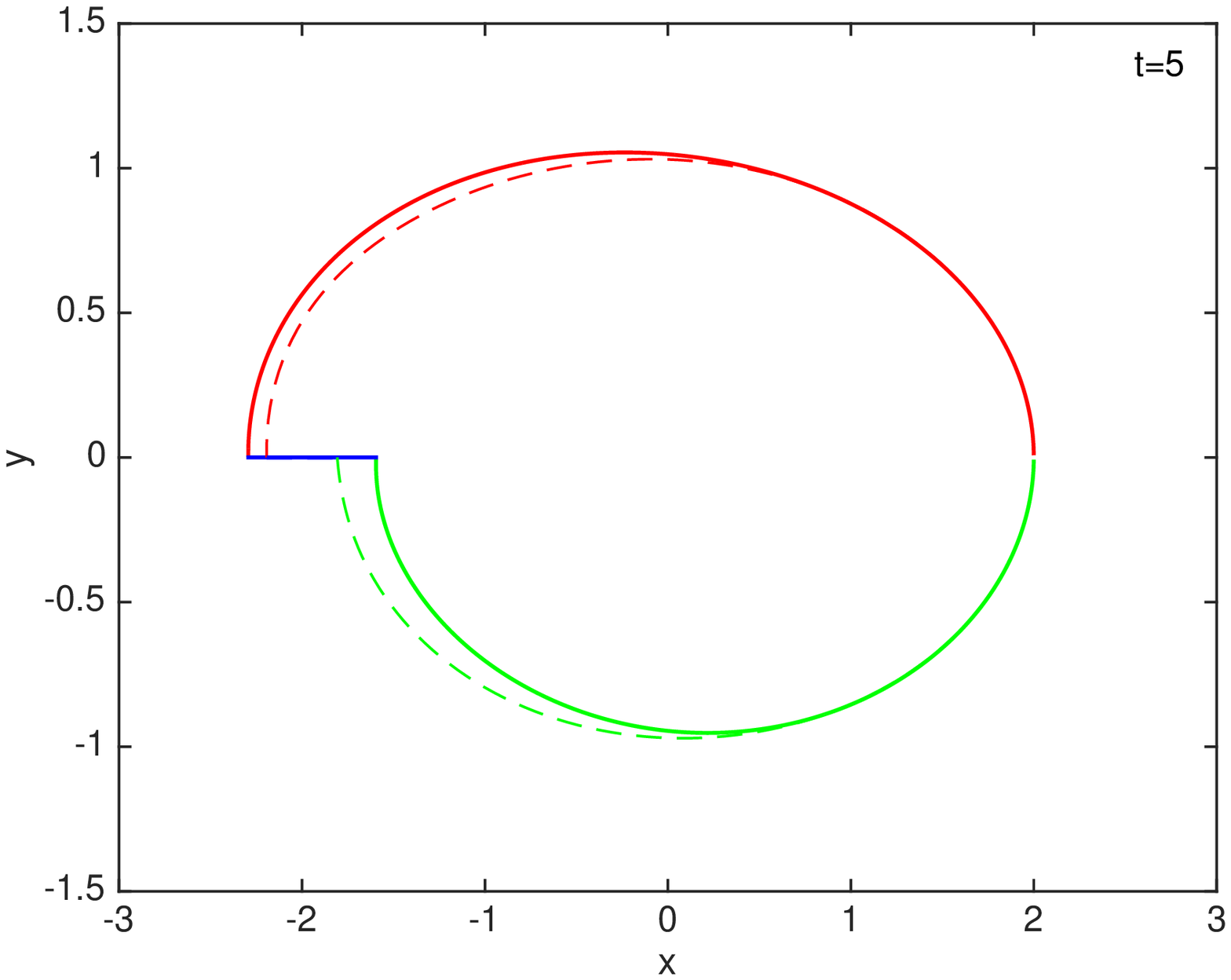} 
\includegraphics[width=0.5 \textwidth,  height=0.18 \textheight]{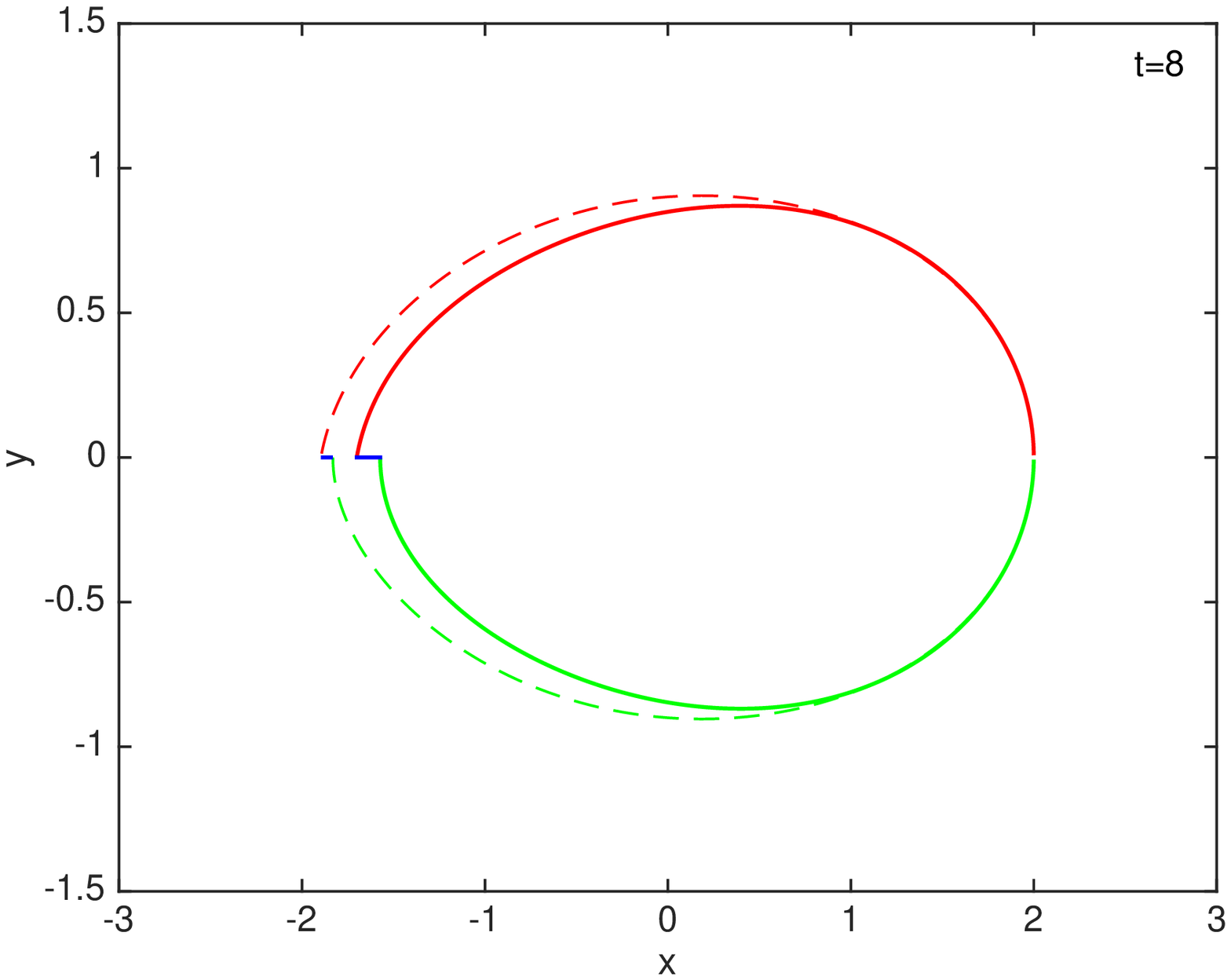} 
\caption{Pseudo-streaklines for the agitated elliptic vortex flow obtained by numerical simulation, at several instances in time. The
color-coding of the pseudo-streakline is consistent with Fig.~\ref{fig:pseudostreakline}, showing the upstream streakline
(red), gate (blue) and downstream streakline (green).  The dashed lines are the analytical approximations.}
\label{fig:evpseudostreakline}
\end{figure}

Numerically determining the pseudo-streakline at an instance in time $ t $ entails first releasing particles from $ \vec{a} $ at
some
 time instance in the past, and then allowing the upstream streakline to evolve until it crosses the gate $ {\mathcal G}(\pi,t) $
 on the $ -x $-axis.  The time one needs to start releasing particles is therefore only implicitly defined, which lends some difficulty
 in the computation.  Similarly, particles need to be released from $\vec{b} $ at some time instance in the future of $ t $, and 
 evolved backwards in time until $ {\mathcal G}(\pi,t) $ is intersected at {\em exactly} time $ t $.  In performing numerical
 simulations, an overestimate for the time before $ t $ for the upstream streakline (or after $ t $ for the downstream) is
 estimated first to be $ 1.2 \tau(\pi) $ from (\ref{eq:tautheta}), since in the absence of a velocity agitation the time would be
 exactly $ \tau(\pi) $.  Then the upstream streakline is evolved up to this time beyond $ t $, and portions which extrude across
 $ {\mathcal G}(\pi,t) $ are then clipped.  By following the same idea, the downstream streakline can be obtained, and the
 gate drawn in between these.  Using exactly the parameter values as in Section~\ref{sec:streakexamples}, the pseudo-streaklines
 were obtained using this procedure at different times $ t $, and are pictured in Fig.~\ref{fig:evpseudostreakline} by the
 solid curves.  The dashed curves are obtained from the analytical approximation (\ref{eq:meluellipticex}) and the corresponding
 expression (not shown) for the downstream streakline.  
  The instantaneous flux is going {\em out of} the vortex at all instances pictured, with a large value at $ t = 5 $, smaller at $ t = 3 $,
 and very small at $ t = 8 $.  Numerical simulations at many other time values (not shown) indicate that the flux is always
 negative (i.e., outward), but becomes vanishingly small as $ \left| t - 5 \right| $ gets large.  Can an insight to this be obtained
 from the theoretical transport measure?

\begin{figure}
\includegraphics[width=0.5 \textwidth, height=0.18 \textheight]{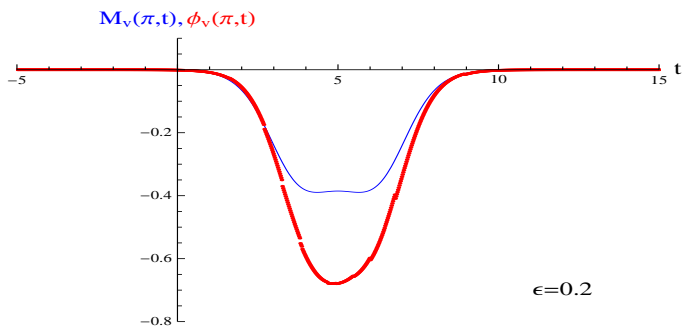} 
\caption{Transport into the elliptic vortex using numerical simulation (red) and the analytical expression
(\ref{eq:melelliptic}), for the agitation conditions specified in Section~\ref{sec:streakexamples} and with $ \eps = 0.2 $.}
\label{fig:evtransport}
\end{figure}

In calculating this, the first observation is that the downstream quantity $ M_v^d(\theta,t) $ for a general incompressible velocity agitation is almost the same as (\ref{eq:meluelliptic}), excepting for the absence of
the leading negative sign, and the fact that the limits are from $ \theta $ to $ \pi $.  Thus, the instantaneous flux function
$ M_v(\theta,t) = M_v^u(\theta,t) - M_v^d(\theta,t) $ becomes
\begin{widetext}
\begin{equation}
M_v(\theta,t) = - \int_0^{2 \pi} \left( m \cos \alpha, l \sin \alpha \right) \cdot \vec{v} \! \left( l \cos \alpha, m \sin \alpha, 
t + \frac{m l}{2} \int_\theta^\alpha \! \sqrt{ \frac{m^2 \tan^2 \beta \! + \! l^2}{l^2 \tan^2 \beta \! + \! m^2}} \, d \beta \right) \, \left( \sqrt{ \frac{m^2 \tan^2 \alpha \! + \! l^2}{l^2 \tan^2 \alpha \! +\!  m^2}} \, \d \alpha
\right) \, .
\label{eq:melelliptic}
\end{equation}
\end{widetext}
For the specific velocity perturbation considered in Section~\ref{sec:streakexamples}, and with the gate chosen to be
at $ \theta = \pi $, the instantaneous flux becomes
\begin{widetext}
\begin{equation}
M_v(\pi,t) = - \eps l \int_0^{2 \pi} \sin \alpha \sin \left[ l (\cos \alpha - 1) \right] \tanh \left[ t - 5 + 
\frac{m l}{2} \int_\pi^\alpha \! \sqrt{ \frac{m^2 \tan^2 \beta \! + \! l^2}{l^2 \tan^2 \beta \! + \! m^2}} \, d \beta \right]
\, \sqrt{ \frac{m^2 \tan^2 \alpha \! + \! l^2}{l^2 \tan^2 \alpha \! +\!  m^2}} \, \d \alpha
 \, .
\label{eq:melellipticex}
\end{equation}
\end{widetext}
This is shown by the blue curve in the top panel of Fig.~\ref{fig:evtransport} for the same parameter values used for the 
numerical simulation.   The theoretical flux is always negative, has a 
time-dependence which is symmetric about $ t = 5 $ (reflecting the term $ \tanh(t-5) $ chosen in the velocity agitation),
and decays to zero.   The implication is that the fluid which was in the interior of the elliptic vortex continues to leak out
at all times, though the leakage is largest near $ t = 5 $ and becomes vanishingly small as $ \left| t -  5 \right| $ gets large.
This will be visible as a tendril or filament of the inner fluid escaping to the outer one, with the tendril wrapping around in the
anti-clockwise direction.  It is interesting that this new approach reveals exactly the qualitative behavior that 
is well-documented for vortices in external shear flows \cite{meuniervillermaux,eddy,bassomgilbert}.  From the mixing perspective, 
diffusion would then act on the tendril, causing the inner fluid to becomes dispersed in the outer one.
The effect of diffusion is not explicit in the theory here, which focusses specifically on the advective (Lagrangian) flow.  However,
in reality this advective process promotes fluid mixing through advection-driven diffusion.

The red curve in Fig.~\ref{fig:evtransport} is obtained from numerical simulations, using the definition (\ref{eq:fluxdef}), for
the same parameter values as the theoretical blue curve.  At this relatively high value of $ \eps $, there is some difference
between the curves.  Taking smaller $ \eps $ values would make the curves approach one another.

A minor point which must be made is the connection to imcompressibility, which is satisfied for the velocity conditions chosen
in this example.  Does the fact that there is a flux exiting the vortex for all time contradict incompressibility?  The answer is no, since the closed 
pseodo-streakline---the purported flow interface under the unsteady conditions---compensates for the expelled fluid by becoming
smaller.  This is indeed visible by comparing the areas enclosed within the closed pseudo-streaklines in Fig.~\ref{fig:evpseudostreakline}; these get smaller as $ t $ increases.  Indeed, the rate of change of the area enclosed by the
pseudo-streakline is {\em precisely} the definition of the instantaneous flux, when $ \tilde{\Gamma} $ is closed.

%%%%%%%%%%%%%%%%%%%%%%%%%
%%%%%%%%%%%%%%%%%%%%%%%%%
\section{Concluding remarks}
\label{sec:conclusions}

The concepts of stable and unstable manifolds---and various finite-time analogues such as finite-time Lyapunov exponent ridges 
\cite{shadden,he,huntley,nelsonjacobs,johnsonmeneveau,bozorgmaghamross,branickiwiggins,raben}, curves of
extremal attraction/repulsion \cite{hallerreview,kelley,halleryuan,karraschfarazmandhaller,onu}, partitions based on Perron-Frobenius operators \cite{froylandpadberg,froylanddetection,froylandchaos}, time-clipped manifolds \cite{eigenvector,droplet},  fronts associated with averaged flow quantities \cite{mezicscience,levnajicmezic} and other Lagrangian Coherent Structure
type approaches \cite{allshousethiffeault,mundel,ma,mabollt,budisicthiffeault}---are important in demarcating flow
barriers in {\em single phase} unsteady flows.  Here, the focus is on {\em two-phase} flows, in which such entities derived
purely by examining the velocity field do not distinguish the flow barrier.  In the steady, as-yet-unmixed  situation, the flow barrier is the  {\em interface} between the two fluids, which is identified physically as opposed to from characteristics of the fluid velocity.
This article has focussed on determining the flow interface, and the transport across it, when an unsteady
velocity agitation is introduced.  The relevance of streaklines has been highlighted, and their usage in defining a nominal
interface (a {\em pseudo-streakline}), and quantifying an instantaneous transport across this has been elucidated.  Under
the condition that the velocity agitation is weak (i.e., has speeds significantly smaller than that of the base steady
velocities), a theory for determining the streaklines, and the transport flux, has been developed.  Validations of the method
in comparison to numerically computed streaklines was performed for the open (two fluids flowing along a channel)
and closed (a blob of one fluid inside a vortex) flow interfaces situations.

The method for quantifying the transport across the flow interface developed in this paper depends on the velocity agitation
being small in comparison to the base flow, which is steady.  However, the method does {\em not} require incompressibility,
or a particular form of time-dependence, and is thus fairly general.  Indeed, it is likely that recent work \cite{impulsive} which
characterizes the effect of an {\em impulsive} velocity (such as obtained by tapping a fluidic device, say) on stable/unstable manifolds, can be modified to determine the streaklines and transport associated with a non-heteroclinic flow interface,
consonant for example with the impulsively strained vortex numerics of \cite{bassomgilbert}.

The presence of a theoretical approximation for the transport offers scope in being able to {\em optimize} it.  Such has been
done in the case where the flow barrier was a heteroclinic manifold (coincident stable and unstable manifold) in the situation
where a sinusoidally varying velocity agitation was applied \cite{optimal,mixer,l2mixer,frequency}.  Given that the transport
expression (\ref{eq:melnikov}) obtained in this non-heteroclinic instance shares some similarities with the heteroclinic theory
of \cite{optimal,mixer,l2mixer,frequency}, there is obvious scope in being able to adapt those ideas to this situation.   Since
there are also analytical expressions  (\ref{eq:upstream}) and (\ref{eq:downstream}) for the streaklines, another possibility
is being able to quantify how complicated these entities get due to advection, as a measure of eventual diffusive transport.
There is considerable evidence of the presence of an optimal frequency of a velocity agitation in order to maximize
mixing \cite{frequency,fu,rodrigo,shin,wangmagnet,lim,wangacoustic,song,suzuki,niu}; would it be possible to determine this in, for example, configurations such as that of the channel
examined here?  Alternatively, can one obtain insight into the best positioning of cross-channels to effect the best mixing 
across the flow interface?  Questions such as these are under investigation, and will be reported on in follow-up work.

It is interesting that streaklines have {\em also} been suggested as having importance in evaluating transport, 
independently by Karrasch in a recently accepted article \cite{karrasch} discovered by the author just before initial submission
of the present paper.  The differential topology and 
donating region viewpoints of that article \cite{karrasch}
do not have an obvious connection to the present paper, which focusses on fluid interchange between two miscible
fluids across their interface.  Nevertheless, further investigation of a possible relationship will be pursued.

The approach presented here offers a different viewpoint on the oft-examined vortex-in-an-external-strain problem
\cite{romkedarjfm,kida,delcastillonegrete,meuniervillermaux,leweke,mula,turner,bassomgilbert,meleshkovanheijst,friedland}, 
in the
sense that it permits a direct computation of the fluid flux into, or out of, the vortex as a function of time, for a given weak
external strain field.  In this sense, it captures Lagrangian transport, as opposed to standard methods which, for example,
picture the time evolution of the frozen-time (Eulerian) vorticity field \cite{bassomgilbert,turner}.  There are of course connections
between the Lagrangian and Eulerian viewpoints, but in cases where the {\em transport} is important, the current approach may
provide new insights.

The closed flow interface situation appears to be of particular interest in modeling how a blob of fluid (an oil/pollutant/nutrient/plankton/chemical patch) which is placed in an anomalous fluid mixes in with its surroundings.  Once again, `standard' methods
for detecting coherent structure boundaries directly from the velocity field are not necessarily applicable, since the flow interface 
is a physical boundary between two fluids as opposed to an entity derivable from the velocity field.  The methods outlined in this
article are a first step towards understanding the Lagrangian transport associated with this from a theoretical perspective.  The
trick is trying to identifying the flow interface in the presence of an unsteady velocity, which can be done here if there is an anchor
point on the interface at which the {\em unsteady} component of the velocity is zero.  If not, there is a difficulty in deciding where
to release particles for the streakline determination.  How one might figure out the pseudo-streakline (i.e., the relevant unsteady flow interface) when there is no such anchor point is not clear.  Note that the process of evolving particles
on the closed steady flow interface (i.e., examining the {\em timeline}) is not effective in assessing transport, since this closed loop
simply remains a closed loop under unsteady velocity agitations, and therefore a transport between the interior and exterior
{\em across this} is zero.  This is of course true for any material curve, which precludes their usage for transport assessment.
A nominal flow interface across which there is transport needs to be enunciated.  The ability to do so using streaklines when
there is at least one anchor point, as outlined here, can hopefully be build on, in the situation of weak strain which is never zero
on the interface.

%%%%%%%%%%%%%%%%%%%%%%%%%
%%%%%%%%%%%%%%%%%%%%%%%%
\appendix
\section{Proof of Theorem~\ref{theorem:upstream} [Upstream streakline]}
\label{app:proof}

The proof here is in the spirit of the proof of Theorem~2.1 in \cite{tangential}, in which an unstable
manifold's displacement is characterized.  However, this situation is different, since it is the upstream
streakline that is required.  Imagine fixing the time $ t $, and also the particle $ p $ in this time-slice
which is at the location $ \bar{\vec{x}}(p) $ in the steady flow (\ref{eq:steady}).  Due to the action
of the unsteady velocity agitation $ \vec{v} $, this particle will be at a nearby location, $ {\mathcal O}(\eps) $ away,
at a location $ \vec{x}_\eps^u(p,t) $.  {\em Thinking of $ (p,t) $ as fixed}, but with $ \tau $ as the time-variable,
define
\begin{equation}
M_\eps^u(p,\tau \!)  \! =  \! \left[ J \vec{u} \! \left( \bar{\vec{x}}(\tau \! - \! t \! + \! p \! ) \right) \! \right] \! \cdot  \! \left[  \vec{x}_\eps^u(p,\tau \! ) 
\! -  \! 
\bar{\vec{x}}(\tau \! - \! t \! + \! p \! )  \! \right]  .
\label{eq:mfdef}
\end{equation}
From (\ref{eq:upstreamsteady}), it is clear that $ \bar{\vec{x}}(\tau-t+p) = \vec{x}_0^u(p,\tau) $, the steady streakline at location
$ p $; this passed through $ \vec{a} $ a time $ p-p^u $ prior to $ \tau $.  For the {\em unsteady} flow streakline as defined 
through (\ref{eq:upstreamunsteady}), $ \vec{x}_\eps^u(p,\tau) $
represents the location of the same particle, and thus $ \vec{x}_\eps^u(p,\tau) - \bar{\vec{x}}(\tau \! - \! t \! + \! p \! ) $ is the
difference occurring as the result of including $ \vec{v} $.  This difference is $ {\mathcal O}(\eps) $ since $ \vec{v} $ is
$ {\mathcal O}(\eps) $, and then so is $ M_\eps^u(p,\tau) $.  Note moreover that 
\begin{equation}
\frac{M_\eps^u(p,t)}{\left| \vec{u} \left( \bar{\vec{x}}(p) \right) \right|} = \frac{J \vec{u} \left( \bar{\vec{x}}(p  ) \right)}{\left| \vec{u} \left( \bar{\vec{x}}(p) \right) \right|} \cdot  \left[  \vec{x}_\eps^u(p,t ) -  
\bar{\vec{x}}(p )  \right] 
\label{eq:mfproject}
\end{equation}
is the projection of the displacement of the streakline in the direction normal to  $\Gamma $ at $ \bar{\vec{x}}(p) $.  
Hence, the goal is to determine
$ M_\eps^u(p,t) $ in terms of known quantities from the steady flow.  To do this, it is necessary to differentiate (\ref{eq:mfdef})
with respect to $ \tau $ at fixed $ (p,t) $.  Fortunately, it turns out that this part of the calculation is {\em identical} to that in the 
proof of Theorem~2.1 in \cite{tangential}, and hence by inspection of equation~(3.6) of \cite{tangential} it is possible to
write
\begin{eqnarray}
\frac{\pa M_\eps^u}{\pa \tau} & - &
\left[ \vec{\nabla} \cdot \vec{u} \right] \left( \bar{\vec{x}}(\tau - t+p) \right) M_\eps^u \nonumber \\
& & \hspace*{-0.8cm} =
\left[ J \vec{u} \left( \bar{\vec{x}}(\tau - t+p) \right) \right] \cdot \vec{v}
\left( \bar{\vec{x}}(\tau - t + p), \tau \right) \, ,
\label{eq:mfdiff}
\end{eqnarray}
where $ {\mathcal O}(\eps^2) $ terms have been discarded.
Next, (\ref{eq:mfdiff}) will be multiplied by the integrating factor
\[
\mu(\tau) := \exp \left[ - \int_0^\tau \left[ \vec{\nabla} \cdot \vec{u} \right] \left( \bar{\vec{x}}(\xi-t+p) \right) \,
\d \xi \right] \, ,
\]
and integrated from $ \tau = t+p^u -p $ to $ t + p_+^u - p $.  Before proceeding, these limits will require some explanation.  The lower limit arises
since when inserted into (\ref{eq:mfdef}) this yields
\begin{eqnarray*}
M_\eps^u(p, t \! + \! p^u \! - \! p \! ) & = & \left[ J \vec{u} \! \left( \bar{\vec{x}}(p^u) \right) \! \right]  \cdot  
\left[  \vec{x}_\eps^u(p,t \! + \! p^u \! - \! p \! ) 
-    \bar{\vec{x}}(p^u)  \! \right] \\
& = & \left[ J \vec{u} \! \left( \bar{\vec{x}}(p^u) \right) \! \right]  \cdot \left[ \vec{a} - \vec{a} \right] = 0 \, , 
\end{eqnarray*}
by using the streakline property (\ref{eq:upstreamunsteady}).  Intuitively, this is because the
particle from the steady streakline, {\em and} that from the unsteady streakline, share the property that they both emanated
from the point $ \vec{a} $.  Next, consider the upper limit $  t + p_+^u - p $.  If $ \tilde{\Gamma} $ is open, using 
Table~\ref{table:p} in the situation where $ p < p^d $ (i.e., the streakline has not yet reached $ \vec{b} $), then 
$ t + p_+^u - p = t $.
Inserting this will give $ M_\eps^u(p,t) $, which according to (\ref{eq:mfproject}) is the correct entity sought in the time-slice $ t $.
If, however, the streakline is beyond $ \vec{b} $, it no longer experiences a velocity agitation.  That is, $ \vec{v} = \vec{0} $
for $ p > p^d $.  In this case the upper limit becomes $ t + p_+^u - p = t + p^d - p $, thereby switching off the $ \vec{v} $ term
when it is not present.
On the other hand, if $ \tilde{\Gamma} $ is closed, Table~\ref{table:p} implies that $  t + p_+^u - p = t $ directly, leading
to the quantity $ M_\eps^u(p,t) $.  As the streakline repeatedly traverses an $ {\mathcal O}(\eps) $-close path to
$ \tilde{\Gamma} $, it continually accumulates modifications due to the velocity agitation.
Thus, multiplying (\ref{eq:mfdiff}) by the integrating factor and integrating from 
$ \tau = t+p^u -p $ to $ t + p_+^u - p $ yields
\begin{widetext}
\begin{equation}
M_\eps^u(p,t)  =  \int_{ t+p^u -p}^{t + p_+^u - p} \exp \left[ 
\int_\tau^t  \left[ \vec{\nabla} \cdot \vec{u} \right]
\left( \bar{\vec{x}} (\xi - t + p) \right)
\d \xi \right] \left[ J \vec{u} \left( \bar{\vec{x}} (\tau - t  +  p) \right) \right] \cdot
\vec{v} \left( \bar{\vec{x}} (\tau - t  +  p), \tau \right)  \d \tau \, .
\label{eq:mftemp}
\end{equation}
\end{widetext}
Next, a change of variables $ \eta = \tau - t + p $ is applied, and the integral above becomes exactly (\ref{eq:mupstream}).
However, there is an additional prefactor $  \I_{[p_-^u,P]}(p) $ in (\ref{eq:mupstream}).  This simply `turns on' the function for
$ p $ values above $ p_-^u $ (all the way up to $ P $, where $ P $ can be large).  To understand this, consider first open
$ \tilde{\Gamma} $, in which case according to Table~\ref{table:p}, $ p_-^u = p^u $.   If $ p < p^u $, the understanding of the
streakline, according to (\ref{eq:upstreamunsteady}) would be points which {\em in the future} pass through $ p^u $.
Since there is no velocity agitation before particles reach $ \bar{\vec{x}}(p^u) = \vec{a} $, there is no correction to the
steady streakline in this case.  Hence, the correction term, encoded in $ M_\eps^p $, must be set to zero if $ p < p^u $, 
which is what is accomplished by the prefactor.  Next, if $ \tilde{\Gamma} $ is closed, particles which arrive at $ \vec{a} $ 
are already arriving from a region in which the velocity agitation applies, and thus will incur displacements from the
steady streakline.  Therefore, $ M_\eps^u $ should {\em not} be set to zero for values of $ p < p^u $, which is accomplished
by setting $ p_-^u = - P $.  (Recall that the definitions for the upstream (\ref{eq:upstreamunsteady}) and downstream (\ref{eq:downstreamunsteady}) streaklines were only legitimate for $ p \in [-P,P] $, for $ P $ any finite value.) 
This completes
the derivation of (\ref{eq:mupstream}).  The interpretation (\ref{eq:upstream}), as being the projection in the normal direction
to $ \Gamma $, arises because of (\ref{eq:mfproject}).

%%%%%%%%%%%%%%%%%%%
\section{Proof of Theorem~\ref{theorem:transport} [Instantaneous transport]}
\label{app:transport}

The proof is based on two observations.  First, the displacement from the downstream to the upstream streakline,
measured along the gate, is from Theorems~\ref{theorem:upstream} and \ref{theorem:downstream} simply
$ \left[ M^u(p,t) - M^d(p,t) \right] / \left| \vec{u} \left( \bar{\vec{x}}(p) \right) \right| $ to leading-order.  If this is positive,
it means that the upstream streakline is situated in a positive direction in comparison to the downstream one when
considering the direction $ \hat{\vec{n}}(p) $ along $ {\mathcal G}(t) $.  This enables the determination of whether the
gate provides a channel for instantaneous flux from fluid $ 2 $ to $ 1' $ (if positive), or from $ 1 $ to $ 2' $ (if negative),
as is clear from Fig.~\ref{fig:transport}.  Second, the fluid velocity at all points on $ {\mathcal G}(t) $, in the normal
direction, is to leading-order $ \left| \vec{u}
\left( \bar{\vec{x}}(p) \right) \right| $.  This is because in the absence of a velocity agitation the velocity at $ \bar{\vec{x}}(p) $,
on $ {\mathcal G} $, is  $ \vec{u} \left( \bar{\vec{x}}(p) \right) $, and points normal to $ {\mathcal G} $, and moreover both
$ {\mathcal G}(t) $ and the unsteady velocity agitation have size $ {\mathcal O}(\eps) $.  Multiplying the velocity across
$ {\mathcal G}(t) $ by the length of $ {\mathcal G}(t) $ therefore gives the instantaneous flux across it; to leading order
this is therefore $ M^u(p,t) - M^d(p,t) $.  Thus,
\[
\phi(p,t) = M^u(p,t) - M^d(p,t) + {\mathcal O}(\eps^2)
\]
However, $ p \in [p^u,p^d] $ in this situation, and so $ p_+^u = p $ and $ p_-^d = p $, whether $ \tilde{\Gamma} $ is open
or closed.  Moreover, by taking $ P $ large, it is clear that the indicator function appearing outside the integrals in
(\ref{eq:mupstream}) and (\ref{eq:mdownstream}) is always unity.  Therefore,  $ M^u $ has an
integral from $ p^u $ to $ p $, whereas $ M^d $ has negative an integral from $ p^d $ to $ p $.  Since the integrand is
identical, this simply transforms to an integral from $ p^u $ to $ p^d $, which gives the expression (\ref{eq:melnikov}). \\

%%%%%%%%%%%%%%%%%%%%%%%%%%%%%%%%%%
\begin{acknowledgments}
Support from the Australian Research Council  through Future Fellowship grant FT130100484 is gratefully acknowledged.
\end{acknowledgments}



\begin{thebibliography}{92}%
\makeatletter
\providecommand \@ifxundefined [1]{%
 \@ifx{#1\undefined}
}%
\providecommand \@ifnum [1]{%
 \ifnum #1\expandafter \@firstoftwo
 \else \expandafter \@secondoftwo
 \fi
}%
\providecommand \@ifx [1]{%
 \ifx #1\expandafter \@firstoftwo
 \else \expandafter \@secondoftwo
 \fi
}%
\providecommand \natexlab [1]{#1}%
\providecommand \enquote  [1]{``#1''}%
\providecommand \bibnamefont  [1]{#1}%
\providecommand \bibfnamefont [1]{#1}%
\providecommand \citenamefont [1]{#1}%
\providecommand \href@noop [0]{\@secondoftwo}%
\providecommand \href [0]{\begingroup \@sanitize@url \@href}%
\providecommand \@href[1]{\@@startlink{#1}\@@href}%
\providecommand \@@href[1]{\endgroup#1\@@endlink}%
\providecommand \@sanitize@url [0]{\catcode `\\12\catcode `\$12\catcode
  `\&12\catcode `\#12\catcode `\^12\catcode `\_12\catcode `\%12\relax}%
\providecommand \@@startlink[1]{}%
\providecommand \@@endlink[0]{}%
\providecommand \url  [0]{\begingroup\@sanitize@url \@url }%
\providecommand \@url [1]{\endgroup\@href {#1}{\urlprefix }}%
\providecommand \urlprefix  [0]{URL }%
\providecommand \Eprint [0]{\href }%
\providecommand \doibase [0]{http://dx.doi.org/}%
\providecommand \selectlanguage [0]{\@gobble}%
\providecommand \bibinfo  [0]{\@secondoftwo}%
\providecommand \bibfield  [0]{\@secondoftwo}%
\providecommand \translation [1]{[#1]}%
\providecommand \BibitemOpen [0]{}%
\providecommand \bibitemStop [0]{}%
\providecommand \bibitemNoStop [0]{.\EOS\space}%
\providecommand \EOS [0]{\spacefactor3000\relax}%
\providecommand \BibitemShut  [1]{\csname bibitem#1\endcsname}%
\let\auto@bib@innerbib\@empty
%</preamble>
\bibitem [{\citenamefont {Haller}(2015)}]{hallerreview}%
  \BibitemOpen
  \bibfield  {author} {\bibinfo {author} {\bibfnamefont {G.}~\bibnamefont
  {Haller}},\ }\href@noop {} {\bibfield  {journal} {\bibinfo  {journal} {Annu.
  Rev. Fluid Mech.}\ }\textbf {\bibinfo {volume} {47}},\ \bibinfo {pages} {137}
  (\bibinfo {year} {2015})}\BibitemShut {NoStop}%
\bibitem [{\citenamefont {Peacock}\ \emph {et~al.}(2015)\citenamefont
  {Peacock}, \citenamefont {Froyland},\ and\ \citenamefont
  {Haller}}]{peacockfroylandhaller}%
  \BibitemOpen
  \bibfield  {author} {\bibinfo {author} {\bibfnamefont {T.}~\bibnamefont
  {Peacock}}, \bibinfo {author} {\bibfnamefont {G.}~\bibnamefont {Froyland}}, \
  and\ \bibinfo {author} {\bibfnamefont {G.}~\bibnamefont {Haller}},\
  }\href@noop {} {\bibfield  {journal} {\bibinfo  {journal} {Chaos}\ }\textbf
  {\bibinfo {volume} {25}},\ \bibinfo {pages} {087201} (\bibinfo {year}
  {2015})}\BibitemShut {NoStop}%
\bibitem [{\citenamefont {Balasuriya}(2016{\natexlab{a}})}]{siam_book}%
  \BibitemOpen
  \bibfield  {author} {\bibinfo {author} {\bibfnamefont {S.}~\bibnamefont
  {Balasuriya}},\ }\href@noop {} {\emph {\bibinfo {title} {{M}elnikov methods
  for barriers and transport in unsteady flows}}},\ SIAM Series on Mathematical
  Modeling and Computation\ (\bibinfo  {publisher} {SIAM Press},\ \bibinfo
  {address} {in press},\ \bibinfo {year} {2016})\BibitemShut {NoStop}%
\bibitem [{\citenamefont {Balasuriya}(2014)}]{open}%
  \BibitemOpen
  \bibfield  {author} {\bibinfo {author} {\bibfnamefont {S.}~\bibnamefont
  {Balasuriya}},\ }in\ \href@noop {} {\emph {\bibinfo {booktitle} {Ergodic
  Theory, Open Dynamics and Structures}}}\ (\bibinfo  {publisher} {Springer},\
  \bibinfo {year} {2014})\ Chap.~\bibinfo {chapter} {1}, pp.\ \bibinfo {pages}
  {1--30}\BibitemShut {NoStop}%
\bibitem [{\citenamefont {Balasuriya}(2015{\natexlab{a}})}]{microfluid_review}%
  \BibitemOpen
  \bibfield  {author} {\bibinfo {author} {\bibfnamefont {S.}~\bibnamefont
  {Balasuriya}},\ }\href@noop {} {\bibfield  {journal} {\bibinfo  {journal} {J.
  Micromech. Microeng.}\ }\textbf {\bibinfo {volume} {25}},\ \bibinfo {pages}
  {094005} (\bibinfo {year} {2015}{\natexlab{a}})}\BibitemShut {NoStop}%
\bibitem [{\citenamefont {Balasuriya}\ and\ \citenamefont
  {Jones}(2001)}]{eddy}%
  \BibitemOpen
  \bibfield  {author} {\bibinfo {author} {\bibfnamefont {S.}~\bibnamefont
  {Balasuriya}}\ and\ \bibinfo {author} {\bibfnamefont {C.}~\bibnamefont
  {Jones}},\ }\href@noop {} {\bibfield  {journal} {\bibinfo  {journal} {Nonlin.
  Proc. Geophys.}\ }\textbf {\bibinfo {volume} {8}},\ \bibinfo {pages} {241}
  (\bibinfo {year} {2001})}\BibitemShut {NoStop}%
\bibitem [{\citenamefont {Kelley}\ \emph {et~al.}(2013)\citenamefont {Kelley},
  \citenamefont {Allshouse},\ and\ \citenamefont {Ouellette}}]{kelley}%
  \BibitemOpen
  \bibfield  {author} {\bibinfo {author} {\bibfnamefont {D.}~\bibnamefont
  {Kelley}}, \bibinfo {author} {\bibfnamefont {M.}~\bibnamefont {Allshouse}}, \
  and\ \bibinfo {author} {\bibfnamefont {N.}~\bibnamefont {Ouellette}},\
  }\href@noop {} {\bibfield  {journal} {\bibinfo  {journal} {Phys. Rev. E}\
  }\textbf {\bibinfo {volume} {88}},\ \bibinfo {pages} {013017} (\bibinfo
  {year} {2013})}\BibitemShut {NoStop}%
\bibitem [{\citenamefont {Guckenheimer}\ and\ \citenamefont
  {Holmes}(1983)}]{guckenheimerholmes}%
  \BibitemOpen
  \bibfield  {author} {\bibinfo {author} {\bibfnamefont {J.}~\bibnamefont
  {Guckenheimer}}\ and\ \bibinfo {author} {\bibfnamefont {P.}~\bibnamefont
  {Holmes}},\ }\href@noop {} {\emph {\bibinfo {title} {Nonlinear Oscillations,
  Dynamical Systems and Bifurcations of Vector Fields}}}\ (\bibinfo
  {publisher} {Springer},\ \bibinfo {address} {New York},\ \bibinfo {year}
  {1983})\BibitemShut {NoStop}%
\bibitem [{\citenamefont {Haller}\ and\ \citenamefont
  {Yuan}(2000)}]{halleryuan}%
  \BibitemOpen
  \bibfield  {author} {\bibinfo {author} {\bibfnamefont {G.}~\bibnamefont
  {Haller}}\ and\ \bibinfo {author} {\bibfnamefont {G.-C.}\ \bibnamefont
  {Yuan}},\ }\href@noop {} {\bibfield  {journal} {\bibinfo  {journal} {Phys.\
  D}\ }\textbf {\bibinfo {volume} {147}},\ \bibinfo {pages} {352} (\bibinfo
  {year} {2000})}\BibitemShut {NoStop}%
\bibitem [{\citenamefont {Karrasch}\ \emph {et~al.}(2015)\citenamefont
  {Karrasch}, \citenamefont {Farazmand},\ and\ \citenamefont
  {Haller}}]{karraschfarazmandhaller}%
  \BibitemOpen
  \bibfield  {author} {\bibinfo {author} {\bibfnamefont {D.}~\bibnamefont
  {Karrasch}}, \bibinfo {author} {\bibfnamefont {M.}~\bibnamefont {Farazmand}},
  \ and\ \bibinfo {author} {\bibfnamefont {G.}~\bibnamefont {Haller}},\
  }\href@noop {} {\bibfield  {journal} {\bibinfo  {journal} {J. Comput. Dyn.}\
  }\textbf {\bibinfo {volume} {2}},\ \bibinfo {pages} {83} (\bibinfo {year}
  {2015})}\BibitemShut {NoStop}%
\bibitem [{\citenamefont {Onu}\ \emph {et~al.}(2015)\citenamefont {Onu},
  \citenamefont {Kuhn},\ and\ \citenamefont {Haller}}]{onu}%
  \BibitemOpen
  \bibfield  {author} {\bibinfo {author} {\bibfnamefont {F.}~\bibnamefont
  {Onu}}, \bibinfo {author} {\bibfnamefont {F.}~\bibnamefont {Kuhn}}, \ and\
  \bibinfo {author} {\bibfnamefont {G.}~\bibnamefont {Haller}},\ }\href@noop {}
  {\bibfield  {journal} {\bibinfo  {journal} {J. Comput. Sci.}\ }\textbf
  {\bibinfo {volume} {7}},\ \bibinfo {pages} {26} (\bibinfo {year}
  {2015})}\BibitemShut {NoStop}%
\bibitem [{\citenamefont {Shadden}\ \emph {et~al.}(2005)\citenamefont
  {Shadden}, \citenamefont {Lekien},\ and\ \citenamefont {Marsden}}]{shadden}%
  \BibitemOpen
  \bibfield  {author} {\bibinfo {author} {\bibfnamefont {S.}~\bibnamefont
  {Shadden}}, \bibinfo {author} {\bibfnamefont {F.}~\bibnamefont {Lekien}}, \
  and\ \bibinfo {author} {\bibfnamefont {J.}~\bibnamefont {Marsden}},\
  }\href@noop {} {\bibfield  {journal} {\bibinfo  {journal} {Phys. D}\ }\textbf
  {\bibinfo {volume} {212}},\ \bibinfo {pages} {271} (\bibinfo {year}
  {2005})}\BibitemShut {NoStop}%
\bibitem [{\citenamefont {He}\ \emph {et~al.}(2016)\citenamefont {He},
  \citenamefont {Pan}, \citenamefont {feng}, \citenamefont {Gao},\ and\
  \citenamefont {Wang}}]{he}%
  \BibitemOpen
  \bibfield  {author} {\bibinfo {author} {\bibfnamefont {G.}~\bibnamefont
  {He}}, \bibinfo {author} {\bibfnamefont {C.}~\bibnamefont {Pan}}, \bibinfo
  {author} {\bibfnamefont {L.}~\bibnamefont {feng}}, \bibinfo {author}
  {\bibfnamefont {Q.}~\bibnamefont {Gao}}, \ and\ \bibinfo {author}
  {\bibfnamefont {J.}~\bibnamefont {Wang}},\ }\href@noop {} {\bibfield
  {journal} {\bibinfo  {journal} {J. Fluid Mech.}\ }\textbf {\bibinfo {volume}
  {792}},\ \bibinfo {pages} {274} (\bibinfo {year} {2016})}\BibitemShut
  {NoStop}%
\bibitem [{\citenamefont {Huntley}\ \emph {et~al.}(2015)\citenamefont
  {Huntley}, \citenamefont {Lipphardt}, \citenamefont {Jacobs},\ and\
  \citenamefont {Kirwan}}]{huntley}%
  \BibitemOpen
  \bibfield  {author} {\bibinfo {author} {\bibfnamefont {H.}~\bibnamefont
  {Huntley}}, \bibinfo {author} {\bibfnamefont {B.}~\bibnamefont {Lipphardt}},
  \bibinfo {author} {\bibfnamefont {G.}~\bibnamefont {Jacobs}}, \ and\ \bibinfo
  {author} {\bibfnamefont {A.}~\bibnamefont {Kirwan}},\ }\href@noop {}
  {\bibfield  {journal} {\bibinfo  {journal} {J. Geophys. Res. Oceans}\
  }\textbf {\bibinfo {volume} {120}},\ \bibinfo {pages} {6622} (\bibinfo {year}
  {2015})}\BibitemShut {NoStop}%
\bibitem [{\citenamefont {Nelson}\ and\ \citenamefont
  {Jacobs}(2015)}]{nelsonjacobs}%
  \BibitemOpen
  \bibfield  {author} {\bibinfo {author} {\bibfnamefont {D.}~\bibnamefont
  {Nelson}}\ and\ \bibinfo {author} {\bibfnamefont {G.}~\bibnamefont
  {Jacobs}},\ }\href@noop {} {\bibfield  {journal} {\bibinfo  {journal} {J.
  Comput. Phys.}\ }\textbf {\bibinfo {volume} {295}},\ \bibinfo {pages} {65}
  (\bibinfo {year} {2015})}\BibitemShut {NoStop}%
\bibitem [{\citenamefont {Johnson}\ and\ \citenamefont
  {Meneveau}(2015)}]{johnsonmeneveau}%
  \BibitemOpen
  \bibfield  {author} {\bibinfo {author} {\bibfnamefont {P.}~\bibnamefont
  {Johnson}}\ and\ \bibinfo {author} {\bibfnamefont {C.}~\bibnamefont
  {Meneveau}},\ }\href@noop {} {\bibfield  {journal} {\bibinfo  {journal}
  {Phys. Fluids}\ }\textbf {\bibinfo {volume} {27}},\ \bibinfo {pages} {085110}
  (\bibinfo {year} {2015})}\BibitemShut {NoStop}%
\bibitem [{\citenamefont {Bozorg{M}agham}\ and\ \citenamefont
  {Ross}(2015)}]{bozorgmaghamross}%
  \BibitemOpen
  \bibfield  {author} {\bibinfo {author} {\bibfnamefont {A.}~\bibnamefont
  {Bozorg{M}agham}}\ and\ \bibinfo {author} {\bibfnamefont {S.}~\bibnamefont
  {Ross}},\ }\href@noop {} {\bibfield  {journal} {\bibinfo  {journal} {Commun.
  Nonlin. Sci. Numer. Simu.}\ }\textbf {\bibinfo {volume} {22}},\ \bibinfo
  {pages} {964} (\bibinfo {year} {2015})}\BibitemShut {NoStop}%
\bibitem [{\citenamefont {Branicki}\ and\ \citenamefont
  {Wiggins}(2010)}]{branickiwiggins}%
  \BibitemOpen
  \bibfield  {author} {\bibinfo {author} {\bibfnamefont {M.}~\bibnamefont
  {Branicki}}\ and\ \bibinfo {author} {\bibfnamefont {S.}~\bibnamefont
  {Wiggins}},\ }\href@noop {} {\bibfield  {journal} {\bibinfo  {journal}
  {Nonlin. Proc. Geophys.}\ }\textbf {\bibinfo {volume} {17}},\ \bibinfo
  {pages} {1} (\bibinfo {year} {2010})}\BibitemShut {NoStop}%
\bibitem [{\citenamefont {Froyland}\ and\ \citenamefont
  {Padberg}(2009)}]{froylandpadberg}%
  \BibitemOpen
  \bibfield  {author} {\bibinfo {author} {\bibfnamefont {G.}~\bibnamefont
  {Froyland}}\ and\ \bibinfo {author} {\bibfnamefont {K.}~\bibnamefont
  {Padberg}},\ }\href@noop {} {\bibfield  {journal} {\bibinfo  {journal} {Phys.
  D}\ }\textbf {\bibinfo {volume} {238}},\ \bibinfo {pages} {1507} (\bibinfo
  {year} {2009})}\BibitemShut {NoStop}%
\bibitem [{\citenamefont {Froyland}\ \emph {et~al.}(2007)\citenamefont
  {Froyland}, \citenamefont {Padberg}, \citenamefont {England},\ and\
  \citenamefont {Treguier}}]{froylanddetection}%
  \BibitemOpen
  \bibfield  {author} {\bibinfo {author} {\bibfnamefont {G.}~\bibnamefont
  {Froyland}}, \bibinfo {author} {\bibfnamefont {K.}~\bibnamefont {Padberg}},
  \bibinfo {author} {\bibfnamefont {M.}~\bibnamefont {England}}, \ and\
  \bibinfo {author} {\bibfnamefont {A.}~\bibnamefont {Treguier}},\ }\href@noop
  {} {\bibfield  {journal} {\bibinfo  {journal} {Phys. Rev. Lett.}\ }\textbf
  {\bibinfo {volume} {98}},\ \bibinfo {pages} {224503} (\bibinfo {year}
  {2007})}\BibitemShut {NoStop}%
\bibitem [{\citenamefont {Froyland}\ \emph {et~al.}(2010)\citenamefont
  {Froyland}, \citenamefont {Santitissadeekorn},\ and\ \citenamefont
  {Monahan}}]{froylandchaos}%
  \BibitemOpen
  \bibfield  {author} {\bibinfo {author} {\bibfnamefont {G.}~\bibnamefont
  {Froyland}}, \bibinfo {author} {\bibfnamefont {N.}~\bibnamefont
  {Santitissadeekorn}}, \ and\ \bibinfo {author} {\bibfnamefont
  {A.}~\bibnamefont {Monahan}},\ }\href@noop {} {\bibfield  {journal} {\bibinfo
   {journal} {Chaos}\ }\textbf {\bibinfo {volume} {20}},\ \bibinfo {pages}
  {043116} (\bibinfo {year} {2010})}\BibitemShut {NoStop}%
\bibitem [{\citenamefont {Balasuriya}(2016{\natexlab{b}})}]{eigenvector}%
  \BibitemOpen
  \bibfield  {author} {\bibinfo {author} {\bibfnamefont {S.}~\bibnamefont
  {Balasuriya}},\ }\href@noop {} {\bibfield  {journal} {\bibinfo  {journal} {J.
  Nonlin. Sci.}\ ,\ \bibinfo {pages} {in press}} (\bibinfo {year}
  {2016}{\natexlab{b}})}\BibitemShut {NoStop}%
\bibitem [{\citenamefont {del~{C}astillo
  {N}egrete}(1998)}]{delcastillonegrete}%
  \BibitemOpen
  \bibfield  {author} {\bibinfo {author} {\bibfnamefont {D.}~\bibnamefont
  {del~{C}astillo {N}egrete}},\ }\href@noop {} {\bibfield  {journal} {\bibinfo
  {journal} {Phys. Fluids}\ }\textbf {\bibinfo {volume} {10}},\ \bibinfo
  {pages} {576} (\bibinfo {year} {1998})}\BibitemShut {NoStop}%
\bibitem [{\citenamefont {Allshouse}\ and\ \citenamefont
  {Thiffeault}(2012)}]{allshousethiffeault}%
  \BibitemOpen
  \bibfield  {author} {\bibinfo {author} {\bibfnamefont {M.}~\bibnamefont
  {Allshouse}}\ and\ \bibinfo {author} {\bibfnamefont {J.-L.}\ \bibnamefont
  {Thiffeault}},\ }\href@noop {} {\bibfield  {journal} {\bibinfo  {journal}
  {Phys. D}\ }\textbf {\bibinfo {volume} {241}},\ \bibinfo {pages} {95}
  (\bibinfo {year} {2012})}\BibitemShut {NoStop}%
\bibitem [{\citenamefont {Mundel}\ \emph {et~al.}(2014)\citenamefont {Mundel},
  \citenamefont {Fredj}, \citenamefont {Gildor},\ and\ \citenamefont
  {Rom-Kedar}}]{mundel}%
  \BibitemOpen
  \bibfield  {author} {\bibinfo {author} {\bibfnamefont {R.}~\bibnamefont
  {Mundel}}, \bibinfo {author} {\bibfnamefont {E.}~\bibnamefont {Fredj}},
  \bibinfo {author} {\bibfnamefont {H.}~\bibnamefont {Gildor}}, \ and\ \bibinfo
  {author} {\bibfnamefont {V.}~\bibnamefont {Rom-Kedar}},\ }\href@noop {}
  {\bibfield  {journal} {\bibinfo  {journal} {Phys. Fluids}\ }\textbf {\bibinfo
  {volume} {26}},\ \bibinfo {pages} {126602} (\bibinfo {year}
  {2014})}\BibitemShut {NoStop}%
\bibitem [{\citenamefont {Ma}\ \emph {et~al.}(2016)\citenamefont {Ma},
  \citenamefont {Ouellette},\ and\ \citenamefont {Bollt}}]{ma}%
  \BibitemOpen
  \bibfield  {author} {\bibinfo {author} {\bibfnamefont {T.}~\bibnamefont
  {Ma}}, \bibinfo {author} {\bibfnamefont {N.}~\bibnamefont {Ouellette}}, \
  and\ \bibinfo {author} {\bibfnamefont {E.}~\bibnamefont {Bollt}},\
  }\href@noop {} {\bibfield  {journal} {\bibinfo  {journal} {Chaos}\ }\textbf
  {\bibinfo {volume} {26}},\ \bibinfo {pages} {023112} (\bibinfo {year}
  {2016})}\BibitemShut {NoStop}%
\bibitem [{\citenamefont {Ma}\ and\ \citenamefont {Bollt}(2014)}]{mabollt}%
  \BibitemOpen
  \bibfield  {author} {\bibinfo {author} {\bibfnamefont {T.}~\bibnamefont
  {Ma}}\ and\ \bibinfo {author} {\bibfnamefont {E.}~\bibnamefont {Bollt}},\
  }\href@noop {} {\bibfield  {journal} {\bibinfo  {journal} {SIAM J. Appl. Dyn.
  Sys.}\ }\textbf {\bibinfo {volume} {13}},\ \bibinfo {pages} {1106} (\bibinfo
  {year} {2014})}\BibitemShut {NoStop}%
\bibitem [{\citenamefont {Budi\u{s}i\'c}\ and\ \citenamefont
  {Thiffeault}(2015)}]{budisicthiffeault}%
  \BibitemOpen
  \bibfield  {author} {\bibinfo {author} {\bibfnamefont {M.}~\bibnamefont
  {Budi\u{s}i\'c}}\ and\ \bibinfo {author} {\bibfnamefont {J.-L.}\ \bibnamefont
  {Thiffeault}},\ }\href@noop {} {\bibfield  {journal} {\bibinfo  {journal}
  {Chaos}\ }\textbf {\bibinfo {volume} {25}},\ \bibinfo {pages} {087407}
  (\bibinfo {year} {2015})}\BibitemShut {NoStop}%
\bibitem [{\citenamefont {Mezi\'c}\ \emph {et~al.}(2010)\citenamefont
  {Mezi\'c}, \citenamefont {Loire}, \citenamefont {Fonoberov},\ and\
  \citenamefont {Hogan}}]{mezicscience}%
  \BibitemOpen
  \bibfield  {author} {\bibinfo {author} {\bibfnamefont {I.}~\bibnamefont
  {Mezi\'c}}, \bibinfo {author} {\bibfnamefont {S.}~\bibnamefont {Loire}},
  \bibinfo {author} {\bibfnamefont {V.}~\bibnamefont {Fonoberov}}, \ and\
  \bibinfo {author} {\bibfnamefont {P.}~\bibnamefont {Hogan}},\ }\href@noop {}
  {\bibfield  {journal} {\bibinfo  {journal} {Science}\ }\textbf {\bibinfo
  {volume} {330}},\ \bibinfo {pages} {486} (\bibinfo {year}
  {2010})}\BibitemShut {NoStop}%
\bibitem [{\citenamefont {Levnaji\'c}\ and\ \citenamefont
  {Mezi\'{c}}(2010)}]{levnajicmezic}%
  \BibitemOpen
  \bibfield  {author} {\bibinfo {author} {\bibfnamefont {Z.}~\bibnamefont
  {Levnaji\'c}}\ and\ \bibinfo {author} {\bibfnamefont {I.}~\bibnamefont
  {Mezi\'{c}}},\ }\href@noop {} {\bibfield  {journal} {\bibinfo  {journal}
  {Chaos}\ }\textbf {\bibinfo {volume} {20}},\ \bibinfo {pages} {033114}
  (\bibinfo {year} {2010})}\BibitemShut {NoStop}%
\bibitem [{\citenamefont {Budi\u{s}i\'c}\ and\ \citenamefont
  {Mezi\'c}(2012)}]{budisicmezic}%
  \BibitemOpen
  \bibfield  {author} {\bibinfo {author} {\bibfnamefont {M.}~\bibnamefont
  {Budi\u{s}i\'c}}\ and\ \bibinfo {author} {\bibfnamefont {I.}~\bibnamefont
  {Mezi\'c}},\ }\href@noop {} {\bibfield  {journal} {\bibinfo  {journal} {Phys.
  D}\ }\textbf {\bibinfo {volume} {241}},\ \bibinfo {pages} {1255} (\bibinfo
  {year} {2012})}\BibitemShut {NoStop}%
\bibitem [{\citenamefont {Balasuriya}(2015{\natexlab{b}})}]{droplet}%
  \BibitemOpen
  \bibfield  {author} {\bibinfo {author} {\bibfnamefont {S.}~\bibnamefont
  {Balasuriya}},\ }\href@noop {} {\bibfield  {journal} {\bibinfo  {journal}
  {Phys. Fluids}\ }\textbf {\bibinfo {volume} {27}},\ \bibinfo {pages} {052005}
  (\bibinfo {year} {2015}{\natexlab{b}})}\BibitemShut {NoStop}%
\bibitem [{\citenamefont {Raben}\ \emph {et~al.}(2014)\citenamefont {Raben},
  \citenamefont {Ross},\ and\ \citenamefont {Vlachos}}]{raben}%
  \BibitemOpen
  \bibfield  {author} {\bibinfo {author} {\bibfnamefont {S.}~\bibnamefont
  {Raben}}, \bibinfo {author} {\bibfnamefont {S.}~\bibnamefont {Ross}}, \ and\
  \bibinfo {author} {\bibfnamefont {P.}~\bibnamefont {Vlachos}},\ }\href@noop
  {} {\bibfield  {journal} {\bibinfo  {journal} {Experiments in Fluids}\
  }\textbf {\bibinfo {volume} {55}},\ \bibinfo {pages} {1824} (\bibinfo {year}
  {2014})}\BibitemShut {NoStop}%
\bibitem [{Note1()}]{Note1}%
  \BibitemOpen
  \bibinfo {note} {Local maxima of the speed are defined by Haller as
  `parabolic Lagrangian Coherent Structures,' for which a theory has been
  developed \cite {hallerreview}.}\BibitemShut {Stop}%
\bibitem [{\citenamefont {Thiffeault}(2012)}]{thiffeault}%
  \BibitemOpen
  \bibfield  {author} {\bibinfo {author} {\bibfnamefont {J.-L.}\ \bibnamefont
  {Thiffeault}},\ }\href@noop {} {\bibfield  {journal} {\bibinfo  {journal}
  {Nonlinearity}\ }\textbf {\bibinfo {volume} {84}},\ \bibinfo {pages} {R1}
  (\bibinfo {year} {2012})}\BibitemShut {NoStop}%
\bibitem [{\citenamefont {Guo}\ \emph {et~al.}(2016)\citenamefont {Guo},
  \citenamefont {Zheng}, \citenamefont {Celia},\ and\ \citenamefont
  {Stone}}]{guo}%
  \BibitemOpen
  \bibfield  {author} {\bibinfo {author} {\bibfnamefont {B.}~\bibnamefont
  {Guo}}, \bibinfo {author} {\bibfnamefont {Z.}~\bibnamefont {Zheng}}, \bibinfo
  {author} {\bibfnamefont {M.}~\bibnamefont {Celia}}, \ and\ \bibinfo {author}
  {\bibfnamefont {H.}~\bibnamefont {Stone}},\ }\href@noop {} {\bibfield
  {journal} {\bibinfo  {journal} {Phys. Fluids}\ }\textbf {\bibinfo {volume}
  {28}},\ \bibinfo {pages} {022107} (\bibinfo {year} {2016})}\BibitemShut
  {NoStop}%
\bibitem [{\citenamefont {Zheng}\ \emph {et~al.}(2015)\citenamefont {Zheng},
  \citenamefont {Rongy},\ and\ \citenamefont {Stone}}]{zheng}%
  \BibitemOpen
  \bibfield  {author} {\bibinfo {author} {\bibfnamefont {Z.}~\bibnamefont
  {Zheng}}, \bibinfo {author} {\bibfnamefont {L.}~\bibnamefont {Rongy}}, \ and\
  \bibinfo {author} {\bibfnamefont {H.}~\bibnamefont {Stone}},\ }\href@noop {}
  {\bibfield  {journal} {\bibinfo  {journal} {Phys. Fluids}\ }\textbf {\bibinfo
  {volume} {27}},\ \bibinfo {pages} {062105} (\bibinfo {year}
  {2015})}\BibitemShut {NoStop}%
\bibitem [{\citenamefont {Meunier}\ and\ \citenamefont
  {Villermaux}(2003)}]{meuniervillermaux}%
  \BibitemOpen
  \bibfield  {author} {\bibinfo {author} {\bibfnamefont {P.}~\bibnamefont
  {Meunier}}\ and\ \bibinfo {author} {\bibfnamefont {E.}~\bibnamefont
  {Villermaux}},\ }\href@noop {} {\bibfield  {journal} {\bibinfo  {journal} {J.
  Fluid Mech.}\ }\textbf {\bibinfo {volume} {476}},\ \bibinfo {pages} {213}
  (\bibinfo {year} {2003})}\BibitemShut {NoStop}%
\bibitem [{\citenamefont {Fernandez}\ \emph {et~al.}(2002)\citenamefont
  {Fernandez}, \citenamefont {Kurowski}, \citenamefont {Petitjeans},\ and\
  \citenamefont {Meiburg}}]{fernandez}%
  \BibitemOpen
  \bibfield  {author} {\bibinfo {author} {\bibfnamefont {J.}~\bibnamefont
  {Fernandez}}, \bibinfo {author} {\bibfnamefont {P.}~\bibnamefont {Kurowski}},
  \bibinfo {author} {\bibfnamefont {P.}~\bibnamefont {Petitjeans}}, \ and\
  \bibinfo {author} {\bibfnamefont {E.}~\bibnamefont {Meiburg}},\ }\href@noop
  {} {\bibfield  {journal} {\bibinfo  {journal} {J. Fluid Mech.}\ }\textbf
  {\bibinfo {volume} {451}},\ \bibinfo {pages} {239} (\bibinfo {year}
  {2002})}\BibitemShut {NoStop}%
\bibitem [{\citenamefont {Camassa}\ \emph {et~al.}(2016)\citenamefont
  {Camassa}, \citenamefont {Lin}, \citenamefont {McLaughlin}, \citenamefont
  {Mertens}, \citenamefont {Tzou}, \citenamefont {Walsh},\ and\ \citenamefont
  {White}}]{camassa}%
  \BibitemOpen
  \bibfield  {author} {\bibinfo {author} {\bibfnamefont {R.}~\bibnamefont
  {Camassa}}, \bibinfo {author} {\bibfnamefont {Z.}~\bibnamefont {Lin}},
  \bibinfo {author} {\bibfnamefont {R.}~\bibnamefont {McLaughlin}}, \bibinfo
  {author} {\bibfnamefont {L.}~\bibnamefont {Mertens}}, \bibinfo {author}
  {\bibfnamefont {C.}~\bibnamefont {Tzou}}, \bibinfo {author} {\bibfnamefont
  {J.}~\bibnamefont {Walsh}}, \ and\ \bibinfo {author} {\bibfnamefont
  {B.}~\bibnamefont {White}},\ }\href@noop {} {\bibfield  {journal} {\bibinfo
  {journal} {J. Fluid Mech.}\ }\textbf {\bibinfo {volume} {790}},\ \bibinfo
  {pages} {71} (\bibinfo {year} {2016})}\BibitemShut {NoStop}%
\bibitem [{\citenamefont {Mathew}\ \emph {et~al.}(2007)\citenamefont {Mathew},
  \citenamefont {Mezi\'c}, \citenamefont {Grivopoulos}, \citenamefont
  {Vaidya},\ and\ \citenamefont {Petzold}}]{mathewoptimal}%
  \BibitemOpen
  \bibfield  {author} {\bibinfo {author} {\bibfnamefont {G.}~\bibnamefont
  {Mathew}}, \bibinfo {author} {\bibfnamefont {I.}~\bibnamefont {Mezi\'c}},
  \bibinfo {author} {\bibfnamefont {S.}~\bibnamefont {Grivopoulos}}, \bibinfo
  {author} {\bibfnamefont {U.}~\bibnamefont {Vaidya}}, \ and\ \bibinfo {author}
  {\bibfnamefont {L.}~\bibnamefont {Petzold}},\ }\href@noop {} {\bibfield
  {journal} {\bibinfo  {journal} {J. Fluid Mech.}\ }\textbf {\bibinfo {volume}
  {580}},\ \bibinfo {pages} {261} (\bibinfo {year} {2007})}\BibitemShut
  {NoStop}%
\bibitem [{\citenamefont {Lin}\ \emph {et~al.}(2011)\citenamefont {Lin},
  \citenamefont {Thiffeault},\ and\ \citenamefont
  {Doering}}]{linthiffeaultdoering}%
  \BibitemOpen
  \bibfield  {author} {\bibinfo {author} {\bibfnamefont {Z.}~\bibnamefont
  {Lin}}, \bibinfo {author} {\bibfnamefont {J.-L.}\ \bibnamefont {Thiffeault}},
  \ and\ \bibinfo {author} {\bibfnamefont {C.}~\bibnamefont {Doering}},\
  }\href@noop {} {\bibfield  {journal} {\bibinfo  {journal} {J. Fluid Mech.}\
  }\textbf {\bibinfo {volume} {675}},\ \bibinfo {pages} {465} (\bibinfo {year}
  {2011})}\BibitemShut {NoStop}%
\bibitem [{\citenamefont {Cortelezzi}\ \emph {et~al.}(2008)\citenamefont
  {Cortelezzi}, \citenamefont {Adrover},\ and\ \citenamefont
  {Giona}}]{cortelezzi}%
  \BibitemOpen
  \bibfield  {author} {\bibinfo {author} {\bibfnamefont {L.}~\bibnamefont
  {Cortelezzi}}, \bibinfo {author} {\bibfnamefont {A.}~\bibnamefont {Adrover}},
  \ and\ \bibinfo {author} {\bibfnamefont {M.}~\bibnamefont {Giona}},\
  }\href@noop {} {\bibfield  {journal} {\bibinfo  {journal} {J. Fluid Mech.}\
  }\textbf {\bibinfo {volume} {597}},\ \bibinfo {pages} {199} (\bibinfo {year}
  {2008})}\BibitemShut {NoStop}%
\bibitem [{\citenamefont {Vikhansky}(2002)}]{vikhanskyoptimal}%
  \BibitemOpen
  \bibfield  {author} {\bibinfo {author} {\bibfnamefont {A.}~\bibnamefont
  {Vikhansky}},\ }\href@noop {} {\bibfield  {journal} {\bibinfo  {journal}
  {Chem. Engin. Sci.}\ }\textbf {\bibinfo {volume} {57}},\ \bibinfo {pages}
  {2719} (\bibinfo {year} {2002})}\BibitemShut {NoStop}%
\bibitem [{\citenamefont {Lin}\ \emph {et~al.}(2012)\citenamefont {Lin},
  \citenamefont {Lunasin}, \citenamefont {Novikov}, \citenamefont {Mazzucato},\
  and\ \citenamefont {Doering}}]{lin}%
  \BibitemOpen
  \bibfield  {author} {\bibinfo {author} {\bibfnamefont {Z.}~\bibnamefont
  {Lin}}, \bibinfo {author} {\bibfnamefont {E.}~\bibnamefont {Lunasin}},
  \bibinfo {author} {\bibfnamefont {A.}~\bibnamefont {Novikov}}, \bibinfo
  {author} {\bibfnamefont {A.}~\bibnamefont {Mazzucato}}, \ and\ \bibinfo
  {author} {\bibfnamefont {C.}~\bibnamefont {Doering}},\ }\href@noop {}
  {\bibfield  {journal} {\bibinfo  {journal} {J. Math. Phys.}\ }\textbf
  {\bibinfo {volume} {53}},\ \bibinfo {pages} {115611} (\bibinfo {year}
  {2012})}\BibitemShut {NoStop}%
\bibitem [{\citenamefont {Balasuriya}(2005{\natexlab{a}})}]{mixer}%
  \BibitemOpen
  \bibfield  {author} {\bibinfo {author} {\bibfnamefont {S.}~\bibnamefont
  {Balasuriya}},\ }\href@noop {} {\bibfield  {journal} {\bibinfo  {journal}
  {Phys.\ Fluids}\ }\textbf {\bibinfo {volume} {17}},\ \bibinfo {pages}
  {118103} (\bibinfo {year} {2005}{\natexlab{a}})}\BibitemShut {NoStop}%
\bibitem [{\citenamefont {Balasuriya}\ and\ \citenamefont
  {Finn}(2012)}]{l2mixer}%
  \BibitemOpen
  \bibfield  {author} {\bibinfo {author} {\bibfnamefont {S.}~\bibnamefont
  {Balasuriya}}\ and\ \bibinfo {author} {\bibfnamefont {M.}~\bibnamefont
  {Finn}},\ }\href@noop {} {\bibfield  {journal} {\bibinfo  {journal} {Phys.
  Rev. Lett.}\ }\textbf {\bibinfo {volume} {108}},\ \bibinfo {pages} {244503}
  (\bibinfo {year} {2012})}\BibitemShut {NoStop}%
\bibitem [{\citenamefont {Balasuriya}(2010)}]{frequency}%
  \BibitemOpen
  \bibfield  {author} {\bibinfo {author} {\bibfnamefont {S.}~\bibnamefont
  {Balasuriya}},\ }\href@noop {} {\bibfield  {journal} {\bibinfo  {journal}
  {Phys. Rev. Lett.}\ }\textbf {\bibinfo {volume} {105}},\ \bibinfo {pages}
  {064501} (\bibinfo {year} {2010})}\BibitemShut {NoStop}%
\bibitem [{\citenamefont {Sussman}\ \emph {et~al.}(1994)\citenamefont
  {Sussman}, \citenamefont {Smereka},\ and\ \citenamefont {Osher}}]{sussman}%
  \BibitemOpen
  \bibfield  {author} {\bibinfo {author} {\bibfnamefont {M.}~\bibnamefont
  {Sussman}}, \bibinfo {author} {\bibfnamefont {P.}~\bibnamefont {Smereka}}, \
  and\ \bibinfo {author} {\bibfnamefont {S.}~\bibnamefont {Osher}},\
  }\href@noop {} {\bibfield  {journal} {\bibinfo  {journal} {J. Comput. Phys.}\
  }\textbf {\bibinfo {volume} {114}},\ \bibinfo {pages} {146} (\bibinfo {year}
  {1994})}\BibitemShut {NoStop}%
\bibitem [{\citenamefont {Lynden-Bell}\ \emph {et~al.}(2005)\citenamefont
  {Lynden-Bell}, \citenamefont {Kohanoff},\ and\ \citenamefont
  {Popolo}}]{lyndenbell}%
  \BibitemOpen
  \bibfield  {author} {\bibinfo {author} {\bibfnamefont {R.}~\bibnamefont
  {Lynden-Bell}}, \bibinfo {author} {\bibfnamefont {J.}~\bibnamefont
  {Kohanoff}}, \ and\ \bibinfo {author} {\bibfnamefont {M.~D.}\ \bibnamefont
  {Popolo}},\ }\href@noop {} {\bibfield  {journal} {\bibinfo  {journal}
  {Faraday Discussions}\ }\textbf {\bibinfo {volume} {129}},\ \bibinfo {pages}
  {57} (\bibinfo {year} {2005})}\BibitemShut {NoStop}%
\bibitem [{\citenamefont {Young}\ \emph {et~al.}(2001)\citenamefont {Young},
  \citenamefont {Tufo}, \citenamefont {Dubey},\ and\ \citenamefont
  {Rosner}}]{young}%
  \BibitemOpen
  \bibfield  {author} {\bibinfo {author} {\bibfnamefont {Y.}~\bibnamefont
  {Young}}, \bibinfo {author} {\bibfnamefont {H.}~\bibnamefont {Tufo}},
  \bibinfo {author} {\bibfnamefont {A.}~\bibnamefont {Dubey}}, \ and\ \bibinfo
  {author} {\bibfnamefont {R.}~\bibnamefont {Rosner}},\ }\href@noop {}
  {\bibfield  {journal} {\bibinfo  {journal} {J. Fluid Mech.}\ }\textbf
  {\bibinfo {volume} {447}},\ \bibinfo {pages} {377} (\bibinfo {year}
  {2001})}\BibitemShut {NoStop}%
\bibitem [{\citenamefont {Sahu}(2016)}]{sahu}%
  \BibitemOpen
  \bibfield  {author} {\bibinfo {author} {\bibfnamefont {K.}~\bibnamefont
  {Sahu}},\ }\href@noop {} {\bibfield  {journal} {\bibinfo  {journal} {J. Fluid
  Mech.}\ }\textbf {\bibinfo {volume} {789}},\ \bibinfo {pages} {830} (\bibinfo
  {year} {2016})}\BibitemShut {NoStop}%
\bibitem [{\citenamefont {Roca}\ \emph {et~al.}(2014)\citenamefont {Roca},
  \citenamefont {Cammilleri}, \citenamefont {Duriez}, \citenamefont
  {Mathelin},\ and\ \citenamefont {Artana}}]{roca}%
  \BibitemOpen
  \bibfield  {author} {\bibinfo {author} {\bibfnamefont {P.}~\bibnamefont
  {Roca}}, \bibinfo {author} {\bibfnamefont {A.}~\bibnamefont {Cammilleri}},
  \bibinfo {author} {\bibfnamefont {T.}~\bibnamefont {Duriez}}, \bibinfo
  {author} {\bibfnamefont {L.}~\bibnamefont {Mathelin}}, \ and\ \bibinfo
  {author} {\bibfnamefont {G.}~\bibnamefont {Artana}},\ }\href@noop {}
  {\bibfield  {journal} {\bibinfo  {journal} {Phys. Fluids}\ }\textbf {\bibinfo
  {volume} {26}},\ \bibinfo {pages} {047102} (\bibinfo {year}
  {2014})}\BibitemShut {NoStop}%
\bibitem [{\citenamefont {Shariff}\ \emph {et~al.}(1991)\citenamefont
  {Shariff}, \citenamefont {Pulliam},\ and\ \citenamefont {Ottino}}]{shariff}%
  \BibitemOpen
  \bibfield  {author} {\bibinfo {author} {\bibfnamefont {K.}~\bibnamefont
  {Shariff}}, \bibinfo {author} {\bibfnamefont {T.}~\bibnamefont {Pulliam}}, \
  and\ \bibinfo {author} {\bibfnamefont {J.}~\bibnamefont {Ottino}},\ }\enquote
  {\bibinfo {title} {A dynamical systems analysis of kinematics in the
  time-periodic wake of a circular cylinder},}\ \ (\bibinfo  {publisher}
  {American Mathematical Society},\ \bibinfo {year} {1991})\ pp.\ \bibinfo
  {pages} {613--646}\BibitemShut {NoStop}%
\bibitem [{\citenamefont {Ziemniak}\ \emph {et~al.}(1994)\citenamefont
  {Ziemniak}, \citenamefont {Jung},\ and\ \citenamefont {T\'{e}l}}]{ziemniak}%
  \BibitemOpen
  \bibfield  {author} {\bibinfo {author} {\bibfnamefont {E.}~\bibnamefont
  {Ziemniak}}, \bibinfo {author} {\bibfnamefont {C.}~\bibnamefont {Jung}}, \
  and\ \bibinfo {author} {\bibfnamefont {T.}~\bibnamefont {T\'{e}l}},\
  }\href@noop {} {\bibfield  {journal} {\bibinfo  {journal} {Phys. D}\ }\textbf
  {\bibinfo {volume} {76}},\ \bibinfo {pages} {123} (\bibinfo {year}
  {1994})}\BibitemShut {NoStop}%
\bibitem [{\citenamefont {Alam}\ \emph {et~al.}(2006)\citenamefont {Alam},
  \citenamefont {Liu},\ and\ \citenamefont {Haller}}]{alam}%
  \BibitemOpen
  \bibfield  {author} {\bibinfo {author} {\bibfnamefont {M.}~\bibnamefont
  {Alam}}, \bibinfo {author} {\bibfnamefont {W.}~\bibnamefont {Liu}}, \ and\
  \bibinfo {author} {\bibfnamefont {G.}~\bibnamefont {Haller}},\ }\href@noop {}
  {\bibfield  {journal} {\bibinfo  {journal} {Phys. Fluids}\ }\textbf {\bibinfo
  {volume} {18}},\ \bibinfo {pages} {043601} (\bibinfo {year}
  {2006})}\BibitemShut {NoStop}%
\bibitem [{\citenamefont {Bottausci}\ \emph {et~al.}(2004)\citenamefont
  {Bottausci}, \citenamefont {Mezi\'c}, \citenamefont {Meinhart},\ and\
  \citenamefont {Cardonne}}]{bottausciRSoc}%
  \BibitemOpen
  \bibfield  {author} {\bibinfo {author} {\bibfnamefont {F.}~\bibnamefont
  {Bottausci}}, \bibinfo {author} {\bibfnamefont {I.}~\bibnamefont {Mezi\'c}},
  \bibinfo {author} {\bibfnamefont {C.}~\bibnamefont {Meinhart}}, \ and\
  \bibinfo {author} {\bibfnamefont {C.}~\bibnamefont {Cardonne}},\ }\href@noop
  {} {\bibfield  {journal} {\bibinfo  {journal} {Proc. Trans. R. Soc. Lond. A}\
  }\textbf {\bibinfo {volume} {362}},\ \bibinfo {pages} {1001} (\bibinfo {year}
  {2004})}\BibitemShut {NoStop}%
\bibitem [{\citenamefont {Bottausci}\ \emph {et~al.}(2007)\citenamefont
  {Bottausci}, \citenamefont {Cardonne}, \citenamefont {Meinhart},\ and\
  \citenamefont {Mezi\'c}}]{bottausci}%
  \BibitemOpen
  \bibfield  {author} {\bibinfo {author} {\bibfnamefont {F.}~\bibnamefont
  {Bottausci}}, \bibinfo {author} {\bibfnamefont {C.}~\bibnamefont {Cardonne}},
  \bibinfo {author} {\bibfnamefont {C.}~\bibnamefont {Meinhart}}, \ and\
  \bibinfo {author} {\bibfnamefont {I.}~\bibnamefont {Mezi\'c}},\ }\href@noop
  {} {\bibfield  {journal} {\bibinfo  {journal} {Lab Chip}\ }\textbf {\bibinfo
  {volume} {7}},\ \bibinfo {pages} {396} (\bibinfo {year} {2007})}\BibitemShut
  {NoStop}%
\bibitem [{\citenamefont {Niu}\ \emph {et~al.}(2006{\natexlab{a}})\citenamefont
  {Niu}, \citenamefont {Liu}, \citenamefont {Wen},\ and\ \citenamefont
  {Sheng}}]{niuactive}%
  \BibitemOpen
  \bibfield  {author} {\bibinfo {author} {\bibfnamefont {X.}~\bibnamefont
  {Niu}}, \bibinfo {author} {\bibfnamefont {L.}~\bibnamefont {Liu}}, \bibinfo
  {author} {\bibfnamefont {W.}~\bibnamefont {Wen}}, \ and\ \bibinfo {author}
  {\bibfnamefont {P.}~\bibnamefont {Sheng}},\ }\href@noop {} {\bibfield
  {journal} {\bibinfo  {journal} {Appl. Phys. Lett.}\ }\textbf {\bibinfo
  {volume} {88}},\ \bibinfo {pages} {153508} (\bibinfo {year}
  {2006}{\natexlab{a}})}\BibitemShut {NoStop}%
\bibitem [{\citenamefont {Tabeling}\ \emph {et~al.}(2004)\citenamefont
  {Tabeling}, \citenamefont {Chabart}, \citenamefont {Dodge}, \citenamefont
  {Jullien},\ and\ \citenamefont {Okkels}}]{tabelingchaotic}%
  \BibitemOpen
  \bibfield  {author} {\bibinfo {author} {\bibfnamefont {P.}~\bibnamefont
  {Tabeling}}, \bibinfo {author} {\bibfnamefont {M.}~\bibnamefont {Chabart}},
  \bibinfo {author} {\bibfnamefont {A.}~\bibnamefont {Dodge}}, \bibinfo
  {author} {\bibfnamefont {C.}~\bibnamefont {Jullien}}, \ and\ \bibinfo
  {author} {\bibfnamefont {F.}~\bibnamefont {Okkels}},\ }\href@noop {}
  {\bibfield  {journal} {\bibinfo  {journal} {Phil. Trans. R. Soc. Lond. A}\
  }\textbf {\bibinfo {volume} {362}},\ \bibinfo {pages} {987} (\bibinfo {year}
  {2004})}\BibitemShut {NoStop}%
\bibitem [{\citenamefont {Lee}\ \emph {et~al.}(2007)\citenamefont {Lee},
  \citenamefont {Shih}, \citenamefont {Tabeling},\ and\ \citenamefont
  {Ho}}]{lee}%
  \BibitemOpen
  \bibfield  {author} {\bibinfo {author} {\bibfnamefont {Y.-K.}\ \bibnamefont
  {Lee}}, \bibinfo {author} {\bibfnamefont {C.}~\bibnamefont {Shih}}, \bibinfo
  {author} {\bibfnamefont {P.}~\bibnamefont {Tabeling}}, \ and\ \bibinfo
  {author} {\bibfnamefont {C.-M.}\ \bibnamefont {Ho}},\ }\href@noop {}
  {\bibfield  {journal} {\bibinfo  {journal} {J. Fluid Mech.}\ }\textbf
  {\bibinfo {volume} {575}},\ \bibinfo {pages} {425} (\bibinfo {year}
  {2007})}\BibitemShut {NoStop}%
\bibitem [{\citenamefont {Kirchhoff}(1876)}]{kirchhoff}%
  \BibitemOpen
  \bibfield  {author} {\bibinfo {author} {\bibfnamefont {G.}~\bibnamefont
  {Kirchhoff}},\ }\href@noop {} {\emph {\bibinfo {title} {Vorlesungen \"{u}ber
  Mathematische Physik}}},\ Vol.~\bibinfo {volume} {1}\ (\bibinfo  {publisher}
  {Teubner},\ \bibinfo {address} {Leipzig},\ \bibinfo {year}
  {1876})\BibitemShut {NoStop}%
\bibitem [{\citenamefont {Wang}(1991)}]{wang}%
  \BibitemOpen
  \bibfield  {author} {\bibinfo {author} {\bibfnamefont {C.~Y.}\ \bibnamefont
  {Wang}},\ }\href@noop {} {\bibfield  {journal} {\bibinfo  {journal} {Annu.
  Rev. Fluid Mech.}\ }\textbf {\bibinfo {volume} {23}},\ \bibinfo {pages} {159}
  (\bibinfo {year} {1991})}\BibitemShut {NoStop}%
\bibitem [{\citenamefont {Mitchell}\ and\ \citenamefont
  {Rossi}(2008)}]{mitchellrossi}%
  \BibitemOpen
  \bibfield  {author} {\bibinfo {author} {\bibfnamefont {T.}~\bibnamefont
  {Mitchell}}\ and\ \bibinfo {author} {\bibfnamefont {L.}~\bibnamefont
  {Rossi}},\ }\href@noop {} {\bibfield  {journal} {\bibinfo  {journal} {Phys.
  Fluids}\ }\textbf {\bibinfo {volume} {20}},\ \bibinfo {pages} {054103}
  (\bibinfo {year} {2008})}\BibitemShut {NoStop}%
\bibitem [{\citenamefont {Friedland}(1999)}]{friedland}%
  \BibitemOpen
  \bibfield  {author} {\bibinfo {author} {\bibfnamefont {L.}~\bibnamefont
  {Friedland}},\ }\href@noop {} {\bibfield  {journal} {\bibinfo  {journal}
  {Phys. Rev. E}\ }\textbf {\bibinfo {volume} {59}},\ \bibinfo {pages} {4106}
  (\bibinfo {year} {1999})}\BibitemShut {NoStop}%
\bibitem [{\citenamefont {Meleshko}\ and\ \citenamefont {van
  Heijst}(1994)}]{meleshkovanheijst}%
  \BibitemOpen
  \bibfield  {author} {\bibinfo {author} {\bibfnamefont {V.}~\bibnamefont
  {Meleshko}}\ and\ \bibinfo {author} {\bibfnamefont {G.}~\bibnamefont {van
  Heijst}},\ }\href@noop {} {\bibfield  {journal} {\bibinfo  {journal} {J.
  Fluid Mech.}\ }\textbf {\bibinfo {volume} {272}},\ \bibinfo {pages} {157}
  (\bibinfo {year} {1994})}\BibitemShut {NoStop}%
\bibitem [{\citenamefont {Balasuriya}(2005{\natexlab{b}})}]{optimal}%
  \BibitemOpen
  \bibfield  {author} {\bibinfo {author} {\bibfnamefont {S.}~\bibnamefont
  {Balasuriya}},\ }\href@noop {} {\bibfield  {journal} {\bibinfo  {journal}
  {Phys. D}\ }\textbf {\bibinfo {volume} {202}},\ \bibinfo {pages} {155}
  (\bibinfo {year} {2005}{\natexlab{b}})}\BibitemShut {NoStop}%
\bibitem [{\citenamefont {Rom-Kedar}\ \emph {et~al.}(1990)\citenamefont
  {Rom-Kedar}, \citenamefont {Leonard},\ and\ \citenamefont
  {Wiggins}}]{romkedarjfm}%
  \BibitemOpen
  \bibfield  {author} {\bibinfo {author} {\bibfnamefont {V.}~\bibnamefont
  {Rom-Kedar}}, \bibinfo {author} {\bibfnamefont {A.}~\bibnamefont {Leonard}},
  \ and\ \bibinfo {author} {\bibfnamefont {S.}~\bibnamefont {Wiggins}},\
  }\href@noop {} {\bibfield  {journal} {\bibinfo  {journal} {J. Fluid Mech.}\
  }\textbf {\bibinfo {volume} {214}},\ \bibinfo {pages} {347} (\bibinfo {year}
  {1990})}\BibitemShut {NoStop}%
\bibitem [{\citenamefont {Kida}(1981)}]{kida}%
  \BibitemOpen
  \bibfield  {author} {\bibinfo {author} {\bibfnamefont {S.}~\bibnamefont
  {Kida}},\ }\href@noop {} {\bibfield  {journal} {\bibinfo  {journal} {J. Phys.
  Soc. Japan}\ }\textbf {\bibinfo {volume} {50}},\ \bibinfo {pages} {3517}
  (\bibinfo {year} {1981})}\BibitemShut {NoStop}%
\bibitem [{\citenamefont {Leweke}\ \emph {et~al.}(2016)\citenamefont {Leweke},
  \citenamefont {Dizes},\ and\ \citenamefont {Williamson}}]{leweke}%
  \BibitemOpen
  \bibfield  {author} {\bibinfo {author} {\bibfnamefont {T.}~\bibnamefont
  {Leweke}}, \bibinfo {author} {\bibfnamefont {S.~L.}\ \bibnamefont {Dizes}}, \
  and\ \bibinfo {author} {\bibfnamefont {C.}~\bibnamefont {Williamson}},\
  }\href@noop {} {\bibfield  {journal} {\bibinfo  {journal} {Annu. Rev. Fluid
  Mech.}\ }\textbf {\bibinfo {volume} {48}},\ \bibinfo {pages} {507} (\bibinfo
  {year} {2016})}\BibitemShut {NoStop}%
\bibitem [{\citenamefont {Mula}\ and\ \citenamefont {Tinney}(2015)}]{mula}%
  \BibitemOpen
  \bibfield  {author} {\bibinfo {author} {\bibfnamefont {S.}~\bibnamefont
  {Mula}}\ and\ \bibinfo {author} {\bibfnamefont {C.}~\bibnamefont {Tinney}},\
  }\href@noop {} {\bibfield  {journal} {\bibinfo  {journal} {J. Fluid Mech.}\
  }\textbf {\bibinfo {volume} {769}},\ \bibinfo {pages} {570} (\bibinfo {year}
  {2015})}\BibitemShut {NoStop}%
\bibitem [{\citenamefont {Turner}(2014)}]{turner}%
  \BibitemOpen
  \bibfield  {author} {\bibinfo {author} {\bibfnamefont {M.}~\bibnamefont
  {Turner}},\ }\href@noop {} {\bibfield  {journal} {\bibinfo  {journal} {Phys.
  Fluids}\ }\textbf {\bibinfo {volume} {26}},\ \bibinfo {pages} {116603}
  (\bibinfo {year} {2014})}\BibitemShut {NoStop}%
\bibitem [{\citenamefont {Bassom}\ and\ \citenamefont
  {Gilbert}(1999)}]{bassomgilbert}%
  \BibitemOpen
  \bibfield  {author} {\bibinfo {author} {\bibfnamefont {A.}~\bibnamefont
  {Bassom}}\ and\ \bibinfo {author} {\bibfnamefont {A.}~\bibnamefont
  {Gilbert}},\ }\href@noop {} {\bibfield  {journal} {\bibinfo  {journal} {J.
  Fluid Mech.}\ }\textbf {\bibinfo {volume} {398}},\ \bibinfo {pages} {245}
  (\bibinfo {year} {1999})}\BibitemShut {NoStop}%
\bibitem [{\citenamefont {Zhmur}\ \emph {et~al.}(2011)\citenamefont {Zhmur},
  \citenamefont {Ryzhov},\ and\ \citenamefont {Koshel}}]{zhmur}%
  \BibitemOpen
  \bibfield  {author} {\bibinfo {author} {\bibfnamefont {V.}~\bibnamefont
  {Zhmur}}, \bibinfo {author} {\bibfnamefont {E.}~\bibnamefont {Ryzhov}}, \
  and\ \bibinfo {author} {\bibfnamefont {K.}~\bibnamefont {Koshel}},\
  }\href@noop {} {\bibfield  {journal} {\bibinfo  {journal} {J. Marine Res.}\
  }\textbf {\bibinfo {volume} {69}},\ \bibinfo {pages} {435} (\bibinfo {year}
  {2011})}\BibitemShut {NoStop}%
\bibitem [{\citenamefont {Balasuriya}(2011)}]{tangential}%
  \BibitemOpen
  \bibfield  {author} {\bibinfo {author} {\bibfnamefont {S.}~\bibnamefont
  {Balasuriya}},\ }\href@noop {} {\bibfield  {journal} {\bibinfo  {journal}
  {SIAM J. Appl. Dyn. Sys.}\ }\textbf {\bibinfo {volume} {10}},\ \bibinfo
  {pages} {1100} (\bibinfo {year} {2011})}\BibitemShut {NoStop}%
\bibitem [{\citenamefont {Wiggins}(1992)}]{wiggins}%
  \BibitemOpen
  \bibfield  {author} {\bibinfo {author} {\bibfnamefont {S.}~\bibnamefont
  {Wiggins}},\ }\href@noop {} {\emph {\bibinfo {title} {Chaotic Transport in
  Dynamical Systems}}}\ (\bibinfo  {publisher} {Springer-Verlag},\ \bibinfo
  {address} {New York},\ \bibinfo {year} {1992})\BibitemShut {NoStop}%
\bibitem [{\citenamefont {Balasuriya}(2005{\natexlab{c}})}]{periodic}%
  \BibitemOpen
  \bibfield  {author} {\bibinfo {author} {\bibfnamefont {S.}~\bibnamefont
  {Balasuriya}},\ }\href@noop {} {\bibfield  {journal} {\bibinfo  {journal}
  {SIAM J.\ Appl.\ Dyn.\ Sys.}\ }\textbf {\bibinfo {volume} {4}},\ \bibinfo
  {pages} {282} (\bibinfo {year} {2005}{\natexlab{c}})}\BibitemShut {NoStop}%
\bibitem [{\citenamefont {Balasuriya}(2006)}]{aperiodic}%
  \BibitemOpen
  \bibfield  {author} {\bibinfo {author} {\bibfnamefont {S.}~\bibnamefont
  {Balasuriya}},\ }\href@noop {} {\bibfield  {journal} {\bibinfo  {journal}
  {Nonlinearity}\ }\textbf {\bibinfo {volume} {19}},\ \bibinfo {pages} {2775}
  (\bibinfo {year} {2006})}\BibitemShut {NoStop}%
\bibitem [{\citenamefont {Balasuriya}(2016{\natexlab{c}})}]{impulsive}%
  \BibitemOpen
  \bibfield  {author} {\bibinfo {author} {\bibfnamefont {S.}~\bibnamefont
  {Balasuriya}},\ }\href@noop {} {\ ,\ \bibinfo {pages} {submitted} (\bibinfo
  {year} {2016}{\natexlab{c}})}\BibitemShut {NoStop}%
\bibitem [{Note2()}]{Note2}%
  \BibitemOpen
  \bibinfo {note} {Ignoring the tangential displacement is reasonable in the
  sense that the streakline has $ p $ as its parameter varying in precisely the
  tangential direction, and when plotting $ \unhbox \voidb@x \hbox {\protect
  \boldmath $x$}_\varepsilon ^u(p,t) $ for a range of $ p $ to obtain the
  streakline curve, any tangential displacement will be barely visible \cite
  {tangential}. This is indeed reflected in the examples presented here. If
  needed, tangential displacements can be quantified \cite {tangential}, but
  this is unwieldy.}\BibitemShut {Stop}%
\bibitem [{\citenamefont {Poje}\ and\ \citenamefont
  {Haller}(1999)}]{pojehaller}%
  \BibitemOpen
  \bibfield  {author} {\bibinfo {author} {\bibfnamefont {A.}~\bibnamefont
  {Poje}}\ and\ \bibinfo {author} {\bibfnamefont {G.}~\bibnamefont {Haller}},\
  }\href@noop {} {\bibfield  {journal} {\bibinfo  {journal} {J.\ Phys.\
  Oceanography}\ }\textbf {\bibinfo {volume} {29}},\ \bibinfo {pages} {1649}
  (\bibinfo {year} {1999})}\BibitemShut {NoStop}%
\bibitem [{\citenamefont {Miller}\ \emph {et~al.}(1997)\citenamefont {Miller},
  \citenamefont {Jones}, \citenamefont {Rogerson},\ and\ \citenamefont
  {Pratt}}]{miller}%
  \BibitemOpen
  \bibfield  {author} {\bibinfo {author} {\bibfnamefont {P.}~\bibnamefont
  {Miller}}, \bibinfo {author} {\bibfnamefont {C.}~\bibnamefont {Jones}},
  \bibinfo {author} {\bibfnamefont {A.}~\bibnamefont {Rogerson}}, \ and\
  \bibinfo {author} {\bibfnamefont {L.}~\bibnamefont {Pratt}},\ }\href@noop {}
  {\bibfield  {journal} {\bibinfo  {journal} {Phys.\ D}\ }\textbf {\bibinfo
  {volume} {110}},\ \bibinfo {pages} {105} (\bibinfo {year}
  {1997})}\BibitemShut {NoStop}%
\bibitem [{\citenamefont {Fu}\ \emph {et~al.}(2005)\citenamefont {Fu},
  \citenamefont {Yang}, \citenamefont {Lin},\ and\ \citenamefont {Chien}}]{fu}%
  \BibitemOpen
  \bibfield  {author} {\bibinfo {author} {\bibfnamefont {L.-M.}\ \bibnamefont
  {Fu}}, \bibinfo {author} {\bibfnamefont {R.-J.}\ \bibnamefont {Yang}},
  \bibinfo {author} {\bibfnamefont {C.-H.}\ \bibnamefont {Lin}}, \ and\
  \bibinfo {author} {\bibfnamefont {Y.-S.}\ \bibnamefont {Chien}},\ }\href@noop
  {} {\bibfield  {journal} {\bibinfo  {journal} {Electrophoresis}\ }\textbf
  {\bibinfo {volume} {5}},\ \bibinfo {pages} {1814} (\bibinfo {year}
  {2005})}\BibitemShut {NoStop}%
\bibitem [{\citenamefont {Rodrigo}\ \emph {et~al.}(2006)\citenamefont
  {Rodrigo}, \citenamefont {Rodrigues}, \citenamefont {Formiga},\ and\
  \citenamefont {Mota}}]{rodrigo}%
  \BibitemOpen
  \bibfield  {author} {\bibinfo {author} {\bibfnamefont {A.}~\bibnamefont
  {Rodrigo}}, \bibinfo {author} {\bibfnamefont {R.}~\bibnamefont {Rodrigues}},
  \bibinfo {author} {\bibfnamefont {N.}~\bibnamefont {Formiga}}, \ and\
  \bibinfo {author} {\bibfnamefont {J.}~\bibnamefont {Mota}},\ }\href@noop {}
  {\bibfield  {journal} {\bibinfo  {journal} {Chem. Eng. Comm.}\ }\textbf
  {\bibinfo {volume} {193}},\ \bibinfo {pages} {743} (\bibinfo {year}
  {2006})}\BibitemShut {NoStop}%
\bibitem [{\citenamefont {Shin}\ \emph {et~al.}(2005)\citenamefont {Shin},
  \citenamefont {Kang},\ and\ \citenamefont {Cho}}]{shin}%
  \BibitemOpen
  \bibfield  {author} {\bibinfo {author} {\bibfnamefont {S.}~\bibnamefont
  {Shin}}, \bibinfo {author} {\bibfnamefont {I.}~\bibnamefont {Kang}}, \ and\
  \bibinfo {author} {\bibfnamefont {Y.-K.}\ \bibnamefont {Cho}},\ }\href@noop
  {} {\bibfield  {journal} {\bibinfo  {journal} {J. Micromech. Microeng.}\
  }\textbf {\bibinfo {volume} {15}},\ \bibinfo {pages} {455} (\bibinfo {year}
  {2005})}\BibitemShut {NoStop}%
\bibitem [{\citenamefont {Wang}\ \emph {et~al.}(2008)\citenamefont {Wang},
  \citenamefont {Zhe}, \citenamefont {Chung},\ and\ \citenamefont
  {Dutta}}]{wangmagnet}%
  \BibitemOpen
  \bibfield  {author} {\bibinfo {author} {\bibfnamefont {Y.}~\bibnamefont
  {Wang}}, \bibinfo {author} {\bibfnamefont {J.}~\bibnamefont {Zhe}}, \bibinfo
  {author} {\bibfnamefont {B.}~\bibnamefont {Chung}}, \ and\ \bibinfo {author}
  {\bibfnamefont {P.}~\bibnamefont {Dutta}},\ }\href@noop {} {\bibfield
  {journal} {\bibinfo  {journal} {Microfluid Nanofluid}\ }\textbf {\bibinfo
  {volume} {4}},\ \bibinfo {pages} {375} (\bibinfo {year} {2008})}\BibitemShut
  {NoStop}%
\bibitem [{\citenamefont {Lim}\ \emph {et~al.}(2010)\citenamefont {Lim},
  \citenamefont {Lam},\ and\ \citenamefont {Yang}}]{lim}%
  \BibitemOpen
  \bibfield  {author} {\bibinfo {author} {\bibfnamefont {C.}~\bibnamefont
  {Lim}}, \bibinfo {author} {\bibfnamefont {Y.}~\bibnamefont {Lam}}, \ and\
  \bibinfo {author} {\bibfnamefont {C.}~\bibnamefont {Yang}},\ }\href@noop {}
  {\bibfield  {journal} {\bibinfo  {journal} {Biomicrofluidics}\ }\textbf
  {\bibinfo {volume} {4}},\ \bibinfo {pages} {014101} (\bibinfo {year}
  {2010})}\BibitemShut {NoStop}%
\bibitem [{\citenamefont {Wang}\ \emph {et~al.}(2009)\citenamefont {Wang},
  \citenamefont {Jiao}, \citenamefont {Huang}, \citenamefont {Yang},\ and\
  \citenamefont {Nguyen}}]{wangacoustic}%
  \BibitemOpen
  \bibfield  {author} {\bibinfo {author} {\bibfnamefont {S.}~\bibnamefont
  {Wang}}, \bibinfo {author} {\bibfnamefont {Z.}~\bibnamefont {Jiao}}, \bibinfo
  {author} {\bibfnamefont {X.}~\bibnamefont {Huang}}, \bibinfo {author}
  {\bibfnamefont {C.}~\bibnamefont {Yang}}, \ and\ \bibinfo {author}
  {\bibfnamefont {N.}~\bibnamefont {Nguyen}},\ }\href@noop {} {\bibfield
  {journal} {\bibinfo  {journal} {Microfluid Nanofluid}\ }\textbf {\bibinfo
  {volume} {6}},\ \bibinfo {pages} {847} (\bibinfo {year} {2009})}\BibitemShut
  {NoStop}%
\bibitem [{\citenamefont {Song}\ \emph {et~al.}(2010)\citenamefont {Song},
  \citenamefont {Cai}, \citenamefont {Noh},\ and\ \citenamefont
  {Bennett}}]{song}%
  \BibitemOpen
  \bibfield  {author} {\bibinfo {author} {\bibfnamefont {H.}~\bibnamefont
  {Song}}, \bibinfo {author} {\bibfnamefont {Z.}~\bibnamefont {Cai}}, \bibinfo
  {author} {\bibfnamefont {H.}~\bibnamefont {Noh}}, \ and\ \bibinfo {author}
  {\bibfnamefont {D.}~\bibnamefont {Bennett}},\ }\href@noop {} {\bibfield
  {journal} {\bibinfo  {journal} {Lab Chip}\ }\textbf {\bibinfo {volume}
  {10}},\ \bibinfo {pages} {734} (\bibinfo {year} {2010})}\BibitemShut
  {NoStop}%
\bibitem [{\citenamefont {Suzuki}\ \emph {et~al.}(2004)\citenamefont {Suzuki},
  \citenamefont {Ho},\ and\ \citenamefont {Kasagi}}]{suzuki}%
  \BibitemOpen
  \bibfield  {author} {\bibinfo {author} {\bibfnamefont {H.}~\bibnamefont
  {Suzuki}}, \bibinfo {author} {\bibfnamefont {C.-M.}\ \bibnamefont {Ho}}, \
  and\ \bibinfo {author} {\bibfnamefont {N.}~\bibnamefont {Kasagi}},\
  }\href@noop {} {\bibfield  {journal} {\bibinfo  {journal} {J.
  Microelectromechanical Sys.}\ }\textbf {\bibinfo {volume} {13}},\ \bibinfo
  {pages} {779} (\bibinfo {year} {2004})}\BibitemShut {NoStop}%
\bibitem [{\citenamefont {Niu}\ \emph {et~al.}(2006{\natexlab{b}})\citenamefont
  {Niu}, \citenamefont {Liu}, \citenamefont {Wen},\ and\ \citenamefont
  {Sheng}}]{niu}%
  \BibitemOpen
  \bibfield  {author} {\bibinfo {author} {\bibfnamefont {X.}~\bibnamefont
  {Niu}}, \bibinfo {author} {\bibfnamefont {L.}~\bibnamefont {Liu}}, \bibinfo
  {author} {\bibfnamefont {W.}~\bibnamefont {Wen}}, \ and\ \bibinfo {author}
  {\bibfnamefont {P.}~\bibnamefont {Sheng}},\ }\href@noop {} {\bibfield
  {journal} {\bibinfo  {journal} {Appl. Phys. Lett.}\ }\textbf {\bibinfo
  {volume} {88}},\ \bibinfo {pages} {153508} (\bibinfo {year}
  {2006}{\natexlab{b}})}\BibitemShut {NoStop}%
\bibitem [{\citenamefont {Karrasch}(2016)}]{karrasch}%
  \BibitemOpen
  \bibfield  {author} {\bibinfo {author} {\bibfnamefont {D.}~\bibnamefont
  {Karrasch}},\ }\href@noop {} {\bibfield  {journal} {\bibinfo  {journal} {SIAM
  J. Appl. Math.}\ ,\ \bibinfo {pages} {accepted}} (\bibinfo {year}
  {2016})}\BibitemShut {NoStop}%
\end{thebibliography}
\end{document}